\documentclass[twocolumn]{aastex631}

\usepackage{amsmath}
\usepackage{hyperref}
\usepackage{booktabs}
\usepackage{longtable}
\usepackage{savesym}
\savesymbol{tablenum}
\usepackage{placeins}
\usepackage{siunitx}
\usepackage[T1]{tipa}
\usepackage{CJK}

\usepackage{threeparttable}
\restoresymbol{SIX}{tablenum}
\DeclareSIUnit{\parsec}{pc}
\DeclareSIUnit{\jansky}{Jy}
\DeclareSIUnit{\dmunit}{pc cm^{-3}}

\DeclareTextFontCommand{\textipa}{%
  \fontfamily{cmss}\tipaencoding
}

\newcommand{\kkoname}{k'ni\textipa{P}atn k'l$\left._\mathrm{\smile}\right.$stk'masqt}

\newcommand{\dmh}{\ensuremath{{\rm DM}_\textrm{h}}}
\newcommand{\pox}{\ensuremath{P(O|x)}}

\newcommand{\msbold}[1]{#1}
\received{\today}%
\revised{\today}
\accepted{\today}

\submitjournal{ApJS}

\shorttitle{First Host Galaxies from CHIME/FRB Outriggers}
\shortauthors{CHIME/FRB Collaboration et al.}
\graphicspath{{./}{figures/}}

\begin{document}
\begin{CJK*}{UTF8}{gbsn}

\title{A Catalog of Local Universe Fast Radio Bursts from CHIME/FRB and the KKO Outrigger}

\author[0000-0001-6523-9029]{The CHIME/FRB Collaboration: Mandana Amiri}
  \affiliation{Department of Physics and Astronomy, University of British Columbia, 6224 Agricultural Road, Vancouver, BC V6T 1Z1 Canada}
\author[0009-0005-5370-7653]{Daniel Amouyal}
  \affiliation{Department of Physics, McGill University, 3600 rue University, Montr\'eal, QC H3A 2T8, Canada}
  \affiliation{Trottier Space Institute, McGill University, 3550 rue University, Montr\'eal, QC H3A 2A7, Canada}
\author[0000-0001-5908-3152]{Bridget C.~Andersen}
  \affiliation{Department of Physics, McGill University, 3600 rue University, Montr\'eal, QC H3A 2T8, Canada}
  \affiliation{Trottier Space Institute, McGill University, 3550 rue University, Montr\'eal, QC H3A 2A7, Canada}
\author[0000-0002-3980-815X]{Shion Andrew}
  \affiliation{MIT Kavli Institute for Astrophysics and Space Research, Massachusetts Institute of Technology, 77 Massachusetts Ave, Cambridge, MA 02139, USA}
  \affiliation{Department of Physics, Massachusetts Institute of Technology, 77 Massachusetts Ave, Cambridge, MA 02139, USA}
\author[0000-0003-3772-2798]{Kevin Bandura}
  \affiliation{Lane Department of Computer Science and Electrical Engineering, 1220 Evansdale Drive, PO Box 6109, Morgantown, WV 26506, USA}
  \affiliation{Center for Gravitational Waves and Cosmology, West Virginia University, Chestnut Ridge Research Building, Morgantown, WV 26505, USA}
\author[0000-0002-3615-3514]{Mohit Bhardwaj}
  \affiliation{McWilliams Center for Cosmology \& Astrophysics, Department of Physics, Carnegie Mellon University, Pittsburgh, PA 15213, USA}
\author[0000-0001-8537-9299]{P.~J.~Boyle}
  \affiliation{Department of Physics, McGill University, 3600 rue University, Montr\'eal, QC H3A 2T8, Canada}
\author[0000-0002-1800-8233]{Charanjot Brar}
  \affiliation{National Research Council of Canada, Herzberg Astronomy and Astrophysics, 5071 West Saanich Road, Victoria, BC V9E2E7, Canada}
\author[0009-0007-0757-9800]{Alyssa Cassity}
  \affiliation{Department of Physics and Astronomy, University of British Columbia, 6224 Agricultural Road, Vancouver, BC V6T 1Z1 Canada}
\author[0000-0002-2878-1502]{Shami Chatterjee}
  \affiliation{Cornell Center for Astrophysics and Planetary Science, Cornell University, Ithaca, NY 14853, USA}
\author[0000-0002-8376-1563]{Alice P.~Curtin}
  \affiliation{Department of Physics, McGill University, 3600 rue University, Montr\'eal, QC H3A 2T8, Canada}
  \affiliation{Trottier Space Institute, McGill University, 3550 rue University, Montr\'eal, QC H3A 2A7, Canada}
\author[0000-0001-7166-6422]{Matt Dobbs}
  \affiliation{Department of Physics, McGill University, 3600 rue University, Montr\'eal, QC H3A 2T8, Canada}
  \affiliation{Trottier Space Institute, McGill University, 3550 rue University, Montr\'eal, QC H3A 2A7, Canada}
\author[0000-0003-4098-5222]{Fengqiu Adam Dong}
  \affiliation{Department of Physics and Astronomy, University of British Columbia, 6224 Agricultural Road, Vancouver, BC V6T 1Z1 Canada}
  \affiliation{National Radio Astronomy Observatory, 520 Edgemont Rd, Charlottesville, VA 22903, USA}
\author[0000-0002-9363-8606]{Yuxin Dong (董雨欣)}
  \affiliation{Department of Physics and Astronomy, Northwestern University, Evanston, IL 60208, USA}
  \affiliation{Center for Interdisciplinary Exploration and Research in Astronomy, Northwestern University, 1800 Sherman Avenue, Evanston, IL 60201, USA }
\author[0000-0003-3734-8177]{Gwendolyn M.~Eadie}
  \affiliation{David A. Dunlap Department of Astronomy and Astrophysics, 50 St. George Street, University of Toronto, ON M5S 3H4, Canada}
  \affiliation{Department of Statistical Science, University of Toronto, Ontario Power Building, 700 University Avenue, 9th Floor, Toronto, ON M5G 1Z5, Toronto, Ontario, Canada}
\author[0000-0003-0307-9984]{Tarraneh Eftekhari}
  \affiliation{Center for Interdisciplinary Exploration and Research in Astronomy, Northwestern University, 1800 Sherman Avenue, Evanston, IL 60201, USA }
  \affiliation{NHFP Einstein Fellow}
\author[0000-0002-7374-935X]{Wen-fai Fong}
  \affiliation{Department of Physics and Astronomy, Northwestern University, Evanston, IL 60208, USA}
  \affiliation{Center for Interdisciplinary Exploration and Research in Astronomy, Northwestern University, 1800 Sherman Avenue, Evanston, IL 60201, USA }
\author[0000-0001-8384-5049]{Emmanuel Fonseca}
  \affiliation{Department of Physics and Astronomy, West Virginia University, PO Box 6315, Morgantown, WV 26506, USA }
  \affiliation{Center for Gravitational Waves and Cosmology, West Virginia University, Chestnut Ridge Research Building, Morgantown, WV 26505, USA}
\author[0000-0002-3382-9558]{B.~M.~Gaensler}
  \affiliation{Department of Astronomy and Astrophysics, University of California Santa Cruz, 1156 High Street, Santa Cruz, CA 95060, USA}
  \affiliation{Dunlap Institute for Astronomy and Astrophysics, 50 St. George Street, University of Toronto, ON M5S 3H4, Canada}
  \affiliation{David A. Dunlap Department of Astronomy and Astrophysics, 50 St. George Street, University of Toronto, ON M5S 3H4, Canada}
\author[0000-0002-1760-0868]{Mark Halpern}
  \affiliation{Department of Physics and Astronomy, University of British Columbia, 6224 Agricultural Road, Vancouver, BC V6T 1Z1 Canada}
\author[0000-0003-2317-1446]{Jason~W.~T.~Hessels}
  \affiliation{Department of Physics, McGill University, 3600 rue University, Montr\'eal, QC H3A 2T8, Canada}
  \affiliation{Trottier Space Institute, McGill University, 3550 rue University, Montr\'eal, QC H3A 2A7, Canada}
  \affiliation{Anton Pannekoek Institute for Astronomy, University of Amsterdam, Science Park 904, 1098 XH Amsterdam, The Netherlands}
  \affiliation{ASTRON, Netherlands Institute for Radio Astronomy, Oude Hoogeveensedijk 4, 7991 PD Dwingeloo, The Netherlands}
\author[0009-0002-1199-8876]{Hans Hopkins}
  \affiliation{Perimeter Institute of Theoretical Physics, 31 Caroline Street North, Waterloo, ON N2L 2Y5, Canada}
\author[0000-0003-2405-2967]{Adaeze L.~Ibik}
  \affiliation{Dunlap Institute for Astronomy and Astrophysics, 50 St. George Street, University of Toronto, ON M5S 3H4, Canada}
  \affiliation{David A. Dunlap Department of Astronomy and Astrophysics, 50 St. George Street, University of Toronto, ON M5S 3H4, Canada}
\author[0000-0003-3457-4670]{Ronniy C.~Joseph}
  \affiliation{Department of Physics, McGill University, 3600 rue University, Montr\'eal, QC H3A 2T8, Canada}
  \affiliation{Trottier Space Institute, McGill University, 3550 rue University, Montr\'eal, QC H3A 2A7, Canada}
\author[0000-0003-4810-7803]{Jane Kaczmarek}
  \affiliation{CSIRO Space \& Astronomy, Parkes Observatory, P.O. Box 276, Parkes NSW 2870, Australia}
\author[0009-0007-5296-4046]{Lordrick Kahinga}
  \affiliation{Department of Astronomy and Astrophysics, University of California Santa Cruz, 1156 High Street, Santa Cruz, CA 95060, USA}
\author[0000-0001-9345-0307]{Victoria Kaspi}
  \affiliation{Department of Physics, McGill University, 3600 rue University, Montr\'eal, QC H3A 2T8, Canada}
  \affiliation{Trottier Space Institute, McGill University, 3550 rue University, Montr\'eal, QC H3A 2A7, Canada}
\author[0009-0005-7115-3447]{Kholoud Khairy}
  \affiliation{Lane Department of Computer Science and Electrical Engineering, 1220 Evansdale Drive, PO Box 6109, Morgantown, WV 26506, USA}
  \affiliation{Center for Gravitational Waves and Cosmology, West Virginia University, Chestnut Ridge Research Building, Morgantown, WV 26505, USA}
\author[0000-0002-5740-7747]{Charles D.~Kilpatrick}
  \affiliation{Center for Interdisciplinary Exploration and Research in Astronomy, Northwestern University, 1800 Sherman Avenue, Evanston, IL 60201, USA }
\author[0000-0003-2116-3573]{Adam E.~Lanman}
  \affiliation{MIT Kavli Institute for Astrophysics and Space Research, Massachusetts Institute of Technology, 77 Massachusetts Ave, Cambridge, MA 02139, USA}
  \affiliation{Department of Physics, Massachusetts Institute of Technology, 77 Massachusetts Ave, Cambridge, MA 02139, USA}
\author[0000-0002-5857-4264 ]{Mattias Lazda}
  \affiliation{Dunlap Institute for Astronomy and Astrophysics, 50 St. George Street, University of Toronto, ON M5S 3H4, Canada}
  \affiliation{David A. Dunlap Department of Astronomy and Astrophysics, 50 St. George Street, University of Toronto, ON M5S 3H4, Canada}
\author[0000-0002-4209-7408]{Calvin Leung}
  \affiliation{Department of Astronomy, University of California, Berkeley, CA 94720, United States}
  \affiliation{NHFP Einstein Fellow}
\author[0000-0002-7164-9507]{Robert Main}
  \affiliation{Department of Physics, McGill University, 3600 rue University, Montr\'eal, QC H3A 2T8, Canada}
\author[0000-0003-4584-8841]{Lluis Mas-Ribas}
  \affiliation{Department of Astronomy and Astrophysics, University of California Santa Cruz, 1156 High Street, Santa Cruz, CA 95060, USA}
  \affiliation{University of California Observatories, 1156 High Street, Santa Cruz, CA 95060, USA}
\author[0000-0002-4279-6946]{Kiyoshi W.~Masui}
  \affiliation{MIT Kavli Institute for Astrophysics and Space Research, Massachusetts Institute of Technology, 77 Massachusetts Ave, Cambridge, MA 02139, USA}
  \affiliation{Department of Physics, Massachusetts Institute of Technology, 77 Massachusetts Ave, Cambridge, MA 02139, USA}
\author[0000-0001-7348-6900]{Ryan Mckinven}
  \affiliation{Department of Physics, McGill University, 3600 rue University, Montr\'eal, QC H3A 2T8, Canada}
  \affiliation{Trottier Space Institute, McGill University, 3550 rue University, Montr\'eal, QC H3A 2A7, Canada}
\author[0000-0002-0772-9326]{Juan Mena-Parra}
  \affiliation{Dunlap Institute for Astronomy and Astrophysics, 50 St. George Street, University of Toronto, ON M5S 3H4, Canada}
  \affiliation{David A. Dunlap Department of Astronomy and Astrophysics, 50 St. George Street, University of Toronto, ON M5S 3H4, Canada}
\author[0000-0001-8845-1225]{Bradley W.~Meyers}
  \affiliation{International Centre for Radio Astronomy Research (ICRAR), Curtin University, Bentley WA 6102 Australia}
\author[0000-0002-2551-7554]{Daniele Michilli}
  \affiliation{MIT Kavli Institute for Astrophysics and Space Research, Massachusetts Institute of Technology, 77 Massachusetts Ave, Cambridge, MA 02139, USA}
  \affiliation{Department of Physics, Massachusetts Institute of Technology, 77 Massachusetts Ave, Cambridge, MA 02139, USA}
\author[0000-0001-8292-0051]{Nikola Milutinovic}
  \affiliation{Department of Physics and Astronomy, University of British Columbia, 6224 Agricultural Road, Vancouver, BC V6T 1Z1 Canada}
\author[0000-0003-0510-0740]{Kenzie Nimmo}
  \affiliation{MIT Kavli Institute for Astrophysics and Space Research, Massachusetts Institute of Technology, 77 Massachusetts Ave, Cambridge, MA 02139, USA}
\author[0000-0002-5254-243X]{Gavin Noble}
  \affiliation{David A. Dunlap Department of Astronomy and Astrophysics, 50 St. George Street, University of Toronto, ON M5S 3H4, Canada}
  \affiliation{Dunlap Institute for Astronomy and Astrophysics, 50 St. George Street, University of Toronto, ON M5S 3H4, Canada}
\author[0000-0002-8897-1973]{Ayush Pandhi}
  \affiliation{David A. Dunlap Department of Astronomy and Astrophysics, 50 St. George Street, University of Toronto, ON M5S 3H4, Canada}
  \affiliation{Dunlap Institute for Astronomy and Astrophysics, 50 St. George Street, University of Toronto, ON M5S 3H4, Canada}
\author[0009-0008-7264-1778]{Swarali Shivraj Patil}
  \affiliation{Department of Physics and Astronomy, West Virginia University, PO Box 6315, Morgantown, WV 26506, USA }
  \affiliation{Center for Gravitational Waves and Cosmology, West Virginia University, Chestnut Ridge Research Building, Morgantown, WV 26505, USA}
\author[0000-0002-8912-0732]{Aaron B.~Pearlman}
  \affiliation{Department of Physics, McGill University, 3600 rue University, Montr\'eal, QC H3A 2T8, Canada}
  \affiliation{Trottier Space Institute, McGill University, 3550 rue University, Montr\'eal, QC H3A 2A7, Canada}
  \affiliation{Banting Fellow}
  \affiliation{McGill Space Institute Fellow}
  \affiliation{FRQNT Postdoctoral Fellow}
\author[0000-0002-9822-8008]{Emily Petroff}
  \affiliation{Department of Physics, McGill University, 3600 rue University, Montr\'eal, QC H3A 2T8, Canada}
  \affiliation{Trottier Space Institute, McGill University, 3550 rue University, Montr\'eal, QC H3A 2A7, Canada}
  \affiliation{Perimeter Institute of Theoretical Physics, 31 Caroline Street North, Waterloo, ON N2L 2Y5, Canada}
\author[0000-0002-4795-697X]{Ziggy Pleunis}
  \affiliation{Anton Pannekoek Institute for Astronomy, University of Amsterdam, Science Park 904, 1098 XH Amsterdam, The Netherlands}
  \affiliation{ASTRON, Netherlands Institute for Radio Astronomy, Oude Hoogeveensedijk 4, 7991 PD Dwingeloo, The Netherlands}
\author[0000-0002-7738-6875]{J.~Xavier Prochaska}
  \affiliation{Department of Astronomy and Astrophysics, University of California Santa Cruz, 1156 High Street, Santa Cruz, CA 95060, USA}
  \affiliation{Kavli Institute for the Physics and Mathematics of the Universe (Kavli IPMU), 5-1-5 Kashiwanoha, Kashiwa, 277-8583, Japan}
  \affiliation{Division of Science, National Astronomical Observatory of Japan, 2-21-1 Osawa, Mitaka, Tokyo 181-8588, Japan}
\author[0000-0001-7694-6650]{Masoud Rafiei-Ravandi}
  \affiliation{Department of Physics, McGill University, 3600 rue University, Montr\'eal, QC H3A 2T8, Canada}
\author[0000-0003-1842-6096]{Mubdi Rahman}
  \affiliation{Sidrat Research, 124 Merton Street, Suite 507, Toronto, Ontario, M4S 2Z2, Canada}
\author[0000-0003-3463-7918]{Andre Renard}
  \affiliation{Dunlap Institute for Astronomy and Astrophysics, 50 St. George Street, University of Toronto, ON M5S 3H4, Canada}
\author[0000-0002-4623-5329]{Mawson W.~Sammons}
  \affiliation{Department of Physics, McGill University, 3600 rue University, Montr\'eal, QC H3A 2T8, Canada}
  \affiliation{Trottier Space Institute, McGill University, 3550 rue University, Montr\'eal, QC H3A 2A7, Canada}
\author[0000-0003-3154-3676]{Ketan R.~Sand}
  \affiliation{Department of Physics, McGill University, 3600 rue University, Montr\'eal, QC H3A 2T8, Canada}
  \affiliation{Trottier Space Institute, McGill University, 3550 rue University, Montr\'eal, QC H3A 2A7, Canada}
\author[0000-0002-7374-7119]{Paul Scholz}
  \affiliation{Department of Physics and Astronomy, York University, 4700 Keele Street, Toronto, ON MJ3 1P3, Canada}
  \affiliation{Dunlap Institute for Astronomy and Astrophysics, 50 St. George Street, University of Toronto, ON M5S 3H4, Canada}
\author[0000-0002-4823-1946]{Vishwangi Shah}
  \affiliation{Department of Physics, McGill University, 3600 rue University, Montr\'eal, QC H3A 2T8, Canada}
  \affiliation{Trottier Space Institute, McGill University, 3550 rue University, Montr\'eal, QC H3A 2A7, Canada}
\author[0000-0002-6823-2073]{Kaitlyn Shin}
  \affiliation{MIT Kavli Institute for Astrophysics and Space Research, Massachusetts Institute of Technology, 77 Massachusetts Ave, Cambridge, MA 02139, USA}
  \affiliation{Department of Physics, Massachusetts Institute of Technology, 77 Massachusetts Ave, Cambridge, MA 02139, USA}
\author[0000-0003-2631-6217]{Seth~R.~Siegel}
  \affiliation{Perimeter Institute of Theoretical Physics, 31 Caroline Street North, Waterloo, ON N2L 2Y5, Canada}
  \affiliation{Department of Physics, McGill University, 3600 rue University, Montr\'eal, QC H3A 2T8, Canada}
  \affiliation{Trottier Space Institute, McGill University, 3550 rue University, Montr\'eal, QC H3A 2A7, Canada}
\author[0000-0003-3801-1496]{Sunil Simha}
  \affiliation{Center for Interdisciplinary Exploration and Research in Astronomy, Northwestern University, 1800 Sherman Avenue, Evanston, IL 60201, USA }
  \affiliation{Department of Astronomy and Astrophysics, University of Chicago, William Eckhardt Research Center, 5640 S Ellis Ave, Chicago, IL 60637}
\author[0000-0002-2088-3125]{Kendrick Smith}
  \affiliation{Perimeter Institute of Theoretical Physics, 31 Caroline Street North, Waterloo, ON N2L 2Y5, Canada}
\author[0000-0001-9784-8670]{Ingrid Stairs}
  \affiliation{Department of Physics and Astronomy, University of British Columbia, 6224 Agricultural Road, Vancouver, BC V6T 1Z1 Canada}
\author[0000-0003-4535-9378]{Keith Vanderlinde}
  \affiliation{David A. Dunlap Department of Astronomy and Astrophysics, 50 St. George Street, University of Toronto, ON M5S 3H4, Canada}
  \affiliation{Dunlap Institute for Astronomy and Astrophysics, 50 St. George Street, University of Toronto, ON M5S 3H4, Canada}
\author[0000-0002-1491-3738]{Haochen Wang}
  \affiliation{MIT Kavli Institute for Astrophysics and Space Research, Massachusetts Institute of Technology, 77 Massachusetts Ave, Cambridge, MA 02139, USA}
  \affiliation{Department of Physics, Massachusetts Institute of Technology, 77 Massachusetts Ave, Cambridge, MA 02139, USA}
\author[0000-0001-7314-9496]{Dallas Wulf}
  \affiliation{Department of Physics, McGill University, 3600 rue University, Montr\'eal, QC H3A 2T8, Canada}
  \affiliation{Trottier Space Institute, McGill University, 3550 rue University, Montr\'eal, QC H3A 2A7, Canada}
\author[0000-0002-7076-8643]{Tarik J.~Zegmott}
  \affiliation{Department of Physics, McGill University, 3600 rue University, Montr\'eal, QC H3A 2T8, Canada}
\newcommand{\allacks}{
B.\,C.\,A. is supported by an FRQNT Doctoral Research Award.
FRB research at WVU is supported by an NSF grant (2006548, 2018490).
M.B is a McWilliams fellow and an International Astronomical Union Gruber fellow. M.B. also receives support from the McWilliams seed grant.
A.P.C is a Vanier Canada Graduate Scholar.
M.D.\ is supported by a CRC Chair, NSERC Discovery Grant, and CIFAR.
F.A.D is supported by a Jansky Fellowship.
Y.D. is supported by the National Science Foundation Graduate Research Fellowship under grant No. DGE-2234667.
G.M.E. holds a Collaborative Research Team grant from the Canadian Statistical Sciences Institute (CANSSI), which is supported by Natural Sciences and Engineering Research Council of Canada (NSERC), and an NSERC Discovery Grant RGPIN2020-04554.
T. E. is supported by NASA through the NASA Hubble Fellowship grant HST-HF2-51504.001-A awarded by the Space Telescope Science Institute, which is operated by the Association of Universities for Research in Astronomy, Inc., for NASA, under contract NAS5-26555.
W.F. gratefully acknowledges support by the David and Lucile Packard Foundation, the Alfred P. Sloan Foundation, the Research Corporation for Science Advancement through Cottrell Scholar Award \#28284, and the National Science Foundation under grant Nos. AST-2206494, AST-2308182, and CAREER grant No. AST-2047919.
E.F. and S.S.P. are supported by NSF grant AST-2407399.
J.W.T.H. and the AstroFlash research group acknowledge support from a Canada Excellence Research Chair in Transient Astrophysics (CERC-2022-00009); the European Research Council (ERC) under the European Union’s Horizon 2020 research and innovation programme (`EuroFlash'; Grant agreement No. 101098079); and an NWO-Vici grant (`AstroFlash'; VI.C.192.045).
V.M.K. holds the Lorne Trottier Chair in Astrophysics \& Cosmology, a Distinguished James McGill Professorship, and receives support from an NSERC Discovery grant (RGPIN 228738-13).
C. L. is supported by NASA through the NASA Hubble Fellowship grant HST-HF2-51536.001-A awarded by the Space Telescope Science Institute, which is operated by the Association of Universities for Research in Astronomy, Inc., under NASA contract NAS5-26555.
K.W.M. holds the Adam J. Burgasser Chair in Astrophysics and is supported by NSF grants (2008031, 2018490).
J.M.P. acknowledges the support of an NSERC Discovery Grant (RGPIN-2023-05373).
K.N. is an MIT Kavli Fellow.
A.P. is funded by the NSERC Canada Graduate Scholarships -- Doctoral program.
A.B.P. is a Banting Fellow, a McGill Space Institute~(MSI) Fellow, and a Fonds de Recherche du Quebec -- Nature et Technologies~(FRQNT) postdoctoral fellow.
Z.P. is supported by an NWO Veni fellowship (VI.Veni.222.295).
F4 acknowledgement
M.W.S. acknowledges support from the Trottier Space Institute Fellowship program.
K.R.S is supported by FRQNT doctoral research award.
P.S. acknowledges the support of an NSERC Discovery Grant (RGPIN-2024-06266).
V.S. is supported by a FRQNT Doctoral Research Award.
K.S. is supported by the NSF Graduate Research Fellowship Program.
S.S. is supported by the joint Northwestern University and University of Chicago Brinson Fellowship.
FRB Research at UBC is supported by an NSERC Discoverry Grant and by the Canadian Insitute for Advanced Research.
}

\correspondingauthor{Calvin Leung}
\email{calvin\_leung@berkeley.edu}
\collaboration{99}{(CHIME/FRB Collaboration)}



\begin{abstract}
We present the first catalog of fast radio burst (FRB) host galaxies from CHIME/FRB Outriggers, selected uniformly in the radio and the optical by localizing 81 new bursts to $2\arcsec \times \sim 60\arcsec$ accuracy using CHIME and the~\kkoname\ Outrigger (KKO) station, located 66 km from CHIME. Of the 81 localized bursts, we use the Probabilistic Association of Transients to their Hosts (PATH) algorithm to securely identify 21 new FRB host galaxies, and compile spectroscopic redshifts for 19 systems, 15 of which are newly obtained via spectroscopic observations. The most nearby source is FRB 20231229A, at a distance of 90 Mpc. One burst in our sample is from a previously reported repeating source in a galaxy merger (FRB 20190303A). Three new FRB host galaxies (FRBs 20230203A, 20230703A, and 20231206A) are found towards X-ray and optically selected galaxy clusters, potentially doubling the sample of known galaxy cluster FRBs. A search for radio counterparts reveals that FRB 20231128A is associated with a luminous persistent radio source (PRS) candidate with high significance ($P_{cc} \sim 10^{-2}$). If its compactness is confirmed, it would be the nearest known compact PRS at $z = 0.1079$. Our catalog significantly increases the statistics of the Macquart relation at low redshifts ($z < 0.2$). In the near future, the completed CHIME/FRB Outriggers array will produce hundreds of FRBs localized with very long baseline interferometry (VLBI). This will significantly expand the known sample and pave the way for future telescopes relying on VLBI for FRB localization.
\end{abstract}


\section{Introduction}\label{sec:intro}
Fast radio bursts (FRBs) are short-duration radio pulses of unknown origin~\citep{cordes2019fast,petroff2022fast}. While thousands of FRBs have been discovered~\citep{collaboration2021first}, their origins remain unknown and their potential as cosmological probes remains largely untapped, limited by a lack of robust burst localization and host galaxy identification. 

Since the localization of the first repeating FRB~\citep{chatterjee2017direct,tendulkar2017host}, host galaxies of FRBs have played a crucial role in constraining their origin. One of the earliest studies of FRB host galaxies used five FRB hosts localized by the Australian Square Kilometre Array Pathfinder~\citep{macquart2020census} to \msbold{claim that the progenitors of FRBs were similar to those of core-collapse and Type Ia supernovae, as well as short GRBs}~\citep{heintz2020host,bhandari2020host} by comparing the distribution of spatial offsets between FRBs and their host galaxies. This small sample was expanded to 15~\citep{bhandari2022characterizing}, and further to 23 systems including for the first time a significant number of repeating sources~\citep{gordon2023demographics}.~\citet{bhardwaj2024host} added 4 new hosts to create a volume-limited sample of 18 nearby systems. Recently, the Deep Synoptic Array has discovered 26 new host galaxies~\citep{kocz2019dsa,law2024deep,sharma2024preferential,connor2024gas} out to $z = 1.3$. In addition, the recent completion of the ASKAP Fast Transient incoherent sum survey~\citep{shannon2024commensal} has added an optically complete sample of 43 FRBs and 30 redshifts ranging from $z = 0.023$ to $1.02$, many of which have been characterized in the sample of~\citet{gordon2023demographics} \citep[see also][]{gordon2023fast}.

This paper expands on these efforts to grow the sample of host galaxies using the Canadian Hydrogen Intensity Mapping Experiment~\citep[CHIME/FRB,][]{collaboration2018chime} and the~\kkoname\ Outrigger~\citep[KKO,][]{lanman2024chime}, the first of its three outrigger stations which will use Very Long Baseline Interferometry (VLBI) to pinpoint FRBs. Early efforts with CHIME have already pinpointed a repeating source (FRB 20240209A) to the outskirts of a quiescent galaxy~\citep{shah2024repeating,eftekhari2024massive}. Here, we present the localization of 81 FRB sources using single-pulse VLBI, with a uniform VLBI calibration strategy, yielding accuracies of $\sim 2\arcsec \times 60\arcsec$ (\S\ref{sec:localization}). We systematically investigate the impact of calibrator choice on astrometric accuracy using a sample of single-pulse localizations from pulsars, finding that on this short baseline, our astrometric accuracy is unaffected by our calibrator choice (Appendix~\ref{sec:accuracy}).

Even though we use a single VLBI outrigger in this work, our localizations are sufficiently accurate and precise for 21 of the 81 bursts to be confidently associated with bright, nearby host galaxies (\S\ref{sec:association}), as quantified by the Probabilistic Association of Transients to their Hosts (PATH) framework~\citep{aggarwal2021probabilistic}, a standard technique in the field. After making host associations, in \S\ref{sec:followup} we describe our spectroscopic follow-up observations, which yield new redshifts for 15 of the 21 hosts, but leave detailed spectral characterization for future work. In \S\ref{sec:crossmatch} we describe a systematic search for multi-wavelength counterparts.
 
Finally, in \S\ref{sec:sample}, we describe the 21 hosts in the sample and highlight a few interesting cases therein. While the localizations presented here are generally not precise enough for spatially resolved followup~\citep[e.g.,][]{mannings2021high,woodland2023environments}, our sample of nearby galaxies is an ideal target for population studies of FRB hosts, which we defer to later work. 

\section{FRB Detection and Localization}\label{sec:localization}
Between December 9, 2023 and February 10, 2024, the CHIME/FRB search engine~\citep{collaboration2018chime} detected $269$ bright (S/N $> 20$) FRB candidates by searching the dynamic spectra of its 1024 detection beams~\citep{ng2017chime,masui2019algorithms} for dispersed single pulses~\citep{collaboration2018chime}. In this section we describe the search and voltage capture system used to take the data, the morphology of the bursts as seen in the data at microsecond time resolution, and describe the process by which we localize the bursts. This consists of two stages: a CHIME-only interferometric localization, and a CHIME-KKO localization using VLBI between the two stations.

\subsection{Burst detection and data acquisition}
In tandem with the FRB search engine, the correlator constantly buffers the channelized voltage (hereafter, baseband) data for all antennas in a set of ring buffers. These ring buffers await a trigger from the FRB search engine. Since the ring buffers hold data from each antenna individually, they enable the real time system to capture the full field of view of CHIME for any $\sim 1$ second snapshot within the buffer, which holds the last 38 seconds of real-time data. Following the real-time detection of FRB candidates, triggers were sent to the CHIME correlator to capture $\sim 100-500$ ms of data surrounding the FRB in time for offline data analysis. In most of these cases, data were also successfully captured at the KKO outrigger~\citep{lanman2024chime}, which was in a commissioning phase until about September 2023. 

Following the real-time detection and data capture, FRB candidates are confirmed by two humans inspecting the dynamic spectra as viewed through the detection beam. Once the candidate is confirmed, an offline pipeline forms a new set of beams distinct from those used in the real-time search engine. This offline pipeline using the baseband data (called the ``baseband localization'' pipeline) aligns the synthesized beam with the true position of the burst, which recovers signal strength with respect to the detection beam. Furthermore it allows each burst to be localized to arcminute-level accuracy~\citep{michilli2021analysis}.

\subsection{Microsecond timescale burst morphologies}
\begin{figure*}
\centering
\includegraphics[width=0.22\textwidth]{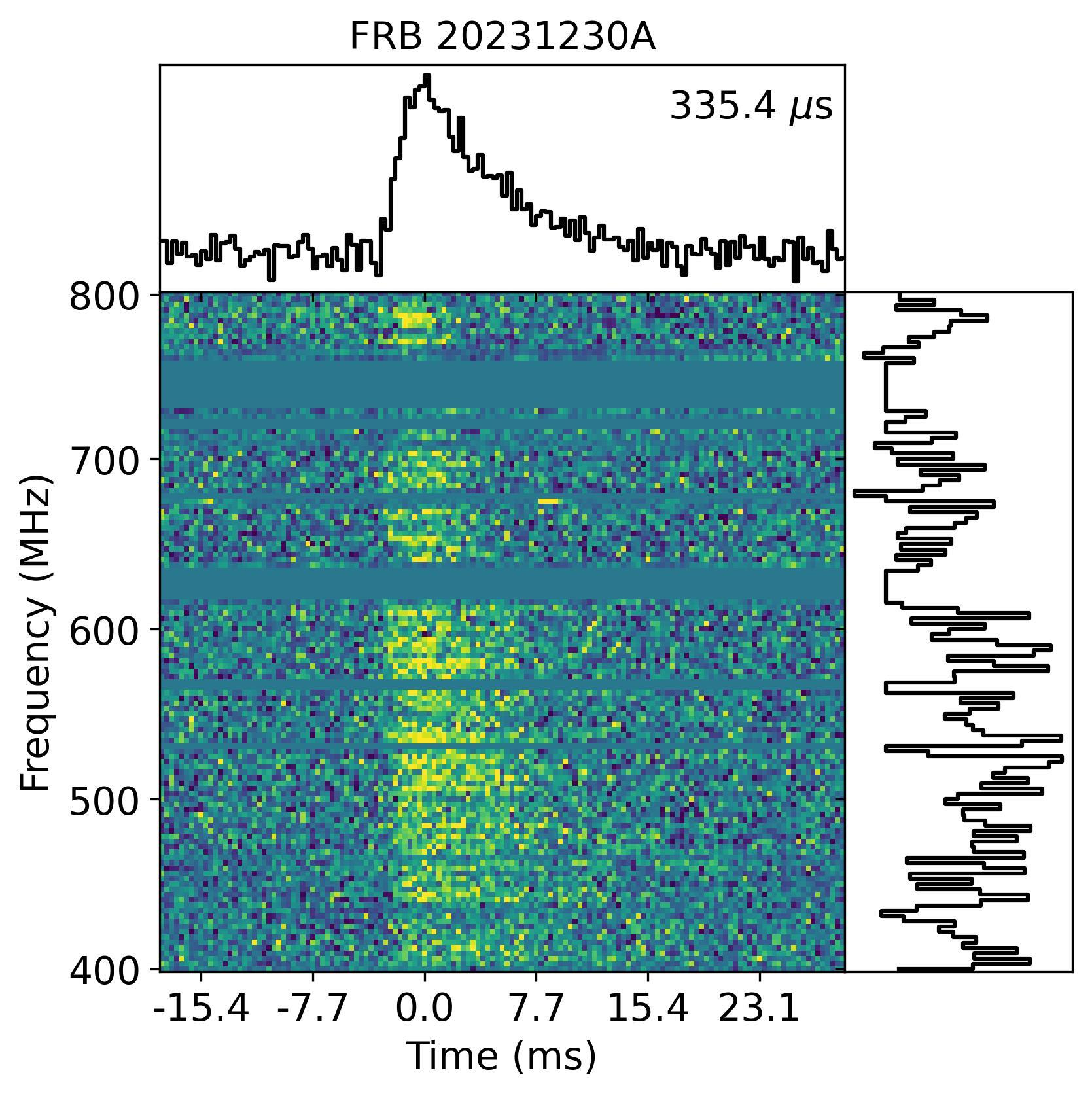}
\includegraphics[width=0.22\textwidth]{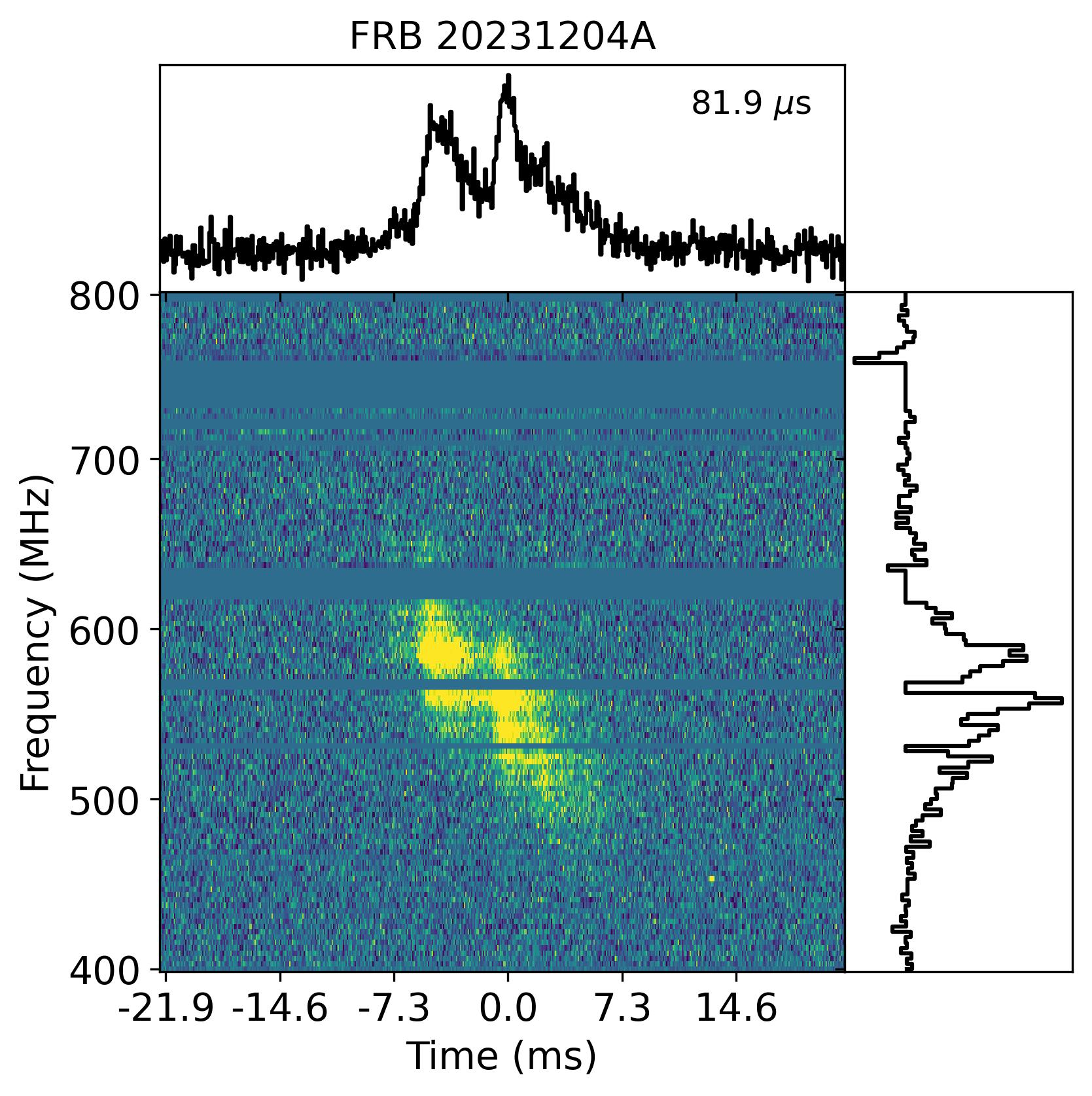}
\includegraphics[width=0.22\textwidth]{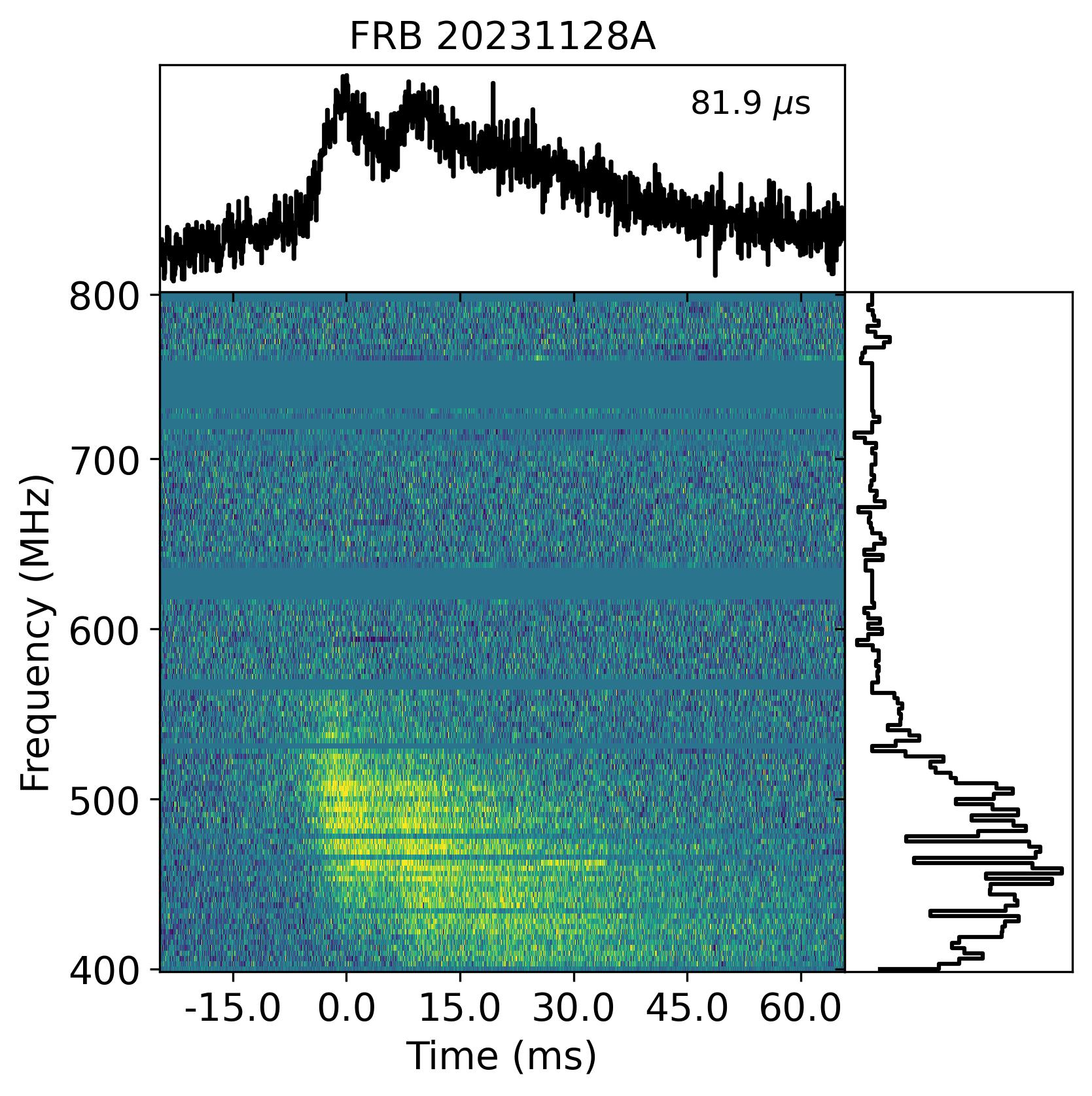}
\includegraphics[width=0.22\textwidth]{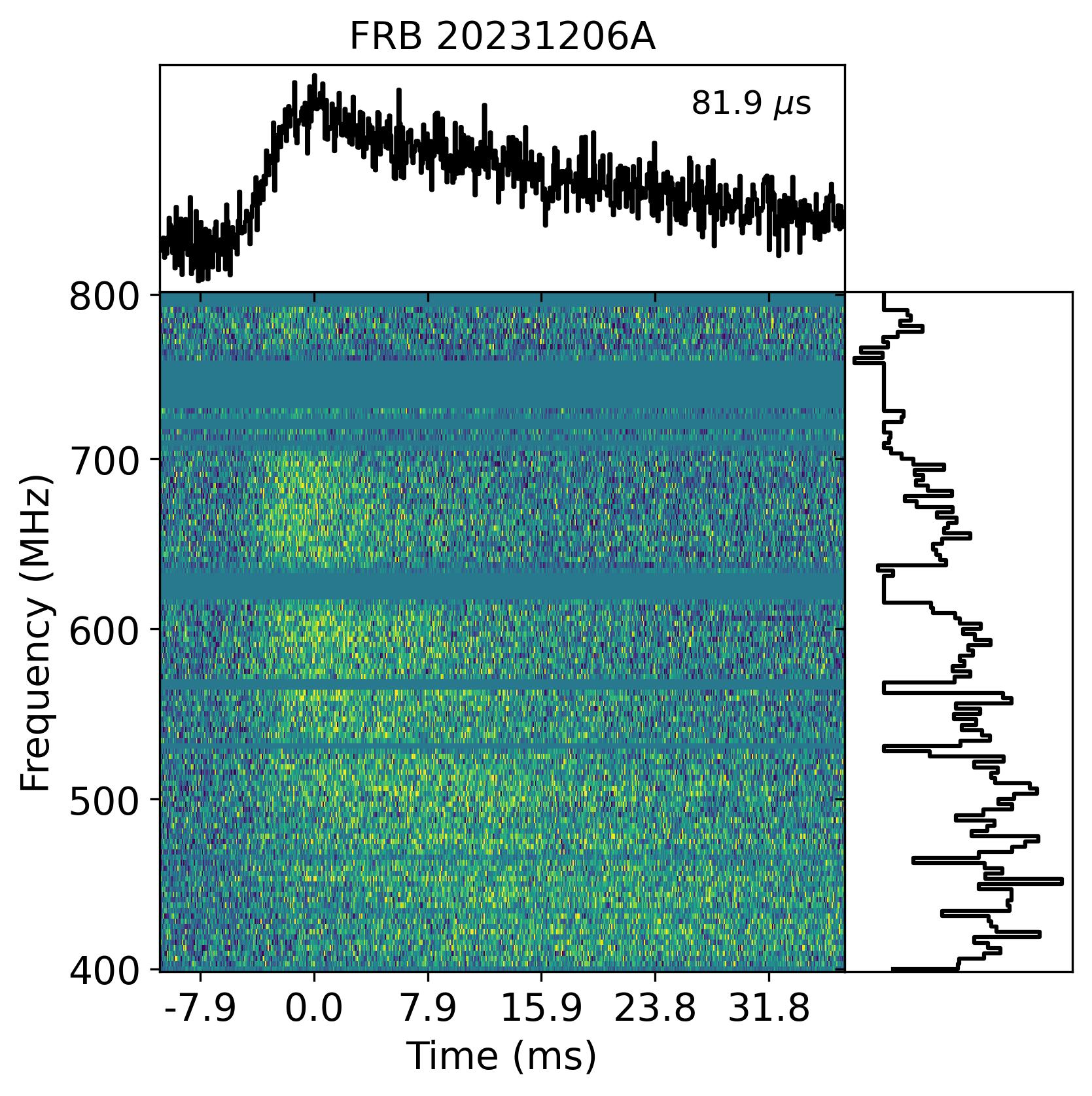}
\includegraphics[width=0.22\textwidth]{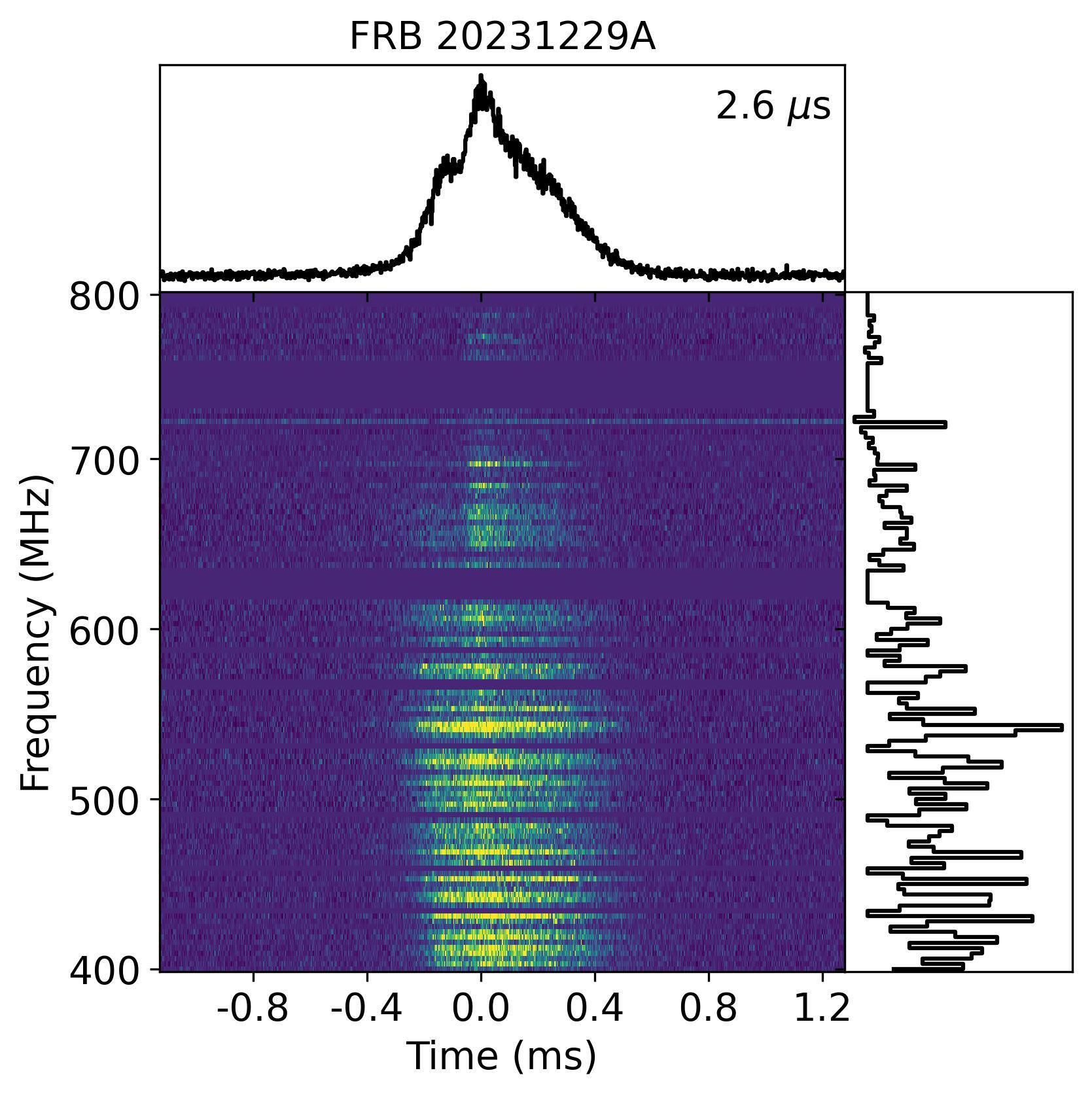}
\includegraphics[width=0.22\textwidth]{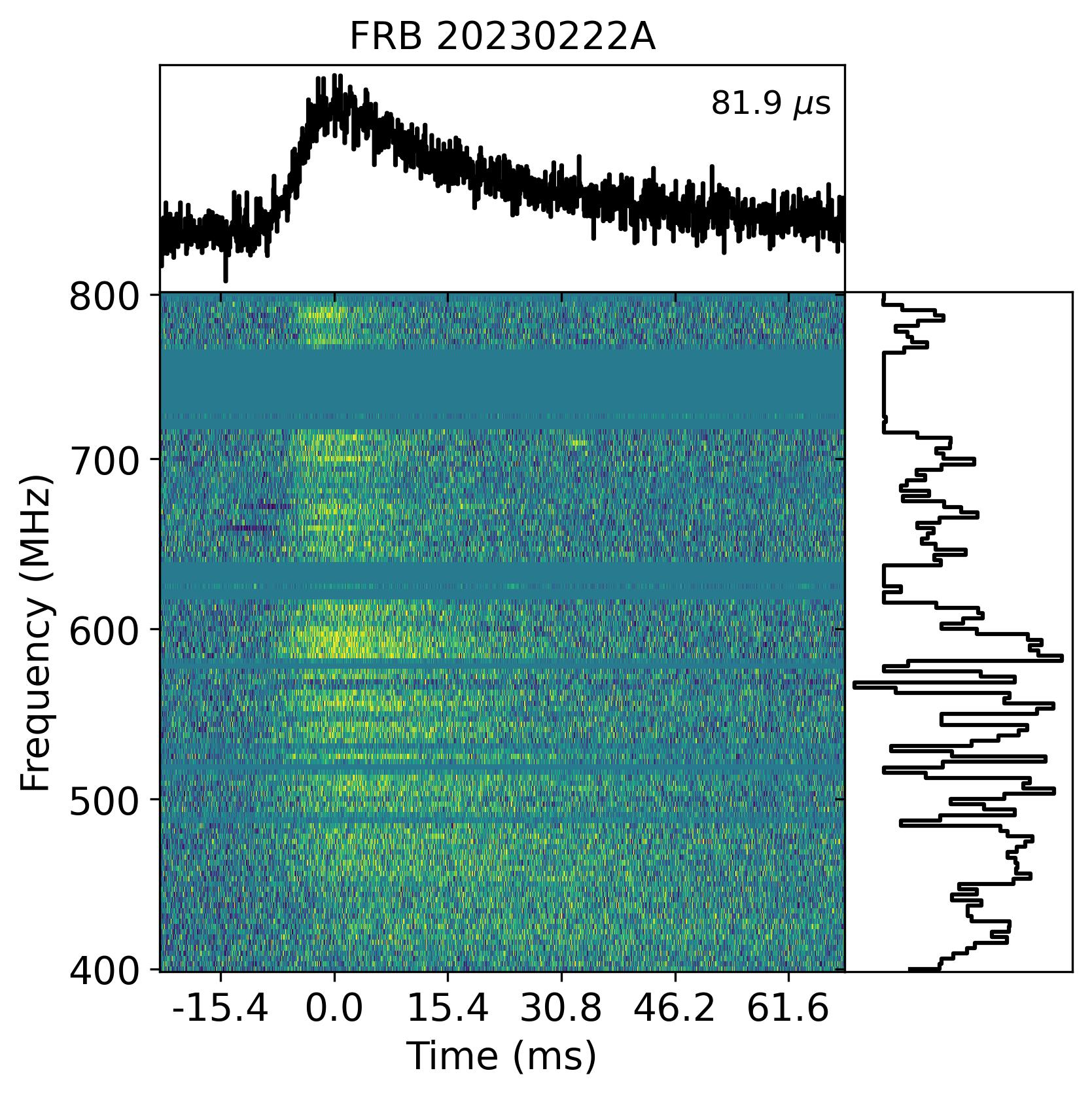}
\includegraphics[width=0.22\textwidth]{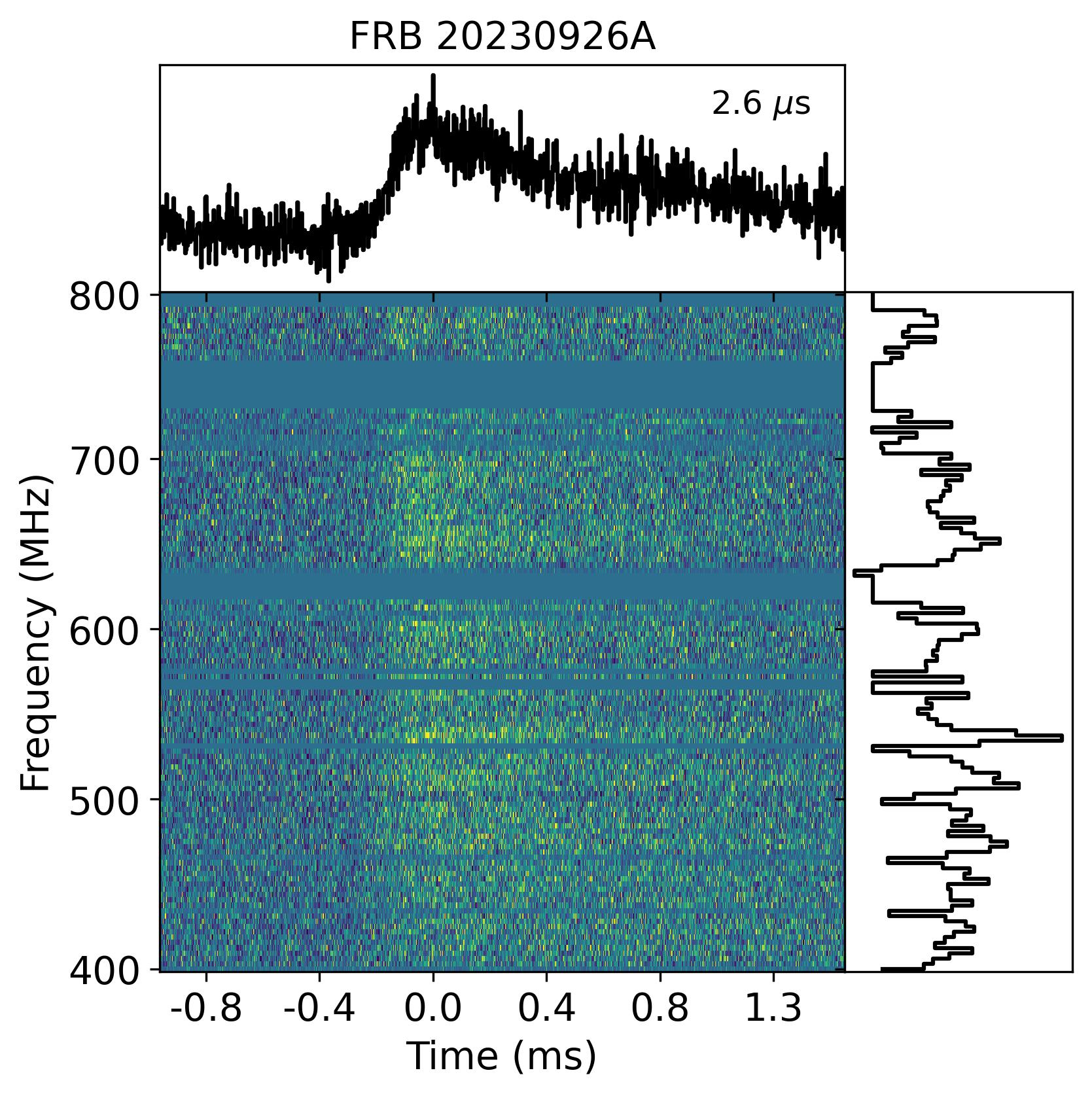}
\includegraphics[width=0.22\textwidth]{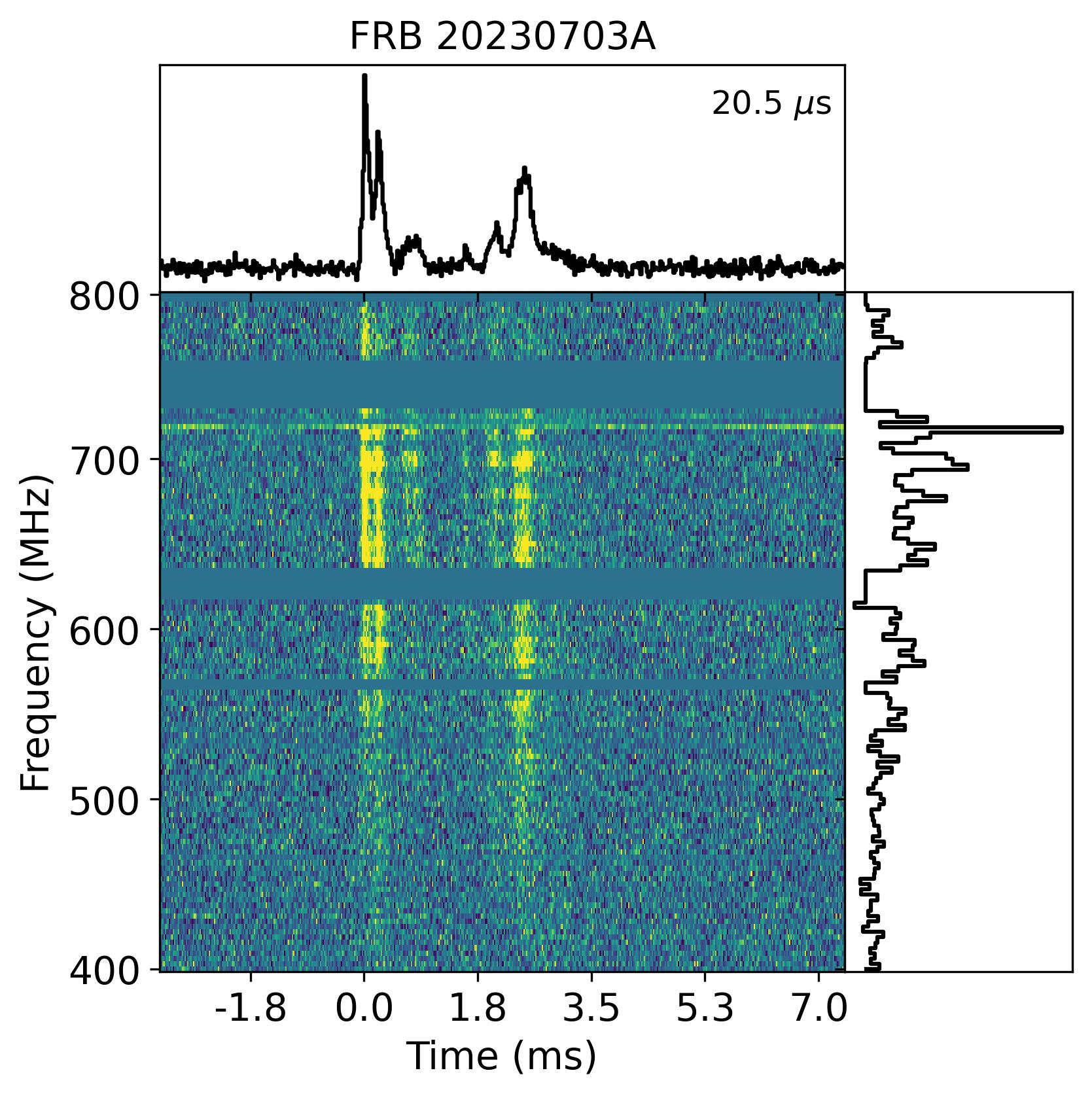}
\includegraphics[width=0.22\textwidth]{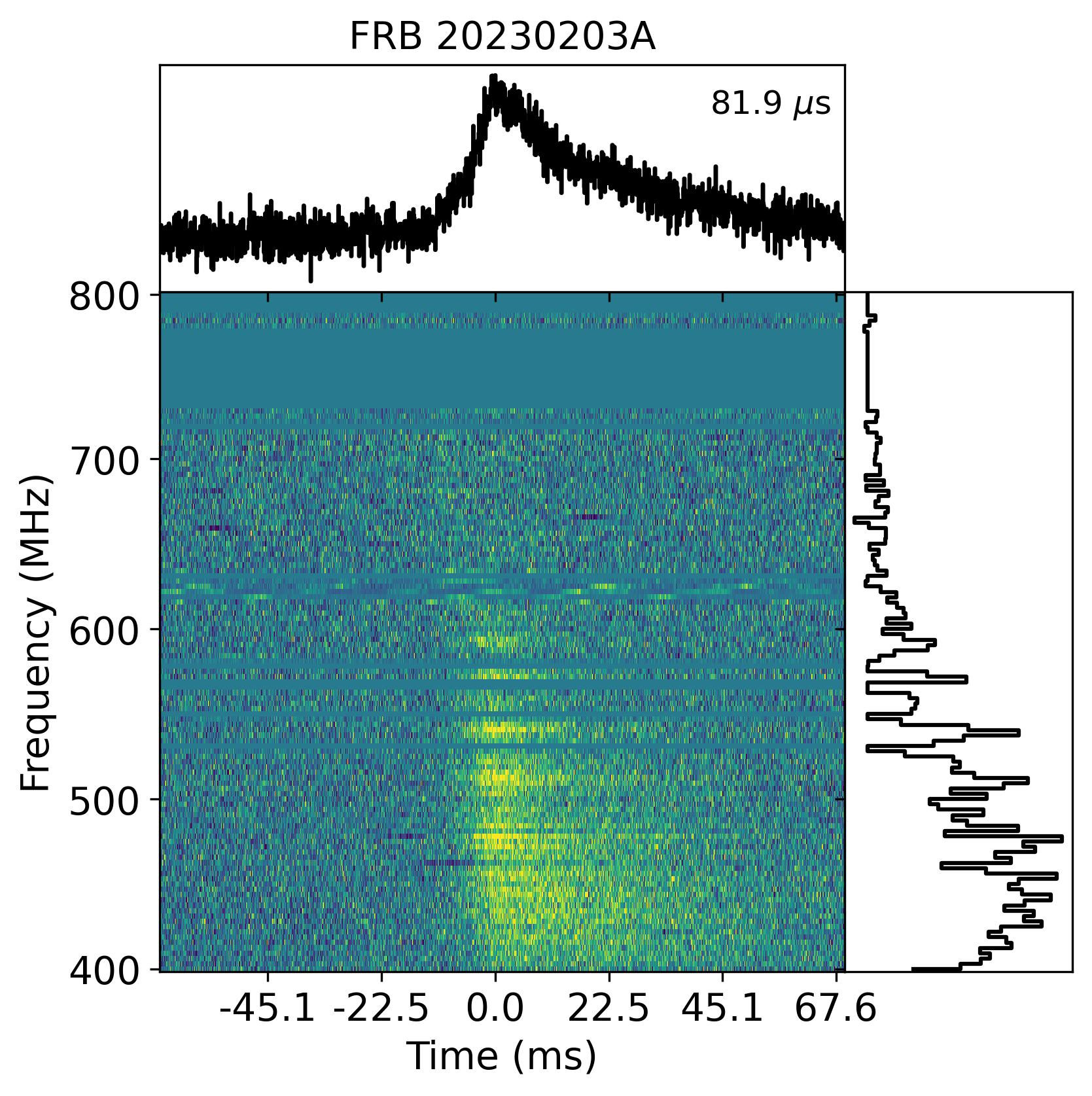}
\includegraphics[width=0.22\textwidth]{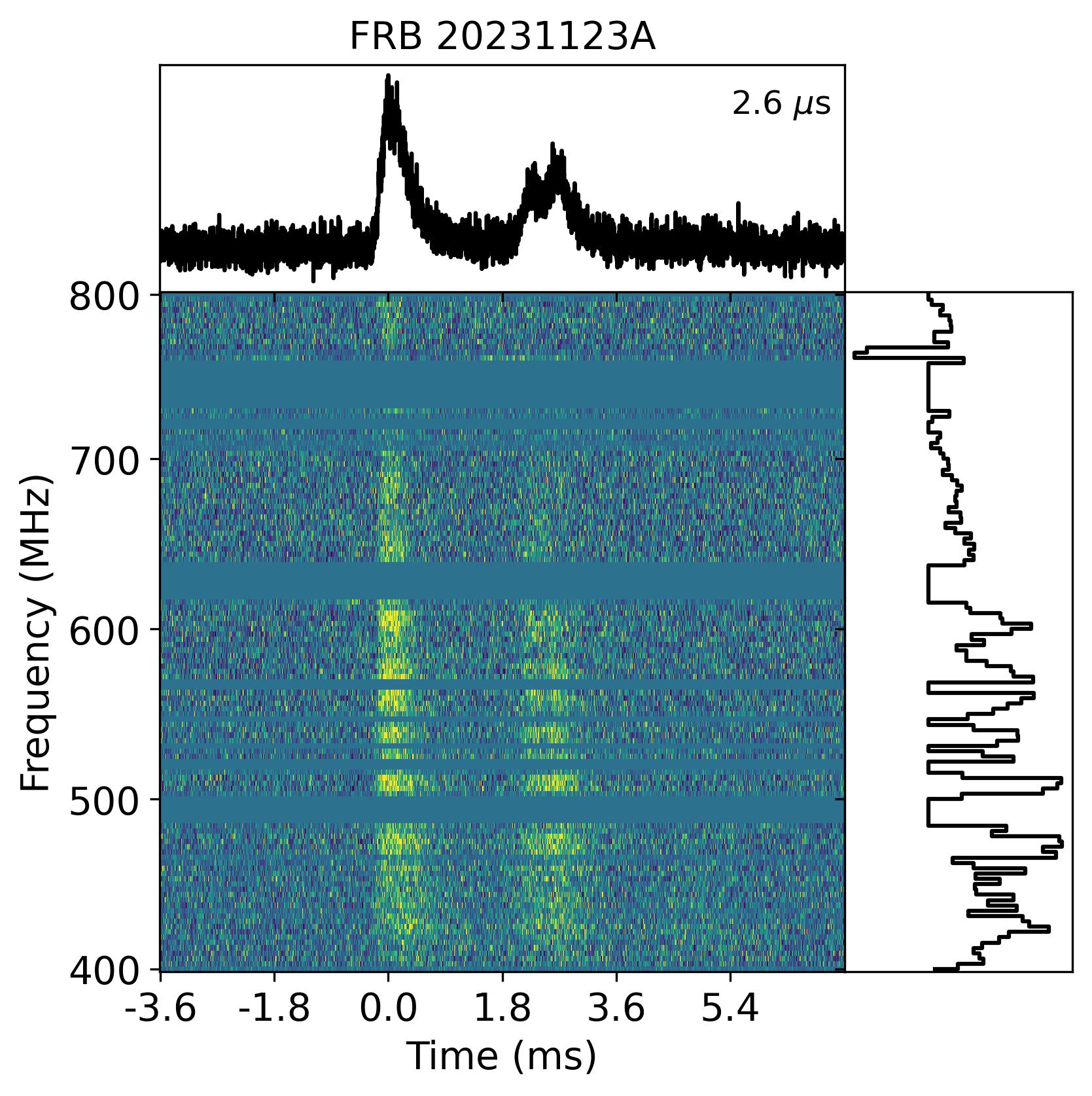}
\includegraphics[width=0.22\textwidth]{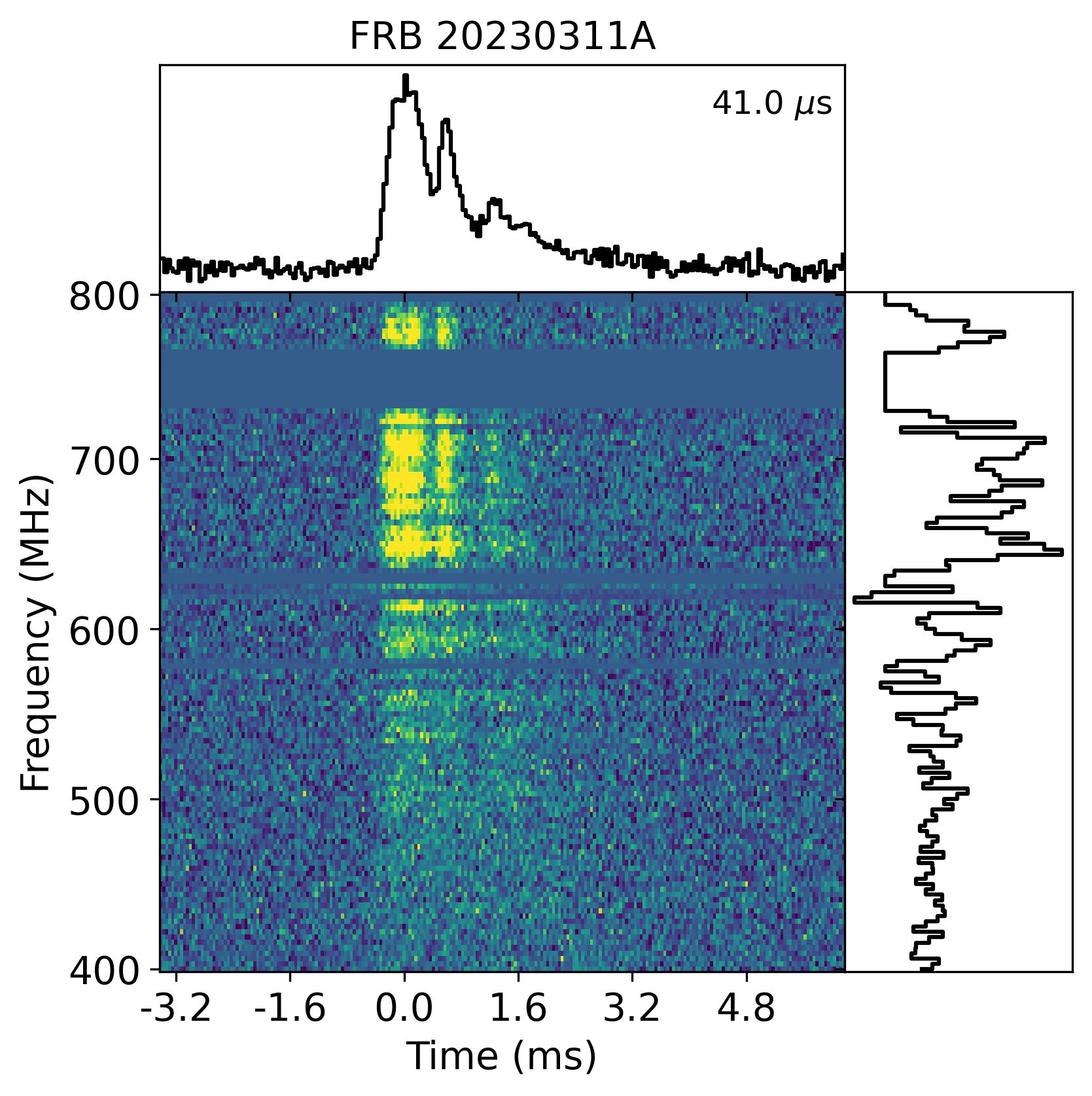}
\includegraphics[width=0.22\textwidth]{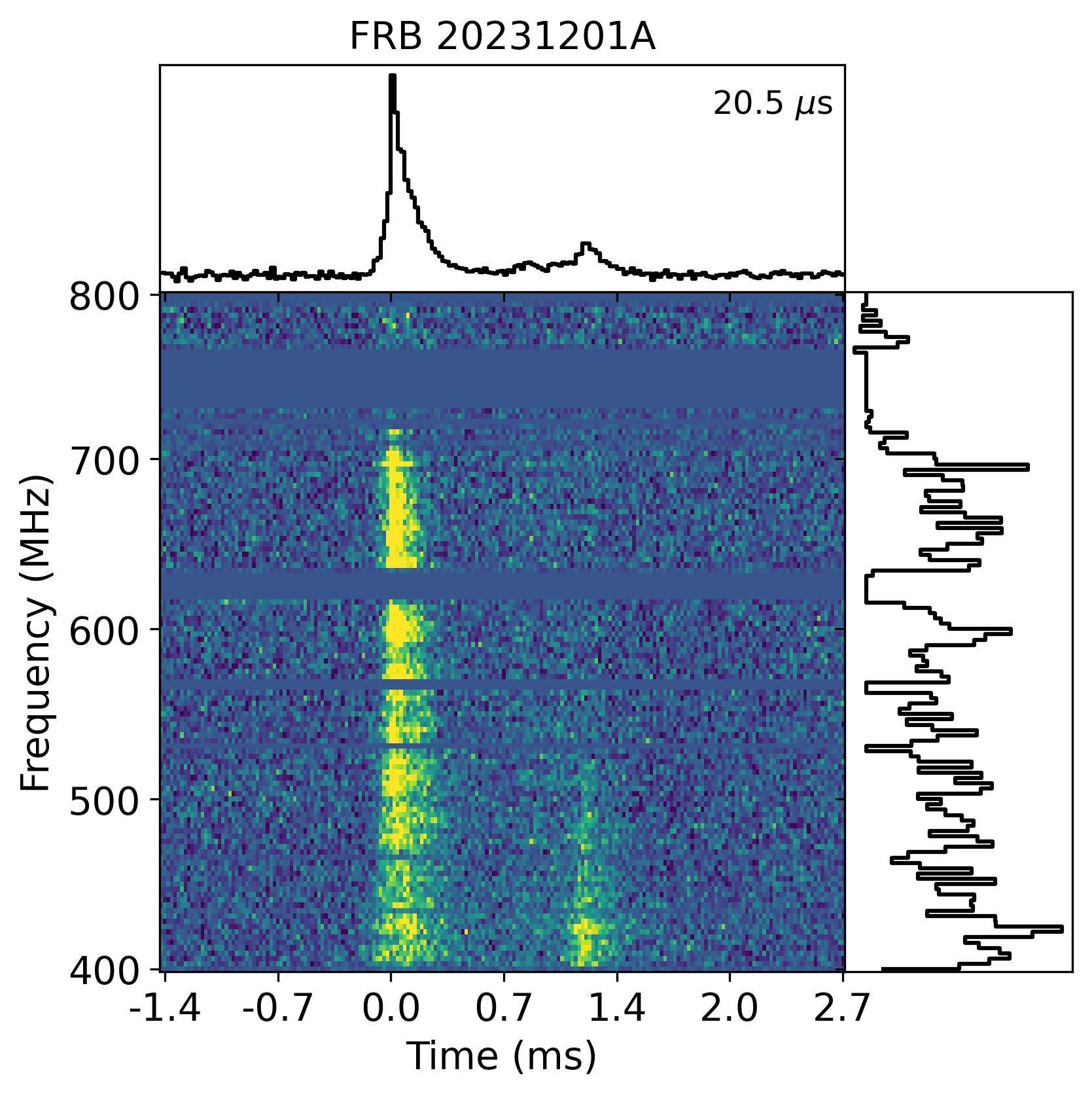}
\includegraphics[width=0.22\textwidth]{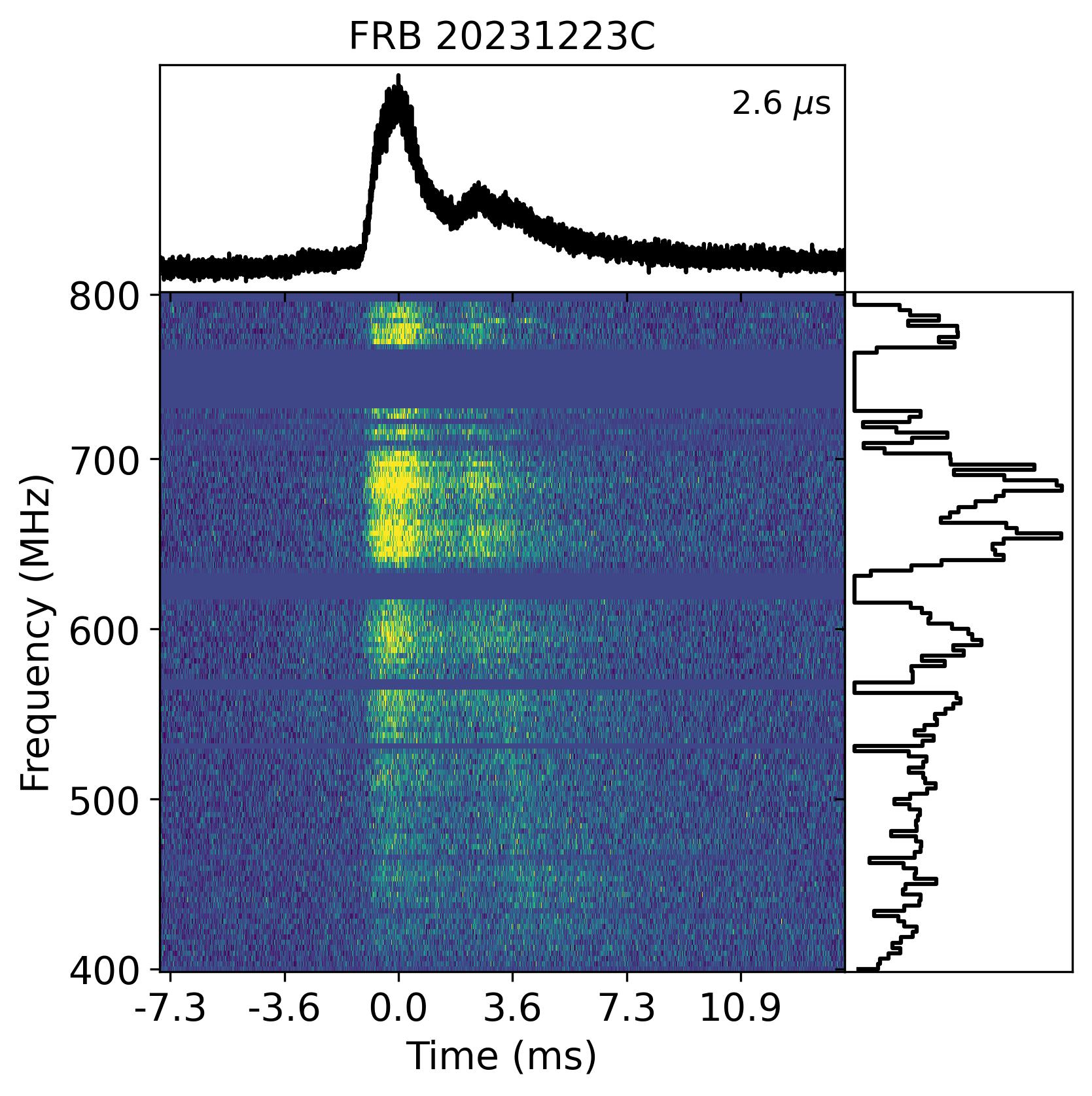}
\includegraphics[width=0.22\textwidth]{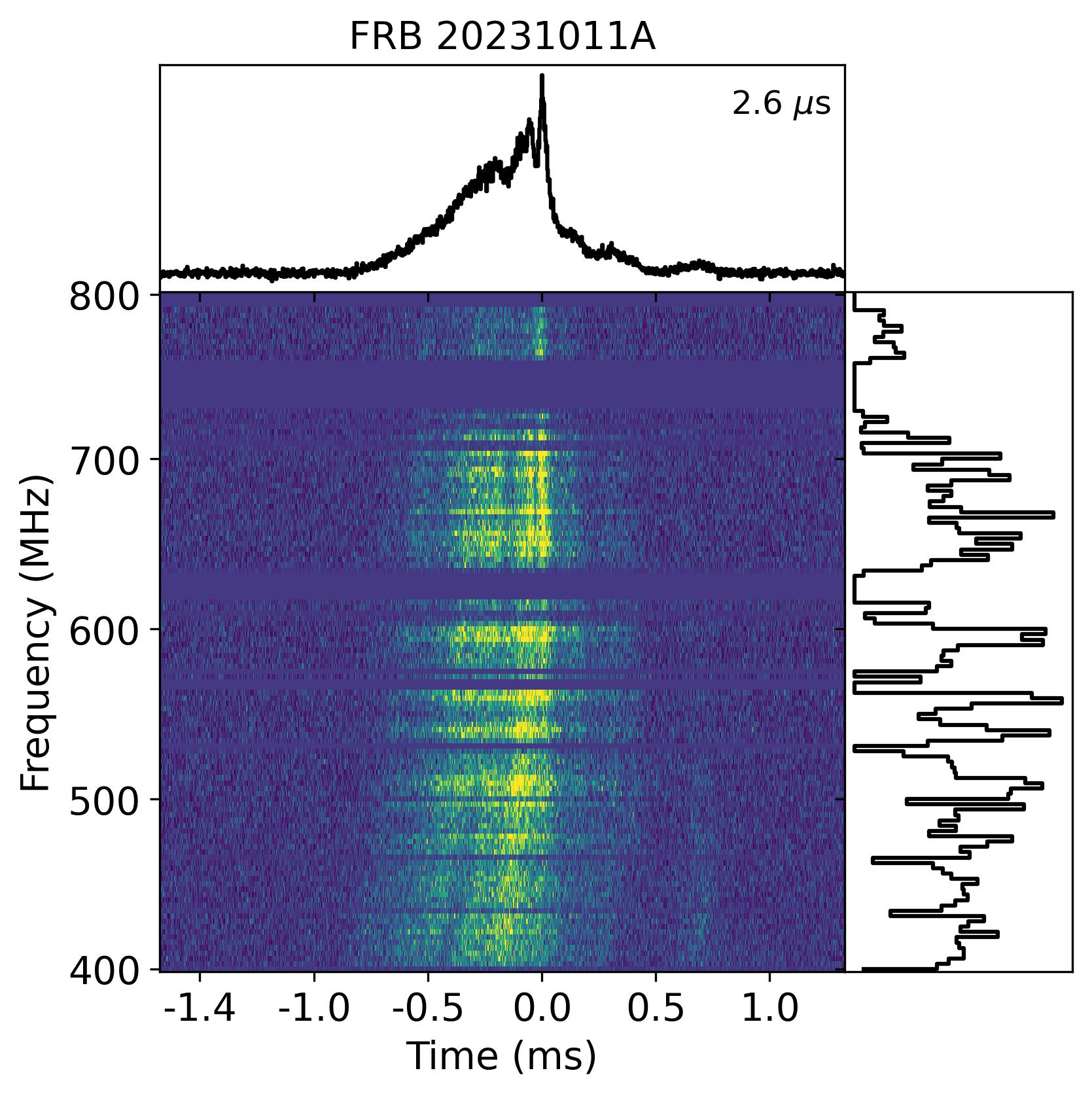}
\includegraphics[width=0.22\textwidth]{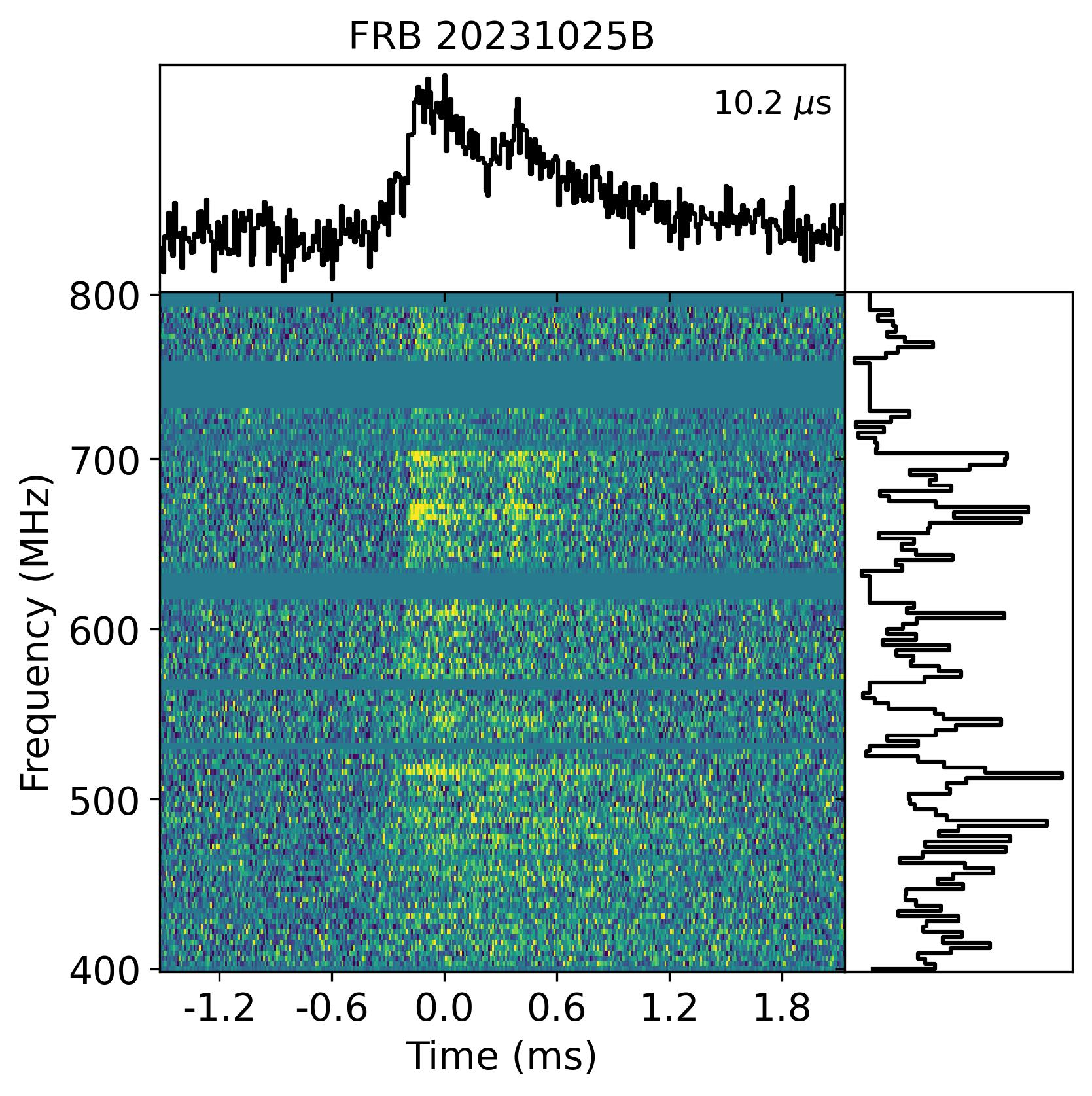}
\includegraphics[width=0.22\textwidth]{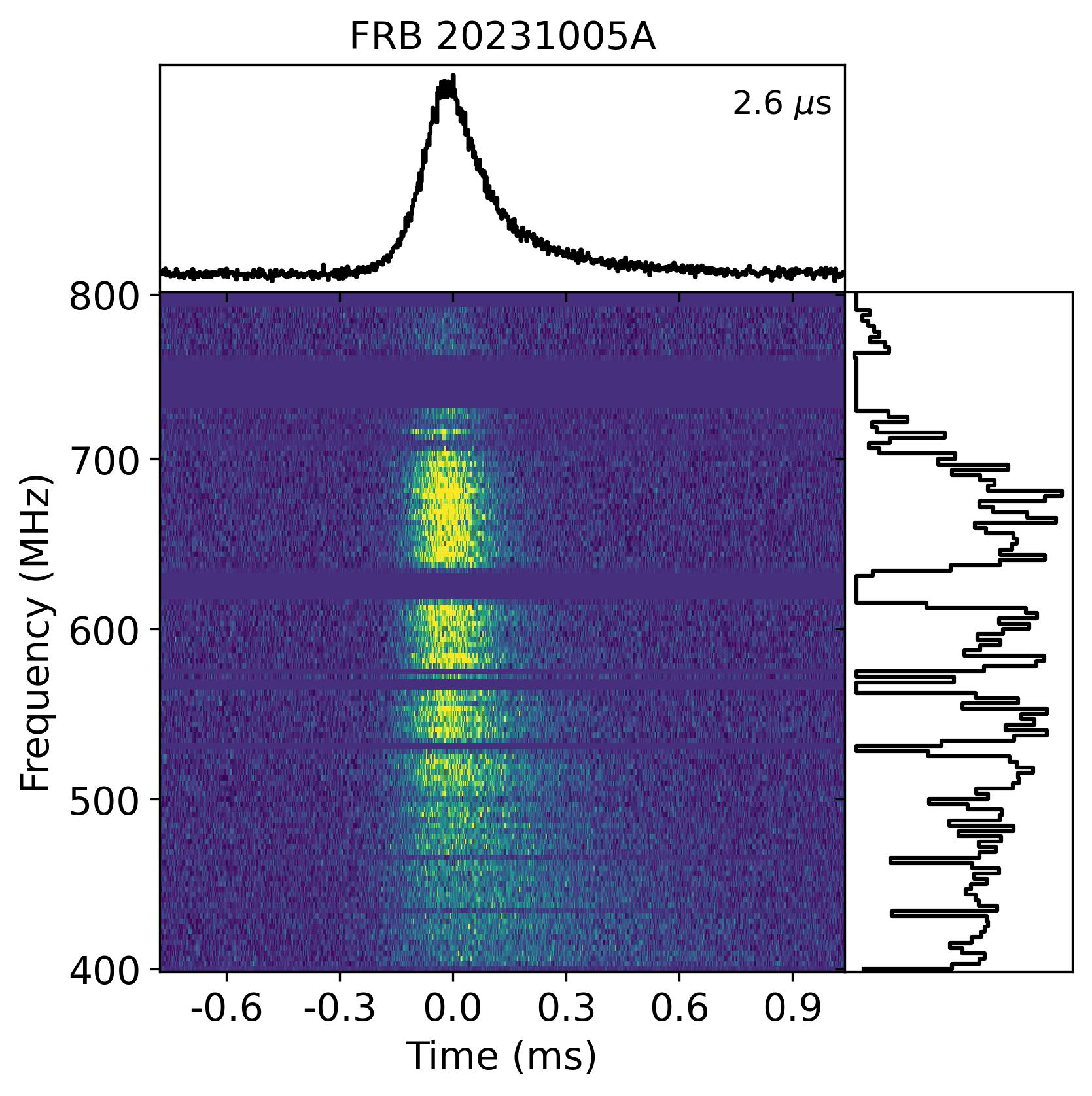}
\includegraphics[width=0.22\textwidth]{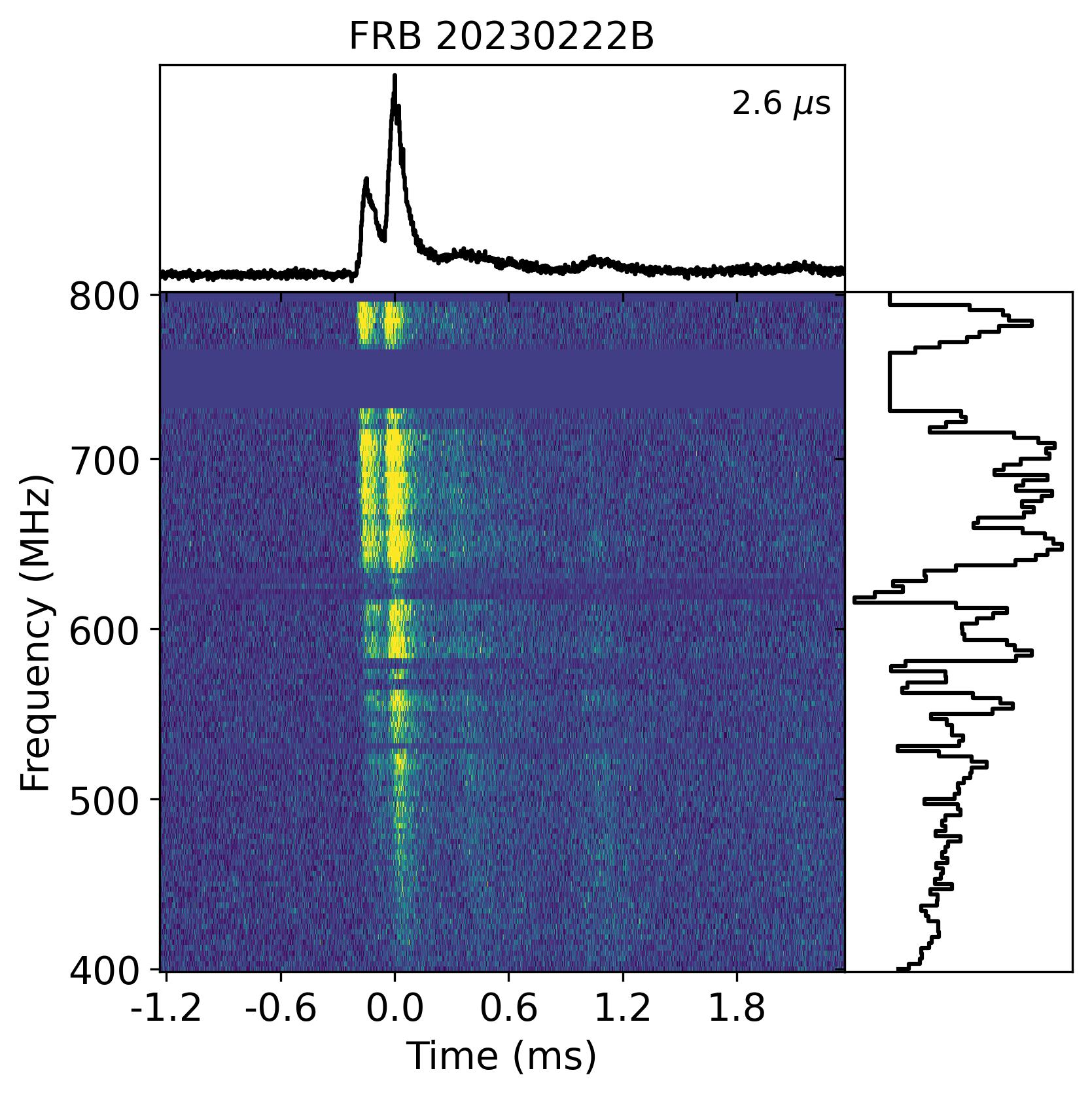}
\includegraphics[width=0.22\textwidth]{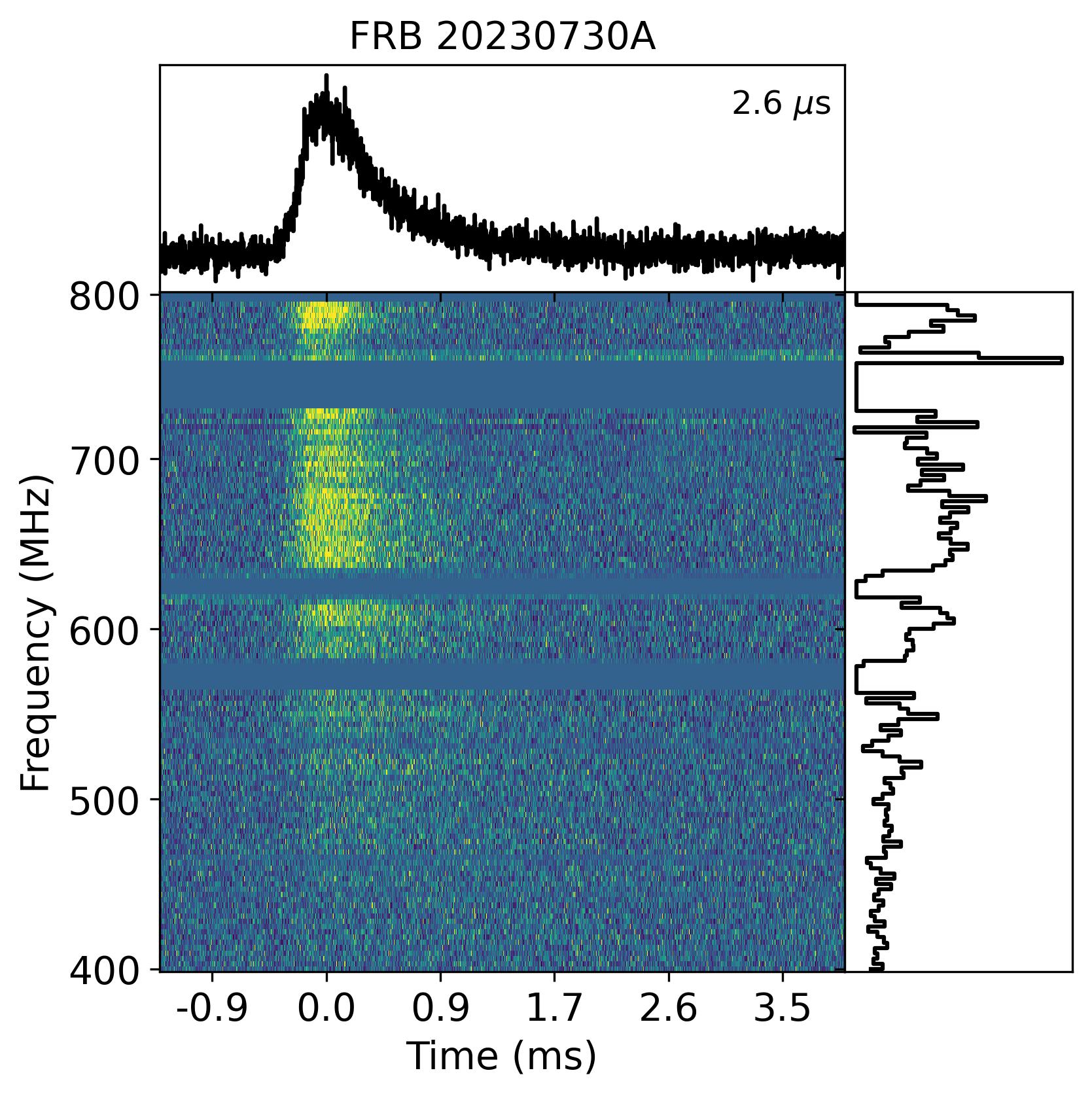}
\includegraphics[width=0.22\textwidth]{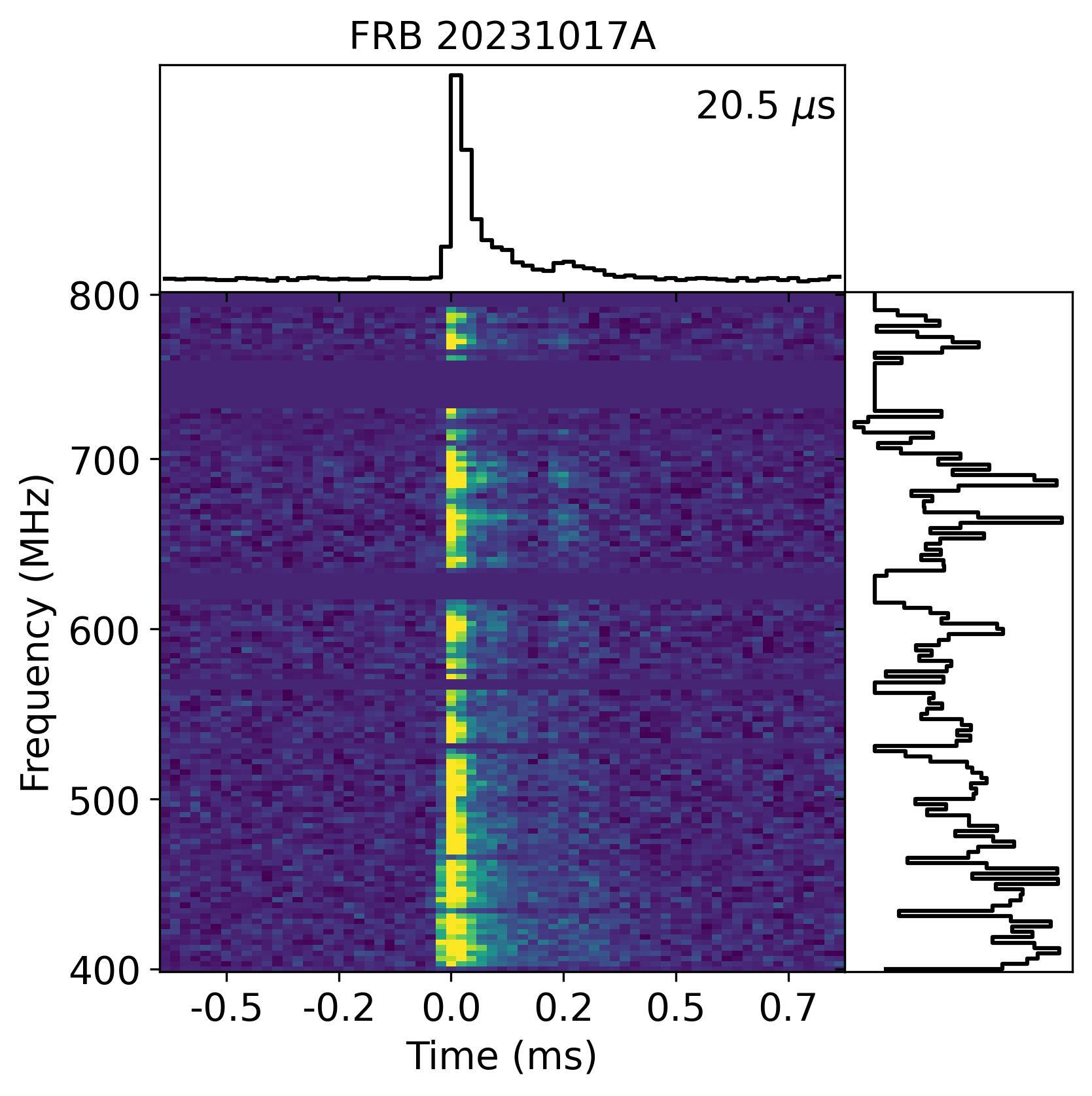}
\caption{Waterfall plots of bursts whose hosts have measured spectroscopic redshifts, ordered by specific peak luminosity (Jy Gpc$^2$) from lowest to highest. We note no obvious trends with increasing isotropic-equivalent burst luminosity but reserve a detailed analysis for future work. FRB\,20231204A and FRB\,20231128A are repeating sources; both show downward-drifting burst morphology characteristic of CHIME/FRB-detected repeating sources. Each plot is labeled with the time resolution to which we downsampled the dynamic spectrum. Masked channels are plotted at the zero point of the colorscale.}
\label{fig:wfall}
\end{figure*}

The baseband data with 2.5$\mu$s time resolution reveal that FRBs observed at CHIME frequencies have intricate morphological characteristics. This was known from previous works~\citep{pleunis2021fast,sand2024morphology137fastradio}, which studied burst morphologies in detail, but lacked distance measurements \msbold{and hence intrinsic information about burst luminosities}. \msbold{Here, we present dynamic spectra, or waterfall plots, for the 19 bursts in our sample for which we report redshifts later in this paper} (see Tables~\ref{tab:gold_sample}, \ref{tab:z_table}) and which are shown in Figure~\ref{fig:wfall}. We present the dynamic spectra in decreasing order of isotropic-equivalent peak specific luminosity (Jy Gpc$^2$) implied by their host redshifts\footnote{In all cosmological calculations we use the \textit{Planck} 2018 cosmology for a flat $\Lambda$CDM universe with $H_0 = 67.7~\mathrm{km~s}^{-1}~\mathrm{Mpc}^{-1}$, $\Omega_m=0.31$ and $\Omega_\Lambda = 0.69$ \citep{collaboration2020planck}.} \msbold{which we obtain using spectroscopic followup} (see \S\ref{sec:followup} and \ref{sec:sample}). \msbold{Of these 19, the bursts from the two repeating sources happen to be those with the lowest isotropic-equivalent peak specific luminosities; this occurs by chance with a probability of $0.02$, given that we localized bursts and associated them to host galaxies agnostic of their repetition properties.} This adds to the preponderance of evidence that burst energetics between repeating and apparently non-repeating sources may not solely be attributable to the narrower bandwidths of the former, \msbold{and that repeating sources are less luminous than non-repeating sources}. \msbold{It is also consistent with findings from the monitoring of single sources~\citep{kirsten2024link,ould2024probe,pelliciari2024northern}} and with indirect evidence from a large-scale comparison of burst morphologies between repeating and non-repeating sources at high time resolution~\citep{curtin2024morphology}. However, we defer detailed work on burst properties, e.g. morphological fitting~\citep{fonseca2024modeling,sand2024morphology137fastradio},  polarimetry~\citep{mckinven2021polarization}, and scintillometry~\citep{schoen2021scintillation,2024arXiv240611053N}, for later work. In the dynamic spectra of bursts from known repeaters (FRB 20231204A and FRB 20231128A), we see a downward-drifting burst substructure, which is known to be characteristic of repeating sources~\citep{pleunis2021fast}. The dynamic spectrum of each burst is downsampled in time to reach S/N $>$ 5, such that the signal is bright enough to determine the structure-maximizing DM \citep{HesselsstructmaxDM} at which the bursts are dedispersed. The time resolution of each dynamic spectrum is denoted in each waterfall plot; all are sampled at a frequency resolution of 390.625 kHz. 

\subsection{Semi-major axis localization using CHIME core}
Our burst localization consists of two stages. In the first stage, we localize our 81 bursts using the CHIME/FRB baseband localization pipeline, which uses data from CHIME only to constrains localization contour to $\sim 1'$. \msbold{Since the 81 bursts are selected on the basis of detectability in VLBI cross-correlation, they are all bright enough to be localized with the baseband pipeline.}

\msbold{We refer the reader to~\citet{michilli2021analysis} for a detailed discussion of the pipeline methodology, but briefly discuss the considerations which limit the astrometric accuracy of the baseband pipeline at the arcminute level. These effects include thermal expansion of the focal line, as well as clock offsets between the eastern and western cylinders comprising the CHIME core.} These effects can be detected and modeled out by observing pulsars with known positions and measuring the astrometric offsets from their true position. When combined with clock offset measurements and measurements of the feed temperatures taken from archived environmental monitoring data at the time of the burst arrival, these effects can be compensated to arcminute accuracy. Nevertheless, we validate the quoted accuracy level of this procedure, which we apply to a sample of 100 pulsars. We find empirically that the quoted localization uncertainties are consistent with the root mean square (RMS) accuracy of the pulsar localizations when compared against the true positions of each pulsar (Appendix~\ref{sec:accuracy}). Follow-up observations and European VLBI Network (EVN) localizations of a handful of CHIME/FRB-discovered repeating FRBs, including FRB 20180916B~\citep{collaboration2020periodic,marcote2020repeating}; FRB 20200120E~\citep{kirsten2022repeating}, FRB 20201124A~\citep{lanman2022sudden,nimmo2022milliarcsecond}, and FRB 20220912A~\citep{mckinven2022nine,hewitt2024milliarcsecond}, act as independent cross-checks of FRBs localized by the baseband pipeline.

\subsection{Semi-minor axis localization using VLBI}
\msbold{Baseband localizations are sufficient to make host galaxy associations for large, nearby galaxies. However, for the vast majority of FRBs detected by CHIME, VLBI localization is crucial to establishing a host association. We therefore define our final sample based on our ability to detect VLBI fringes between CHIME and KKO, due to the smaller collecting area of the latter.} The sample includes the brightest FRBs detected by the CHIME core, which have flux densities ranging from $\sim$3.9--280 Jy when averaged over the CHIME band of 400--800 MHz. Since the sensitivity of KKO varied \msbold{greatly} during the commissioning period, the low-flux cutoff is not well defined. \msbold{We refine the arcminute-scale localization using single-pulse VLBI and the in-beam calibration strategy developed in~\citet{leung2021synoptic}. The remainder of this section describes the methods by which we search for fringes for the FRB and for the VLBI calibrators, select a reliable calibrator for delay calibration, and translate calibrated visibilities into an FRB localization.}

\msbold{To find fringes on the FRB, we phase each antenna within the array and sum over all antennas within each station to synthesize a ``station beam'' at each station toward the baseband position. Next, we define a gate and dispersion measure based on the known width and time of arrival of the pulse. We use the~\texttt{PyFX} correlator to align the wavefront towards some known delay center, which we also choose to be the delay center. We coherently dedisperse and correlate the data towards the delay center between the two stations (effectively synthesizing a ``VLBI beam'' within the station beam). We sum over the coherently-dedispersed data to obtain visibilities for each 390 kHz channel within the 400-800 MHz bandwidth of the array for both the XX and YY polarizations. Finally, we take the Fourier transform of the visibilities over the frequency axis and search for correlated flux over a wide range of delays (from $-1.28$ to $+1.28$ microseconds). FRBs are deemed detected in VLBI if the S/N in both polarizations exceeds 11, or if the S/N in either polarization exceeds 13.}

\msbold{To find calibrators in the field of view of the FRB, we make a list of all VLBI calibrators from the Radio Fundamental Catalog (RFC) and VLBA calibrator catalog~\citep[see e.g.,][ and references therein]{ma1998international,petrov2021wide} within the $\sim 200$ deg$^2$ field of view of CHIME brighter than $\gtrsim$ 0.1 Jy. Similar to the procedure we followed for the FRB, we use the baseband data collected for the FRB to form station beams towards each of these calibrators following~\citet{andrew2024vlbi} and run the~\texttt{PyFX} correlator on the beamformed baseband data~\citep{leung2024vlbi} using the catalog position of each source as the delay center. We detect up to 15 VLBI calibrators per baseband capture with S/N $> 13$ in at least one of the XX and YY polarization combinations, and save visibilities for the detected calibrators to disk. This yields a delay and S/N for each calibrator candidate.}

\msbold{We use a combination of the calibrator delay and S/N to choose a calibrator, which then enables VLBI calibration and localization.} In CHIME-KKO VLBI observations, the uncalibrated delays measured towards the $\sim 15$ calibrators in different parts of the field of view typically agree to within 2 ns of each other. The 2 ns scatter reflects instrumental effects such as a small phase delay between the different polarization combinations and instrumental delays that vary slightly over the wide field of view of CHIME and KKO. In a handful of instances, strong fringes are detected towards delay centers that do not agree within 2 ns of the median taken over all delay centers; we find that these pointings are often close on the sky to brighter calibrators or are sources with radio jets which add structure to otherwise point-like emission~\citep{andrew2024vlbi}. 

\msbold{The presence of these misidentified calibrators for some fields motivates a calibrator-selection strategy which allows us to reliably reject them. We catch misidentified calibrators during delay calibration by disqualifying calibrators whose delays are not within 2 ns of the median delay computed over other calibrators in the baseband dump. 
Then, for all remaining calibrators, we choose the calibrator closest to the target to minimize the effect of the differential ionospheric path, although the latter has already been shown to have an insignificant impact on this 66-kilometer baseline~\citep{lanman2024chime}. We apply this calibrator selection strategy -- which we refer to as a ``closest and good'' strategy -- uniformly across the entire sample of 81 FRB targets; in addition we use a similarly-sized sample of pulsar localizations to show that our astrometric accuracy is insensitive to this choice: two other reasonable calibrator selection strategies yield a similar astrometric accuracy (see Appendix~\ref{sec:accuracy}).}

After a calibrator is selected, we apply delay calibration to the FRB to produce calibrated visibilities. This is repeated for every FRB in the sample. For the majority of bursts, we were able to robustly measure delays in both linear polarizations. In the majority of cases, the delays were coincident between the XX and YY polarization combinations to within 2 ns; we then average the delays between the XX and YY polarizations. When the delays disagreed by more than 2 ns, we observed that the burst was much more strongly detected in one polarization than the other, where the detection was marginal. We attribute this to the intrinsic polarization angle of the burst and the polarization-dependent instrumental sensitivity, particularly at the KKO station, which has 1/16 of the collecting area of CHIME. In this case we use the delay measured in the brighter of the two polarizations (XX or YY). The statistical uncertainty on each delay measurement was much smaller than the systematic uncertainties.

Finally, we use a maximum-likelihood delay-mapping procedure to translate these residual time delays into localization contours (see Eq. 4 in~\citet{marcote2020repeating}; also Eq. A1 in~\citet{leung2024vlbi}). Each VLBI localization contour is combined with the CHIME-only baseband localization ellipse as a prior. The combined contour is convolved with a kernel reflecting a conservative estimate of the localization accuracy. Finally, we calculate ellipse parameters (semi-major/minor axes, orientation angle, and central coordinates) from the posterior probability grid. The distribution of sky positions for the full sample of localized sources is shown in the \msbold{red} histogram in Fig.~\ref{fig:pox_vs_pos}.

To characterize the localization accuracy achieved by this procedure, we apply this procedure to a sample of $62$ single pulses from 18 distinct pulsars. In doing so, we find that CHIME-KKO observations achieve an astrometric accuracy of $< 1\arcsec$ (Appendix~\ref{sec:accuracy}), except at low declinations where a combination of reduced sensitivity and additional radio frequency interference (RFI) at the KKO site increases the overall RMS error. We conservatively prescribe the semi-minor axis astrometric uncertainty of the whole sample to $2\arcsec$ at the $1\sigma$ level. 

\section{Host Association}\label{sec:association}
\msbold{After localizing each burst using a combination of CHIME-only baseband positions and CHIME-KKO VLBI,} we use the PATH framework~\citep{aggarwal2021probabilistic} to assign posterior probabilities \pox\ to each cataloged galaxy $O$ given the combined localization region $x$ from baseband and CHIME-KKO VLBI. In PATH, each localization region is cross-matched with a wide-area galaxy survey. Where possible (55/81 fields), we utilize the Dark Energy Camera Legacy Survey \citep[DECaLS;][]{dey2019overview} to find host galaxy candidates
due to its greater depth of $m_r = 23.5$~mag. However, DECaLS only covers about $60\%$ of the CHIME footprint. In the remaining  (26/81) fields, we default to Pan-STARRS PS1~\citep{chambers2016pan, flewelling2020pan,magnier2020pan,magnier2020pan2, magnier2020pan3,waters2020pan}, which is shallower ($m_r = 23.2$~mag) but covers the entire CHIME/FRB footprint.

Stars are first identified in each field on the basis of their classification in Gaia DR3~\citep{gaia2016gaia, gaia2023gaia} and additional 
survey-specific criteria. To exclude stars, we required that sources satisfy
$m_r > 14$ mag and that \texttt{DECaL\_type} not be ``PSF''\footnote{\msbold{see ~\citet{dey2019overview} and \url{https://www.legacysurvey.org/dr10/description/} for an overview of the morphological classification used by DECaLS}}. For PS1, we also use the
Probabilistic Classifications of Unresolved Point Sources in PanSTARRS1 catalog 
\citep[PS1-PSC;][]{tachibana2018} and required that
\texttt{rKronRad} $ > 0$, $m_r > 15$, $\log_{10}$\texttt{rPSFLikelihood}$ < -2$, and that \texttt{iPSFMag} $-$ \texttt{iKronMag\_1} $> 0.05$.

\begin{figure}
    \centering
    \includegraphics[width=3.0in]{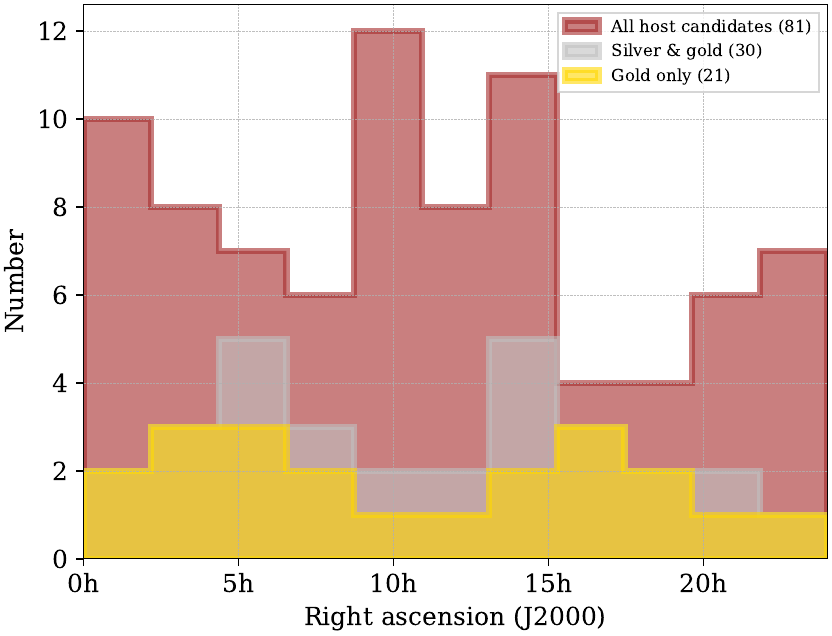}    \includegraphics[width=3.0in]{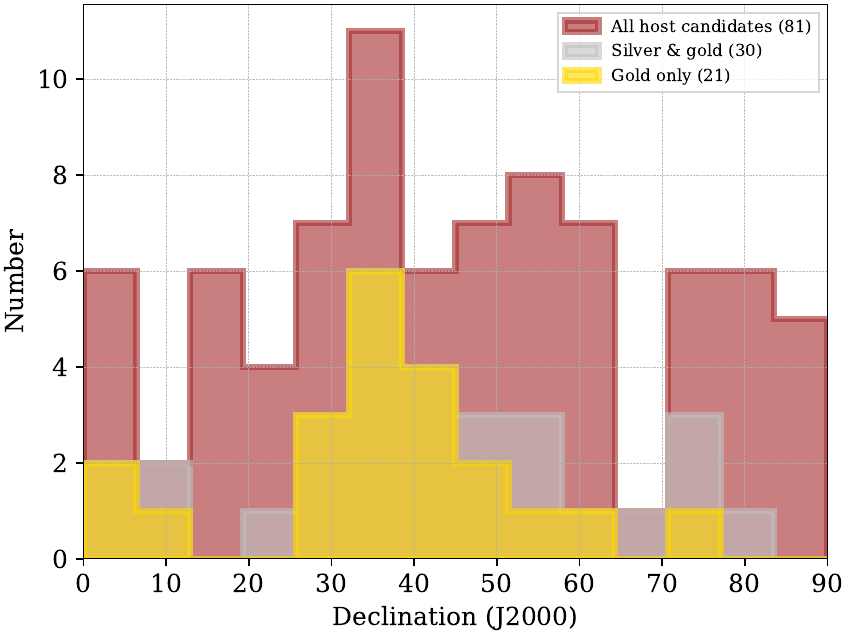}
    \caption{The sky distribution in right ascension (top) and declination (bottom) of host galaxy candidates as a function of PATH posterior probability $\pox$. Hosts with $\pox > 0.9$ have a small false-positive fraction. Of this subset, note that only the host galaxies of FRB\,20230222A and FRB\,20231123A have low declinations; we include both in our sample because the host galaxy candidate is still favored even when inflating the astrometric error ellipse (all others satisfy $\delta \gtrsim +26^\circ$, where the astrometry is more precise; see Appendix).}
    \label{fig:pox_vs_pos}
\end{figure}

\begin{figure}
    \centering
    \includegraphics[width=\linewidth]{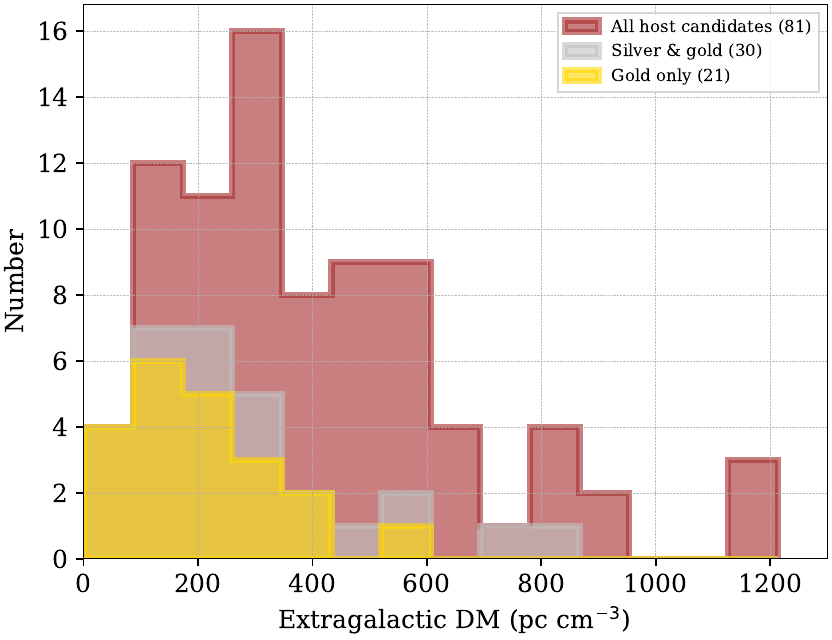}
    \caption{The observed distribution of the extragalactic DM, defined as DM$_\textrm{tot}$ - DM$_\textrm{MW,ISM}$ - DM$_\textrm{MW,halo}$, for FRBs in our sample. Different colors refer to our ``gold'' and ``silver'' samples (see text), which have lower mean extragalactic DMs than that of the full sample.}
    \label{fig:hg_vs_dm}
\end{figure}

After stars satisfying the above criteria are removed from the catalog, each remaining galaxy candidate in the field was assigned a posterior probability $P(O|x)$ of being the true host of the FRB using 
PATH. Finally, each field was visually inspected to identify unaccounted and/or potentially misclassified stars and galaxies that evaded the above criteria. When one of these issues was detected (3/81 cases), PATH was run again for that field after the appropriate removal or addition of the object. The galaxy candidates with the two highest values of $P(O|x)$ are referred to as the ``primary'' and ``secondary'' and are flagged for potential spectroscopic follow-up if the sum of their $P(O|x)$ values exceed 0.9, a choice which we validate through simulations (see below). The distribution of localized source positions and DMs is shown in Fig.~\ref{fig:pox_vs_pos} and~\ref{fig:hg_vs_dm}.

We mostly use the default priors in PATH, in accordance with previous work~\citep[e.g.,][]{law2024deep}. These include a prior on the host galactocentric offset distribution, the angular extent of the galaxy (i.e., the chance-coincidence probability in a frequentist framework), and the prior probability $P(U)$ that an FRB host galaxy remains undetected in an existing galaxy catalog. It is important to note that a highly sensitive 
FRB detector such as CHIME is likely to discover FRBs with hosts that are too faint for the wide-field, archival imaging that
we have used. As a point of reference, the faint dwarf host of FRB 20121102A~\citep{tendulkar2017host} would be detectable in the DECaLS $r$-band only out to $z \sim 0.2$, depending on the Galactic extinction along the line-of-sight. The median DM of our sample (326 pc cm$^{-3}$; see Fig.~\ref{fig:hg_vs_dm}) implies a redshift which significantly exceeds this horizon ($z \approx 0.38$). Even fainter hosts have recently been discovered~\citep{hewitt2024repeating};
the value of $P(U)$ chosen is therefore crucial to making secure associations.

The PATH framework accounts for this potential survey depth mismatch by specifying $P(U)$. This quantity is defined as the prior probability that the host galaxy of an individual FRB in a survey is too faint to be associated with any detected galaxy with high confidence, and uses the FRB redshift distribution, the galaxy luminosity function, the localization ellipse footprint, and the optical survey depth in its calculation. We choose a distinct value of $P(U)$ tailored to our specific combination of CHIME/FRB paired with a wide-field optical survey (Pan-STARRS or DECaLS) using detailed simulations by Andersen et al. in prep. In these simulations, mock FRBs are seeded in host galaxies as faint as $m_r = 28$ mag identified in the Hyper Suprime-Cam Subaru Strategic Survey~\citep{aihara2018first} and subsequently ``re-detected'' in the shallower optical surveys used for our FRB host associations. 
This simulation thus provides an empirical characterization of the false-positive rate of host galaxy associations (i.e., a Type-I error) as a function of galaxy brightness for a fixed threshold value of $P(O|x)$. 

These simulations indicate that the optimal values of $P(U)$ are $0.32$ for the PanSTARRS fields and $P(U) = 0.15$ for the DECaLS fields. We note that these values are higher than those previously used in 
the literature, e.g.,\ $P(U) = 0.1$ is used for Pan-STARRS for the DSA-110 host galaxy sample~\citep{law2024deep}. 
We defer a more extensive justification of these choices to Andersen et al. in prep., but note that this choice is likely conservative because the sample of bursts detectable by the outrigger stations presented here represent the brightest bursts detected by CHIME, and hence preferentially sample the nearby end of the CHIME redshift distribution~\citep{shin2023inferring}. 


\begin{figure}
    \centering
    \includegraphics[width=3.0in]{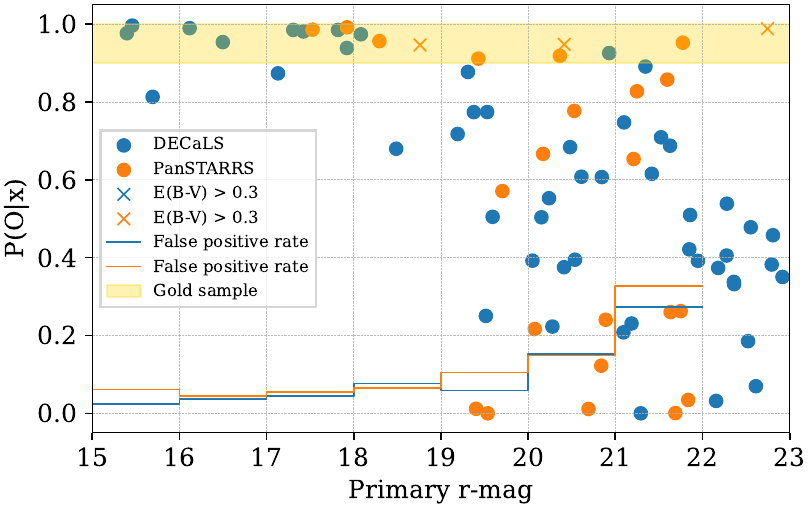}
    \caption{For each sightline, we plot the probability that the galaxy candidate with the highest $P(O|x)$ is the true host as a function of its $r$-band magnitude. We also plot the false-positive fraction rate of claimed associations calculated within a bin in $m_r$, determined from simulations (Andersen, et al. in prep). Points and traces are colored according to their respective optical survey (blue for DECaLS and orange for Pan-STARRS); high-extinction sightlines are denoted as ``E(B-V) > 0.3'' and indicated with x's. We observe that the highest PATH probability achieved by any host candidate in our sample tends to drop below $0.9$ at $\sim 22$ mag. This suggests that our archival imaging, which goes as deep as $23-23.5$ mag, is sufficiently deep for making confident host galaxy associations.}
    \label{fig:pox_vs_rmag}
\end{figure}

Table~\ref{tab:gold_sample} has 22 entries, consisting of 21 of our 81 bursts which can robustly be associated with host galaxy candidates with $P(O|x) > 0.9$. This defines a ``gold'' sample of 21 new host galaxies, \msbold{subject to the false-positive rate discussed in \S\ref{sec:association}}. We also include FRB\,20230311A in that table. \msbold{In an effort to be complete out to large host galaxy offsets/less luminous hosts, we also define a ``silver'' sample of 9 systems (see below) with a less stringent association criteria.}

The silver systems were defined with the criterion that the sum of $P(O|x)$ for the two most probable host candidates exceeds 0.9, so by construction, the silver systems had non-negligible secondaries. It is of interest to discuss the statistics of the silver secondaries in support of future spectroscopic follow-up. The $r$-band magnitudes of the secondary candidate for the nine silver systems ranged from 19.6--23.6~mag. Since it is difficult to make secure host associations towards the fainter end of this magnitude range, we focused on measuring redshifts for just the brightest secondary candidates in the silver sample with the goal of ruling out some secondaries on the basis of their distance. We obtained redshifts for two of the three brightest secondary candidates belonging to FRB\,20230410A ($m_{r,\textrm{sec}} = 19.60$ mag; $z_\textrm{spec,sec} = 0.5713$),
FRB\,20231020A ($m_{r,\textrm{sec}} = 19.9$ mag), and FRB\,20230311A ($m_{r,\textrm{sec}} = 20.2$ mag; $z_\textrm{spec,sec} = 0.1924$). For FRB\,20230410A, the maximum redshift of the host implied by the burst's high extragalactic DM (587 pc cm$^{-3}$) does not allow the secondary candidate to be ruled out. 

Finally, in the case of FRB\,20230311A, the small velocity difference (180 km/s) separating the primary and secondary candidate ($z=0.1918$ and $z=0.1924$ respectively) suggests a physical link between these two systems and a robust association with the combined system. Since these redshifts differ by $\approx 3\%$ we deem that FRB\,20230311A has a secure redshift and include it in Table~\ref{tab:gold_sample} and Figure~\ref{fig:wfall}, which merely requires a secure redshift (not a host galaxy).

Figure~\ref{fig:pox_vs_rmag} shows, for all 81 localization ellipses, $P(O|x)$ as a function of the $r$-band magnitude of the most probable host. We see that the probability $P(O|x)$ drops off as the host candidate gets fainter, far before the limiting magnitude of each optical survey depth is reached. In addition, our simulations indicate that the Type I error rate of host galaxy associations in this sample for a cutoff of $P(O|x) > 0.9$ is $<10\%$ only for $m_r < 21$ and rapidly increases for $m_r > 21$. This can be understood as follows: since the areal density of galaxies fainter than $m_r > 21$ rapidly increases to more than one per localization region~\citep{driver2016galaxy,eftekhari2017associating}, the high chance coincidence probability of encountering a faint galaxy compatible with the FRB localization prevents a confident association from being made at the current level of localization precision. Thus, more precise burst localization, as opposed to deeper imaging of our fields, is the predominant factor in increasing the fraction of highly secure host galaxies. 

Figure~\ref{fig:hg_vs_dm} shows the DM distribution of FRBs which we associate to their hosts. The distribution illustrates that bright, nearby \msbold{bursts can be confidently associated with their hosts} and have systematically lower DM. Conversely, the distant, high-DM FRBs have host galaxies too faint to enable a highly secure association. \msbold{This illustrates that in addition to our sample being flux-limited by radio burst brightness, our host association procedure imposes an additional selection on optical flux.}

\begin{figure*}[h]
\centering
\includegraphics[trim=3.2in 0.4in 3.1in 0.4in, clip,width=0.15\textwidth]{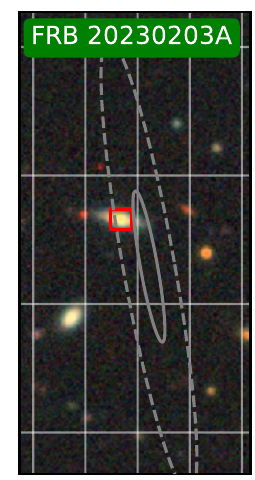}
\includegraphics[trim=3.2in 0.4in 3.1in 0.4in, clip,width=0.15\textwidth]{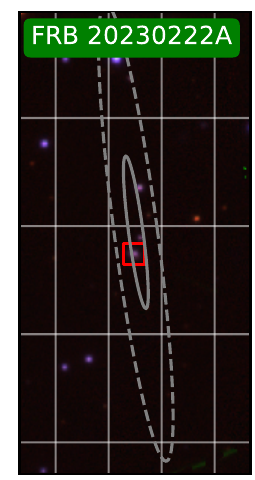}
\includegraphics[trim=3.2in 0.4in 3.1in 0.4in, clip,width=0.15\textwidth]{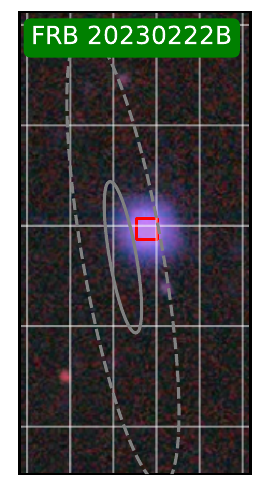}
\includegraphics[trim=3.2in 0.4in 3.1in 0.4in, clip,width=0.15\textwidth]{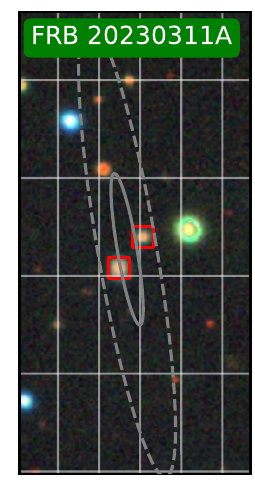}
\includegraphics[trim=3.2in 0.4in 3.1in 0.4in, clip,width=0.15\textwidth]{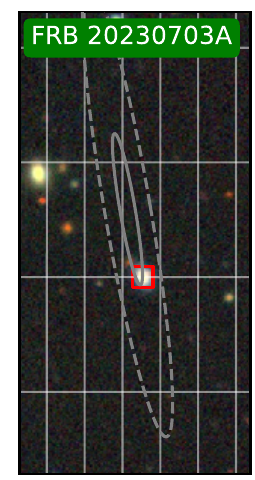}
\includegraphics[trim=3.2in 0.4in 3.1in 0.4in, clip,width=0.15\textwidth]{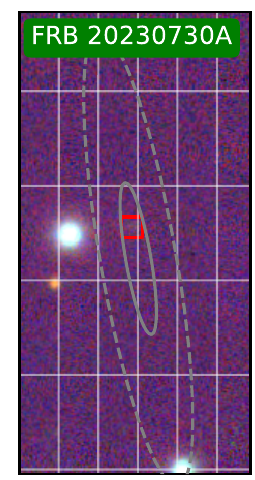}
\includegraphics[trim=3.2in 0.4in 3.1in 0.4in, clip,width=0.15\textwidth]{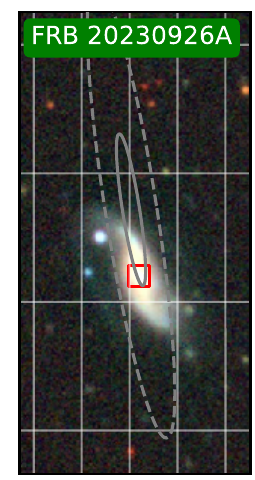}
\includegraphics[trim=3.2in 0.4in 3.1in 0.4in, clip,width=0.15\textwidth]{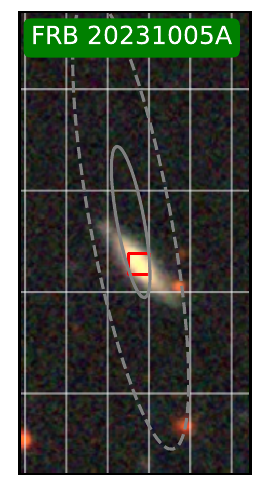}
\includegraphics[trim=3.2in 0.4in 3.1in 0.4in, clip,width=0.15\textwidth]{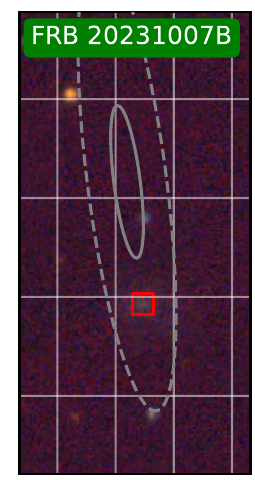}
\includegraphics[trim=3.2in 0.4in 3.1in 0.4in, clip,width=0.15\textwidth]{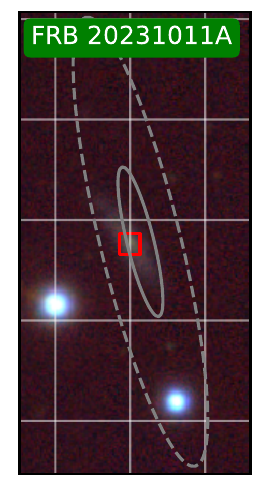}
\includegraphics[trim=3.2in 0.4in 3.1in 0.4in, clip,width=0.15\textwidth]{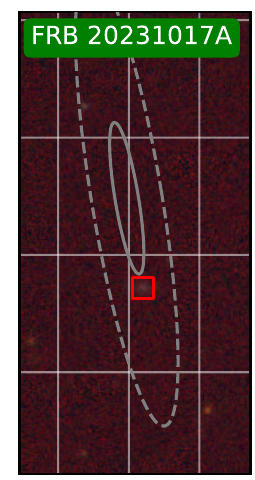}
\includegraphics[trim=3.2in 0.4in 3.1in 0.4in, clip,width=0.15\textwidth]{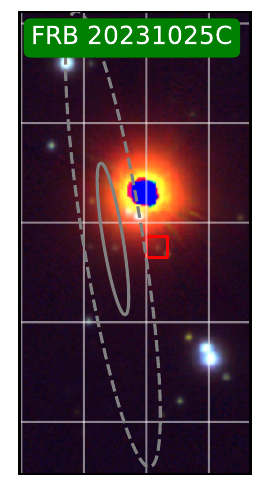}
\includegraphics[trim=3.2in 0.4in 3.1in 0.4in, clip,width=0.15\textwidth]{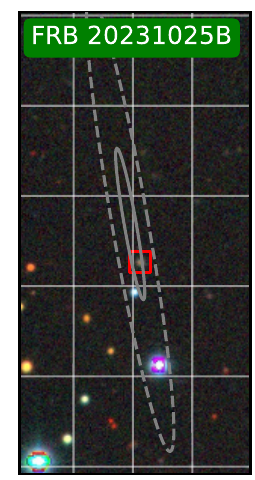}
\includegraphics[trim=3.2in 0.4in 3.1in 0.4in, clip,width=0.15\textwidth]{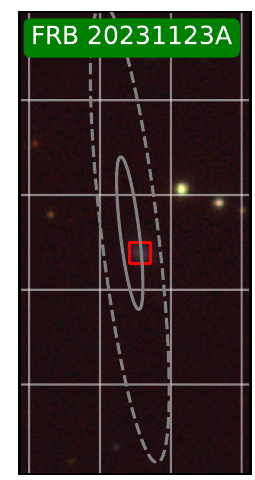}
\includegraphics[trim=3.2in 0.4in 3.1in 0.4in, clip,width=0.15\textwidth]{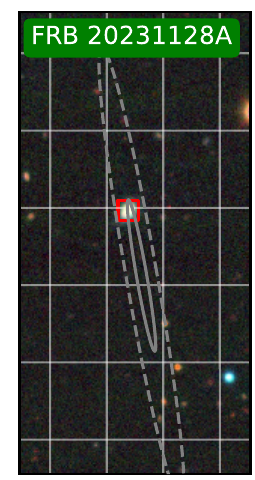}
\includegraphics[trim=3.2in 0.4in 3.1in 0.4in, clip,width=0.15\textwidth]{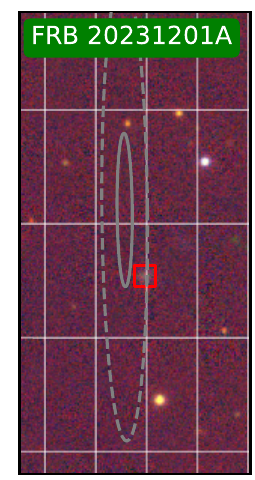}
\includegraphics[trim=3.2in 0.4in 3.1in 0.4in, clip,width=0.15\textwidth]{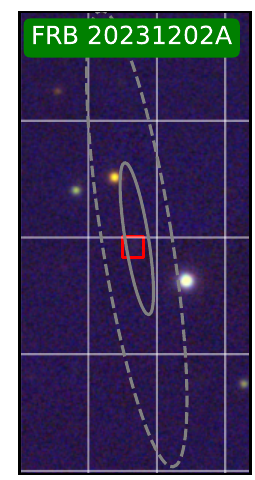}
\includegraphics[trim=3.2in 0.4in 3.1in 0.4in, clip,width=0.15\textwidth]{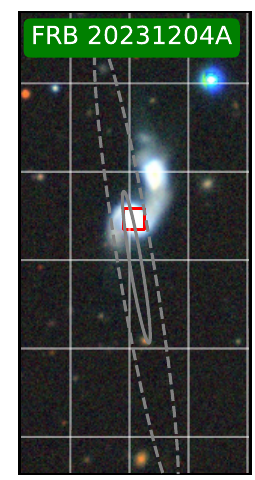}
\includegraphics[trim=3.2in 0.4in 3.1in 0.4in, clip,width=0.15\textwidth]{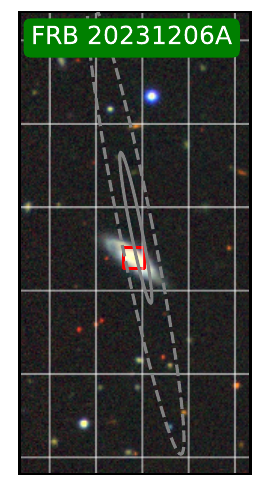}
\includegraphics[trim=3.2in 0.4in 3.1in 0.4in, clip,width=0.15\textwidth]{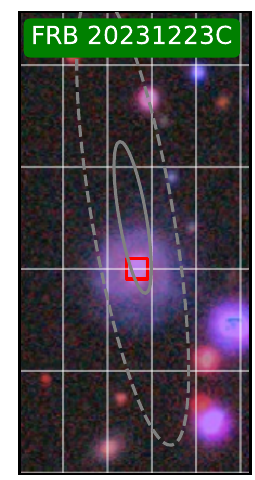}
\includegraphics[trim=3.2in 0.4in 3.1in 0.4in, clip,width=0.15\textwidth]{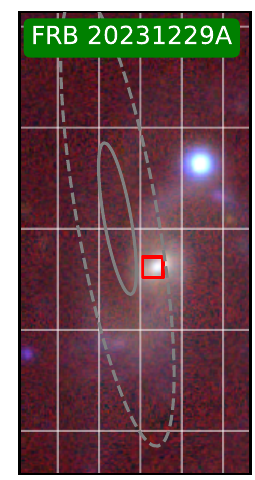}
\includegraphics[trim=3.2in 0.4in 3.1in 0.4in, clip,width=0.15\textwidth]{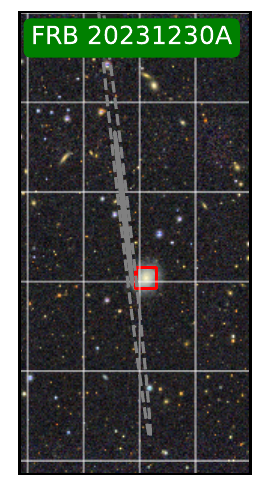}
\caption{Each FRB field with a host galaxy for which the PATH algorithm yields $P(O|x) > 0.9$ (i.e., the gold sample of FRBs in Table~\ref{tab:gold_sample}). Bright stars are visible in many fields; these are excluded automatically or during manual vetting of each host association (see text). The 1$\sigma$ and 3$\sigma$ localization contours are displayed as solid and dashed white ellipses, respectively. The top host candidate for each burst in our gold sample is denoted by a red square. Additionally, we have included the field of FRB\,20230311A (see text). The images of some fields (FRB 20230222B, 20231123A, and 20231223C) lacked one or more color channels in publicly available images, resulting in their tinted appearance.}
\label{fig:path}
\end{figure*}

\begin{figure*}[h]
\centering
\includegraphics[trim=2.5in 0.5in 2.5in 0.5in,clip,width=0.200\textwidth]{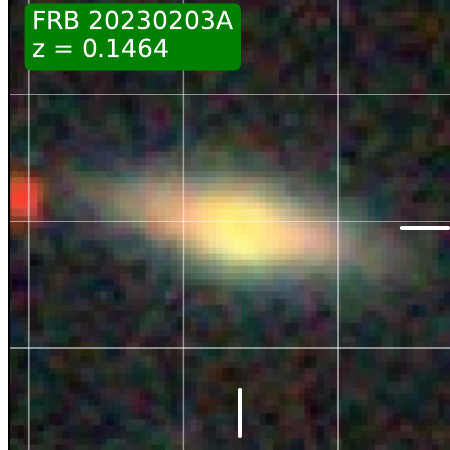}  
\includegraphics[trim=2.5in 0.5in 2.5in 0.5in,clip,width=0.200\textwidth]{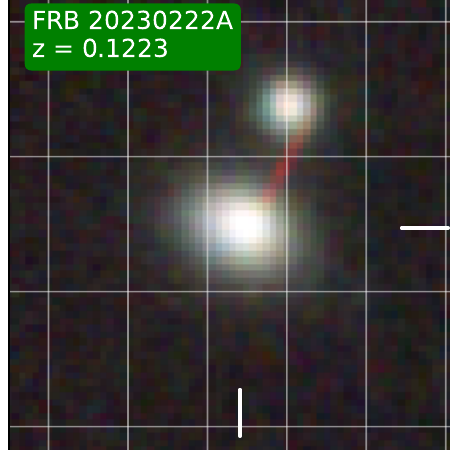} 
\includegraphics[trim=2.5in 0.5in 2.5in 0.5in,clip,width=0.200\textwidth]{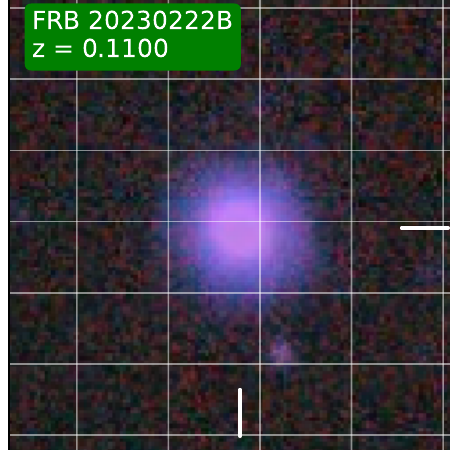} 
\includegraphics[trim=2.5in 0.5in 2.5in 0.5in,clip,width=0.200\textwidth]{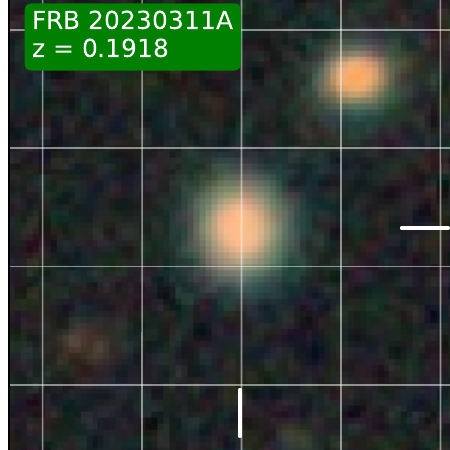} 
\includegraphics[trim=2.5in 0.5in 2.5in 0.5in,clip,width=0.200\textwidth]{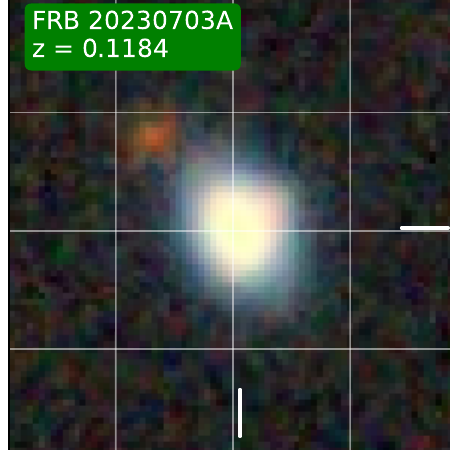} 
\includegraphics[trim=2.5in 0.5in 2.5in 0.5in,clip,width=0.200\textwidth]{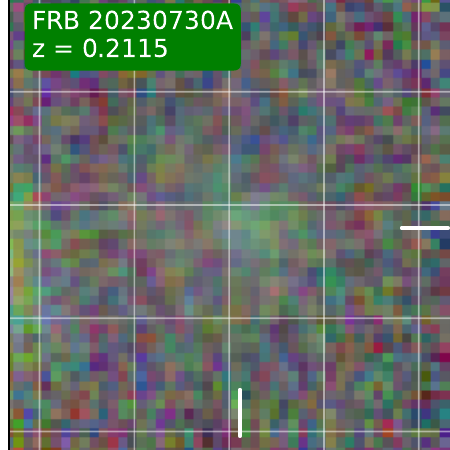} 
\includegraphics[trim=2.5in 0.5in 2.5in 0.5in,clip,width=0.200\textwidth]{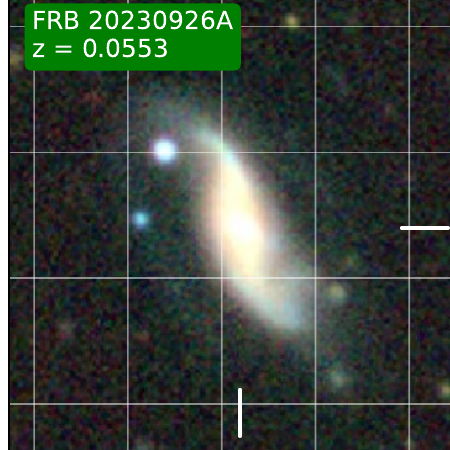} 
\includegraphics[trim=2.5in 0.5in 2.5in 0.5in,clip,width=0.200\textwidth]{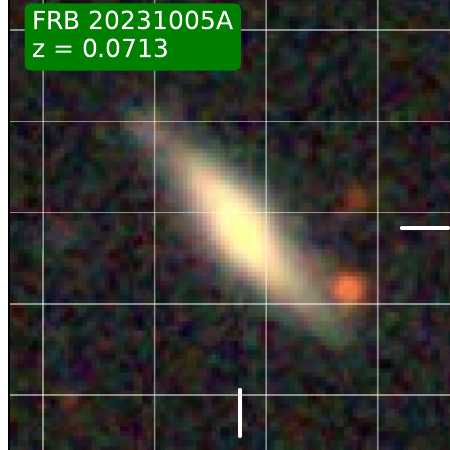} 
\includegraphics[trim=2.5in 0.5in 2.5in 0.5in,clip,width=0.200\textwidth]{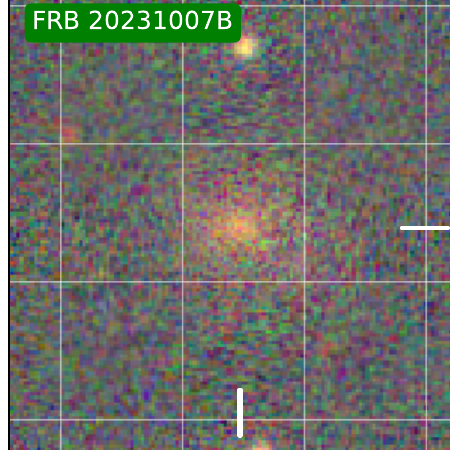} 
\includegraphics[trim=2.5in 0.5in 2.5in 0.5in,clip,width=0.200\textwidth]{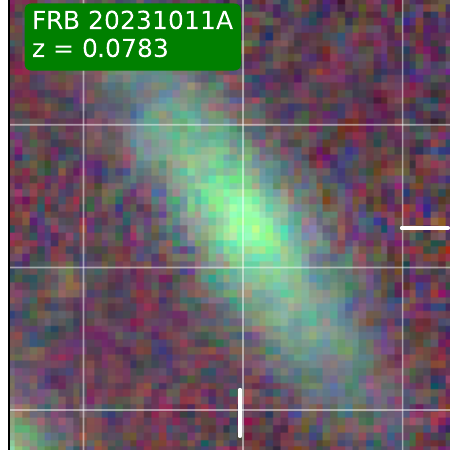} 
\includegraphics[trim=2.5in 0.5in 2.5in 0.5in,clip,width=0.200\textwidth]{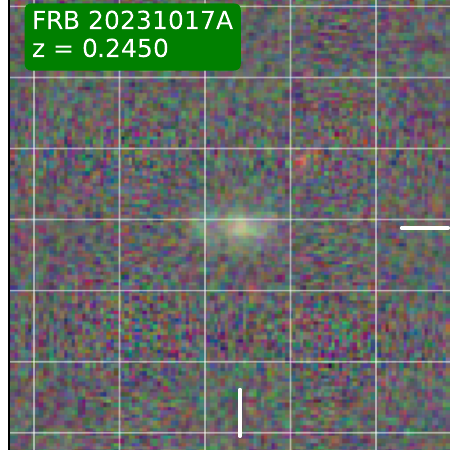} 
\includegraphics[trim=2.5in 0.5in 2.5in 0.5in,clip,width=0.200\textwidth]{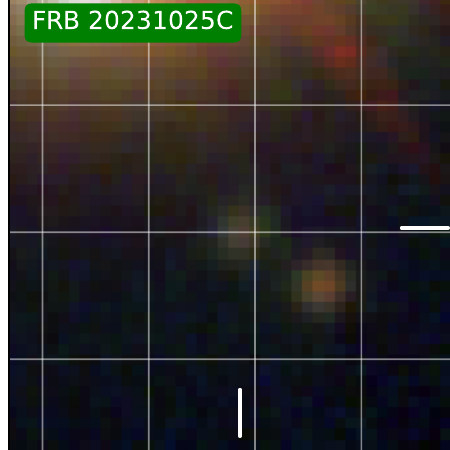} 
\includegraphics[trim=2.5in 0.5in 2.5in 0.5in,clip,width=0.200\textwidth]{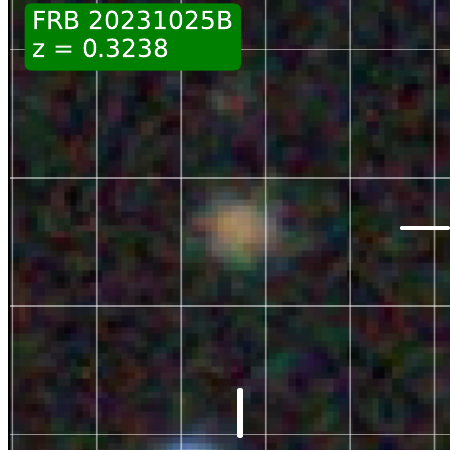}
\includegraphics[trim=2.5in 0.5in 2.5in 0.5in,clip,width=0.200\textwidth]{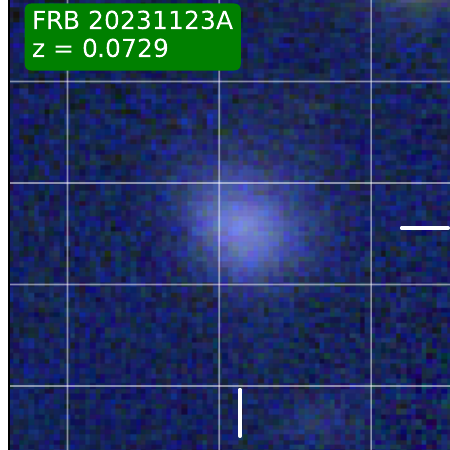}
\includegraphics[trim=2.5in 0.5in 2.5in 0.5in,clip,width=0.200\textwidth]{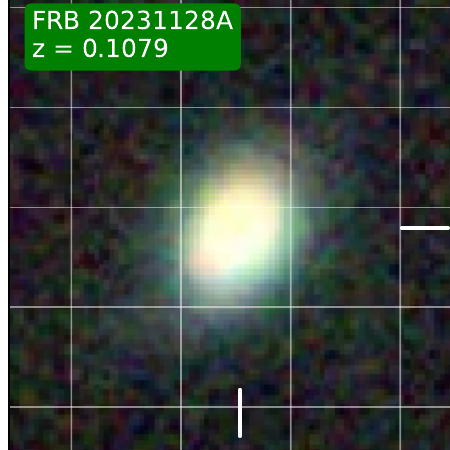}
\includegraphics[trim=2.5in 0.5in 2.5in 0.5in,clip,width=0.200\textwidth]{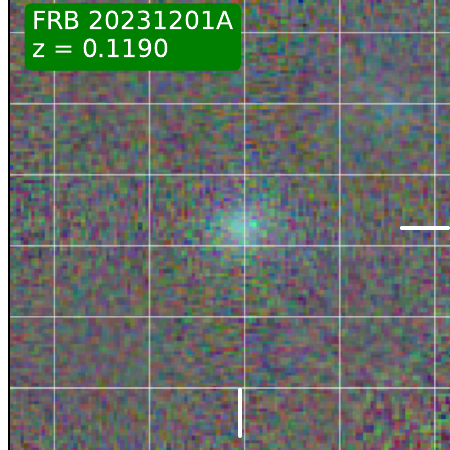}
\includegraphics[trim=2.5in 0.5in 2.5in 0.5in,clip,width=0.200\textwidth]{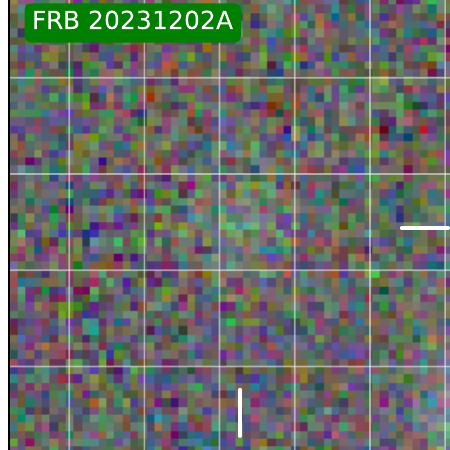}
\includegraphics[trim=2.5in 0.5in 2.5in 0.5in,clip,width=0.200\textwidth]{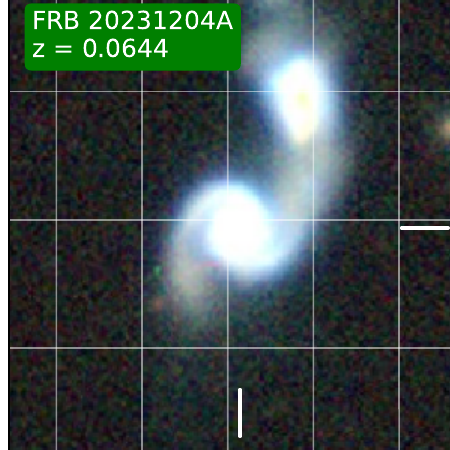}
\includegraphics[trim=2.5in 0.5in 2.5in 0.5in,clip,width=0.200\textwidth]{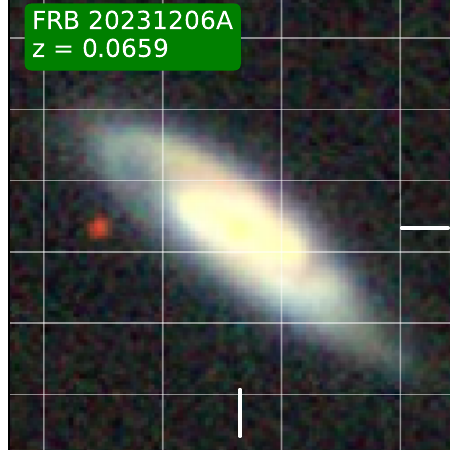}
\includegraphics[trim=2.5in 0.5in 2.5in 0.5in,clip,width=0.200\textwidth]{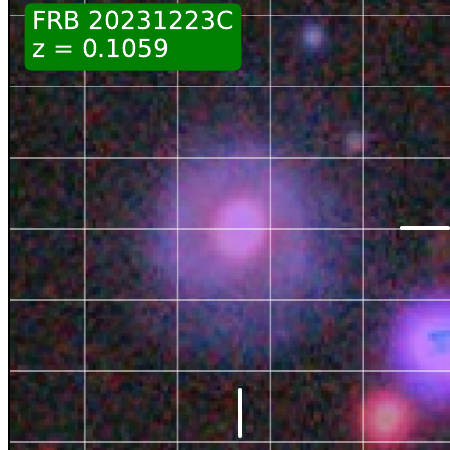}
\includegraphics[trim=2.5in 0.5in 2.5in 0.5in,clip,width=0.200\textwidth]{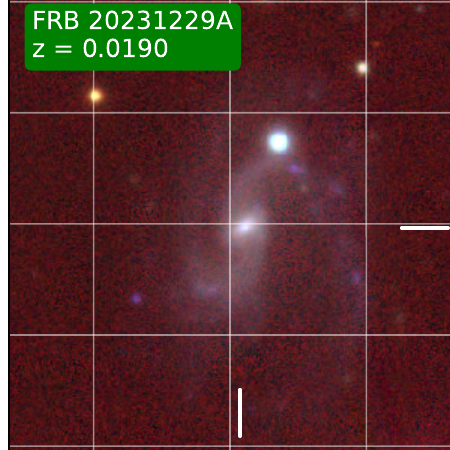}
\includegraphics[trim=2.5in 0.5in 2.5in 0.5in,clip,width=0.200\textwidth]{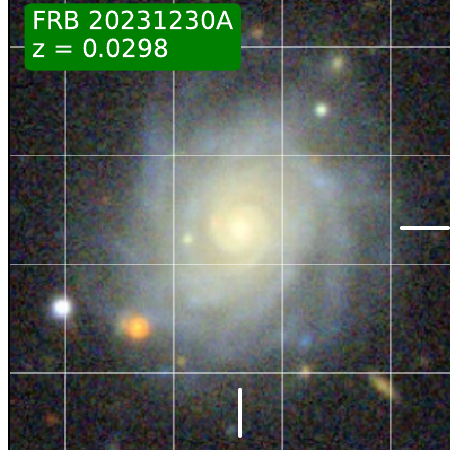}
\caption{Publicly-available images for the 21 host galaxies in our gold sample (those with $P(O|x) > 0.9$) as well as FRB 20230311A (see text), taken from the DECaLS or Pan-STARRS archive. The host of FRB 20230730A is difficult to detect in imaging due to Galactic extinction, but was spectroscopically confirmed. The images of some fields (FRB 20230222B, 20231123A, and 20231223C) lacked one or more color channels in publicly available images, resulting in their tinted appearance.}
\label{fig:hg_images}
\end{figure*}

Since our localization ellipses are on average larger than our host galaxies, we show the full field of each localization in Figure~\ref{fig:path}. The size of each field is set by fixing its height to $3 \times$ the major axis length at a fixed 1:2 aspect ratio. Additionally, we zoom in on each host galaxy candidate in Figure~\ref{fig:hg_images} to highlight the morphology of each host galaxy. 

\section{Spectroscopic Follow-up}\label{sec:followup}
\msbold{We conducted spectroscopic follow-up observations for the secure host candidates in our sample with the primary goal of measuring their redshifts.} Our host candidates have brightnesses of $16.4 < m_r < 22.7$. We prioritized follow-up observations in cases where the host association was considered secure ($P(O|x) > 0.9$), where the redshift was not publicly available. \msbold{To make the most of our limited observing time and maximize sensitivity, we targeted only fields with a low Galactic extinction ($E(B-V) < 0.30$) with the exception of the host galaxy of FRB\,20230730A ($E(B-V) \sim 0.54$), which we observed in the early stages of our follow-up program.} Including this system, we had a total of 16 targets for optical follow-up. In cases where the secondary host candidate (the galaxy in the field with the second-highest $P(O|x)$) was comparable in brightness, we positioned the slit to contain the top two host galaxy candidates within the FoV. For nine host galaxies, we used the Kast Double Spectrograph mounted on the Shane 3-m telescope at Lick Observatory on Mount Hamilton in San Jose, California (PI: Kahinga). For five host galaxies, we used DEIMOS on the Keck II telescope in Mauna Kea (PI: Prochaska), and for one host galaxy we used the \msbold{8.1-m Gemini North telescope in Mauna Kea} as part of our Long and Large Program (PI: Eftekhari). These targets were allocated to these respective follow-up resources on the basis of the faintness/sky positions of the desired targets and the availability of observing time.

For all of our spectroscopic observations we used the Python Spectroscopic Data Reduction Pipeline ~\citep[PypeIt v1.16 for Gemini data and v1.15 otherwise][]{prochaska2020pypeit}, which included spectral data reduction, flux calibration, and co-adding (stacking) the spectra from all exposures. We used standard star observations obtained on each night to generate sensitivity functions, which were used to flux calibrate the spectra. The redshifts were then obtained by manual inspection of the coadded spectra to identify prominent
emission lines. Two or more emission lines were used to confirm each redshift (see Table \ref{tab:z_table}), except for the host galaxy of FRB\,20230730A for which only the H$\alpha$ line was detectable. We attribute this to the high Galactic extinction along this sightline. We find in accordance with expectations that all of our hosts are at low redshifts ($z < 0.33$). The observation dates, targets, host magnitudes, exposure times, and redshifts of galaxies in the gold sample are presented in Table~\ref{tab:z_table}. 

\subsection{Lick Observations}
The majority of redshifts were obtained using follow-up observations at Lick Observatory with the Kast Spectrograph on the Shane 3-m Telescope across 3 semesters (2023B, 2024A and 2024B). 
We used the D57 dichroic along with the $145''$ long and $2''$ wide slit. \msbold{The seeing was $\sim1.2'',1.5'',0.8''$ for the nights in 2023B, 2024A, and 2024B respectively and all targets were observed at airmasses $<1.5$.}

\msbold{In our early (2023B) observations, we covered a wavelength range of $\sim5400 - 8800$\,\AA\ on the red side using the 600/7500 grating and $\sim3000 - 5800$\,\AA\ on the blue side using the 452/3306 grism.
Then, we optimized our wavelength coverage to maximize the chance of covering the H$\alpha$, [NII], and SII emission lines on the red side of the detector, and [OII] and H$\beta$ lines on the blue side.
In later observations (2024A and 2024B), we used the 600/7500 grating tilted to provide wavelength coverage of $\sim5700 - 9000$\,\AA on the red side, and the 600/4310 grism with a shifted CCD to provide a wavelength coverage of $\sim3600 - 5700$\,\AA\ on the blue side.}

\subsection{Keck Observations}
On December 14, 2023 UT we observed the highest \pox\
candidates of FRBs 20230203A, 20230703A, 20230730A, 20230311A, and 20231011A using the DEIMOS spectrograph \citep{deimos} on the Keck~II 10-m telescope. We used custom-made slitmasks with 1$''$-wide
slits
for FRBs 20230311A and 20230730A to observe the primary candidate and additional galaxies in the field. 
We observed the other host galaxy candidates with a $300''$ long $0.7''$ wide slit. 
In all cases, we used the 600ZD grating which provides
a resolution of 3.5\AA~at full width half maximum
along with the GG455 filter. The central wavelength for all observations was set to $7500$\,\AA. 
The seeing was $\sim1''$ throughout the night, and the observing conditions were spectroscopic. All targets were observed at airmasses $<1.5$. 

\subsection{Gemini Observations}
We observed the candidate host galaxies of FRBs 20231017A, 20231025B, and 20231201A using the Gemini Multi-Object Spectrograph \citep[GMOS;][]{GMOS} on the Gemini-North telescope located at Maunakea Observatory, Hawai'i through program GN-2024B-LP-110.  For each observation, a total of four 900 s exposures were taken using the B480 grating using a 1\arcsec\ slit; we obtained two exposures each in a setup centered at 6400 and 6500\,\AA\ with a 15\arcsec\ spatial offset along the slit in both setups.  This provided continuous spectral coverage in the combined data from 4450--8450\AA. \msbold{For each FRB, we oriented the GMOS slitmask at a specific position angle to ensure it covered both the primary and secondary host candidates favored by PATH.} 

\section{Multi-wavelength counterpart cross-matching}\label{sec:crossmatch}
\msbold{Multi-wavelength counterparts may yield crucial insight into the origin of FRBs. The observation of simultaneous X-ray burst from SGR 1935+2154~\citep{collaboration2020bright,bochenek2020fast,mereghetti2020integral} motivates searches for high-energy counterparts from extragalactic FRBs. In addition, so-called ``persistent radio source'' (PRS) counterparts have been associated with several active repeating sources\citep{marcote2017repeating,bhandari2023constraints}. These PRS are potentially significant parts of the FRB puzzle since they are too luminous and compact to be explained by star formation.}

Our sample is a natural dataset in which to search for multi-wavelength counterparts to FRBs, due to the improved localization (median 1$\sigma$ ellipse area of 113 arcsec$^2$), sample size (81 bursts), and low redshifts of the 19 bursts in the sample with hosts (median $z \sim 0.1$). We cross-match the localization ellipses with the locations of known gamma-ray bursts (GRBs), and search for compact PRSs for the bursts within our sample. Future work will present a comprehensive search for coincidences between these sources and known optical transients such as supernovae (Dong et al., in prep.).

\subsection{High-energy counterpart search}
We search for temporal and spatial cross-matches between our 81 FRBs and 2589 long and short GRBs triggers from the GRBWeb database\footnote{\url{https://user-web.icecube.wisc.edu/\~grbweb_public/}} and the General Coordinates Network\footnote{\url{gcn.nasa.gov}} between April 21, 1991 and September 9, 2024. Using only the GRBs with 1$\sigma$ localizations $<1^\circ$, we do not find any joint spatio-temporal (within 1 week in time and within 3$\sigma$ in space) coincidences. We do find three solely spatial cross-matches between FRBs 20231019A, 20230802A, 20230703A and three GRBs: 970306A, 120323A and 920221A*, respectively. To assess the chance coincidence probability we simulate 1000 GRBs and cross-match each with the FRB sample as done by \citet{Curtin2024}. Each mock sample of GRBs has at least three cross matches, implying a high chance coincidence probability, which is consistent with the large uncertainty regions for these GRBs ($\gtrsim0.5^\circ$).  

\subsection{Persistent Radio Source cross-matching}
Thus far, two FRBs have been confidently associated with a PRS~\citep{chatterjee2017direct, niu2021repeating}. FRBs\,20121102A and 20190520B, which have been shown to be compact, luminous, and therefore distinct from emission attributable to star-formation~\citep{marcote2017repeating,bhandari2023constraints}. A third PRS was proposed to be associated with the active repeating FRB\,20201124A \citep{Bruni2024Nature}, but its \msbold{larger physical size, larger spatial offset relative to the FRB, and lower luminosity hint at a physical explanation distinct from the first two.}

\msbold{We perform a systematic search for unresolved radio sources coincident with the CHIME-KKO FRBs in the Faint Images of the Radio Sky at Twenty cm (FIRST; \citealt{FIRST}) and the NRAO VLA Sky Survey (NVSS; \citealt{NVSS}) at 1.4\,GHz, as well as the Very Large Array Sky Survey (VLASS; \citealt{VLASS}) at 3\,GHz. For sources detected in either VLASS or FIRST, we also search the positions in the LOFAR Two-metre Sky Survey (LoTSS) Data Release 2 (DR2; \citealt{LoTSS}) at 144\,MHz. In each survey, we searched for radio point sources above the $5\sigma$ detection threshold, using the Canadian Initiative for Radio Astronomy Data Analysis (CIRADA)\footnote{\url{http://cutouts.cirada.ca/}} portal to access the image cutouts as was done in~\citet{Ibik2024prs}. For sightlines with non-detections, we place upper limits on PRS emission using VLASS data, due to its superior sensitivity and its survey footprint, which overlaps best with the CHIME-KKO dataset.}

In the FIRST and NVSS surveys, we estimated the noise by computing the RMS flux (i.e., $\sigma$) over the entire cutout, whose angular extent is 2' $\times$ 2'.  While this can slightly overestimate the noise if a particularly dominant radio source is present within the field, we found that this method was sufficient for achieving relatively uniform noise estimates across all fields. In VLASS, cutouts of the imaging residuals allowed for a better noise RMS estimate, which improves the sensitivity of our search and deepens our upper limits on persistent radio emission. We identified two potential PRSs detected at $3-5\sigma$ significance in VLASS and discuss each of these sources in \S\ref{sec:FRB20230926A} and \S\ref{sec:FRB20231128A}. In addition, for the 19 FRBs with secure redshifts, we translate the VLASS RMS flux sensitivity into upper limits on the PRS luminosity at the redshift of the FRB at 3\,GHz, as listed in Table \ref{tab:z_table} (excluding FRB\,20231204A, which had no associated VLASS query). 

Due to the low redshift of our host galaxy sample, the majority of our limits fall at or below the luminosities of PRSs associated with FRB\,20121102A and FRB\,20190520B \citep{chatterjee2017direct, niu2021repeating}, adding to the growing body of evidence that those two sources are rare outliers in the FRB population \citep{Law_2022, law2024deep, sharma2024preferential}, with the caveat that fainter PRSs may exist below the VLASS detection threshold \citep[e.g.,][]{Ibik2024prs}. 

Moreover, the limits on a potential PRS are consistent with those found for a sample of FRB repeaters \citep{Ibik2024prs}, and may suggest intrinsically fainter PRSs when compared to the PRSs associated with FRB\,20121102A and FRB\,20190520B, but similar to those associated with FRB\,20201124A \citep{Bruni2024Nature}. \msbold{These luminosity limits are also similar to those presented by~\citet{Eftekhari2020} and \citet{Dong2024radiosources}.} In the neutron star nebula models (see e.g., \citealt{Kashiyama2017, margalit2018concordance}), the lower luminosities can be explained by factors such as older age, larger nebula radius, or intrinsically lower energy output \citep{Margalit2019nebulae}. 

\section{Sources of Interest}\label{sec:sample}
\begin{figure*}
    \centering
    \includegraphics[width=\linewidth]{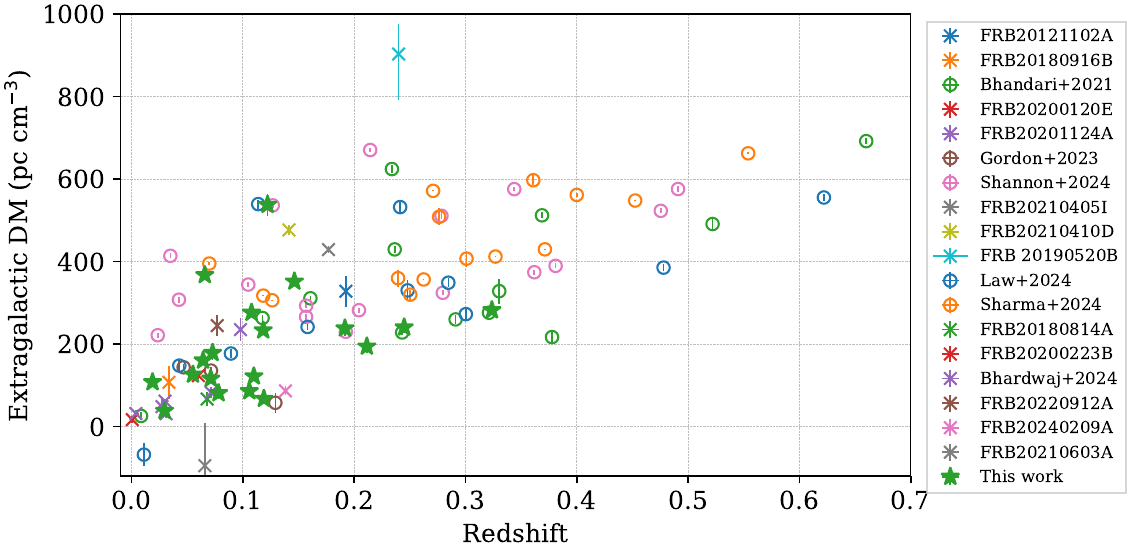}
    \caption{The Macquart relation for a selection of host galaxies in the literature shows the complementary redshift coverage of various FRB surveys at the time of this writing. Galaxy redshifts and FRB DM measurements are taken from~\citep{tendulkar2017host, marcote2020repeating, bhandari2022characterizing,gordon2023demographics,law2024deep,sharma2024preferential,michilli2023subarcminute,bhardwaj2024host,Ibik2024host}.}
    \label{fig:macquart}
\end{figure*}

\msbold{In this section, we highlight some of the interesting sources in our sample, whose redshift distribution is illustrated in Fig.~\ref{fig:macquart}. As a whole, our systems add more coverage to the Macquart relation at low redshifts, roughly doubling the sample of bursts at $z \lesssim 0.2$. The low redshift regime is particularly important since it enables a characterization of the FRB population while minimizing the effect of redshift evolution, and to enable detailed multi-wavelength studies of their hosts.} Two bursts in our sample have potential PRS candidates, one of which has a low chance-coincidence probability. Four of our host galaxies are embedded in larger-scale structures: one is in a massive X-ray cluster, while two others may reside in gas-rich and gas-poor optical cluster environments respectively. A third is a repeating FRB associated with a star-bursting galaxy merger. We comment on the rich potential of FRBs to probe these systems but reserve a detailed treatment for future work.

\subsection{FRB 20231204A: a repeating FRB probing the starburst gas in a galaxy merger}\label{sec:FRB20231204A}
FRB 20231204A is a repeat burst from the source FRB 20190303A, which~\citet{michilli2023subarcminute} previously localized to an interacting galaxy pair. In that work, the localization was too large to identify which of the member galaxies was the host. Here, the updated localization provided by FRB 20231204A indicates that the FRB  originates from the member galaxy to the southwest, which is separated from the northeast member galaxy by only $\sim$ 20 kpc in projection. Long-slit spectroscopy and Baldwin-Phillips-Terlevich (BPT) diagram classification revealed both galaxies in the interacting system to be star-forming, $\sim L^*$ galaxies; We refer the interested reader to~\citet{michilli2023subarcminute,curtin2024morphology,ng2024polarization} for further details regarding its spectroscopic characterization, burst morphologies, and polarimetric monitoring respectively.

\msbold{The existence of an FRB in an interacting galaxy pair enables an interesting measurement of its gas content and ionization state.} We follow~\citet{howk2003ionized,howk2006method}, who combined observations of millisecond pulsars in the globular cluster M3 with far-ultraviolet spectroscopy to constrain the ionization fraction along a high Galactic-latitude sightline in the Milky Way. We apply a simpler version of this idea to this FRB sightline, which skewers its host galaxy at a high host galactic latitude. 

\msbold{This argument relies on the FRB sightline passing through some fraction $f$ of the total HI column density $\Sigma$ towards the FRB. While it is difficult to constrain whether this particular source is behind or in front of its HI column, the FRB is equally likely to be in the foreground ($f = 0$) as it is to be in the background ($f = 1$) for a randomly-oriented sightline. On average, the fraction of the HI column traversed by the FRB sightline will be $\langle f \rangle = 1/2$. In an average over many FRB sightlines, this statistical argument will be more robust. In the meantime, we demonstrate the principle of constraining the sightline-integrated ionization fraction $f_\mathrm{ion}$ by combining the host contribution to the DM excess with an estimate of the HI column which we regard as highly conservative.}
$$\left(\dfrac{f_\mathrm{ion}}{0.78}\right)\left(\dfrac{f}{1/2}\right)\left(\dfrac{ \Sigma}{10^{21}\textrm{ cm}^{-3}}\right) \lesssim \dfrac{\dmh}{112\textrm{ pc cm}^{-3}}.$$

\noindent where the rest-frame host galaxy DM contribution on the right-hand side is defined by

\begin{equation}
\dfrac{\dmh}{1 + z_h} = \textrm{DM}_\mathrm{tot} - \langle \textrm{DM}_\textrm{cosmic}(z_h)\rangle - \textrm{DM}_{\textrm{MW,ISM}} - \textrm{DM}_{\textrm{MW,halo}}.
\label{eq:dmh_definition}
\end{equation}

\noindent with $z_h$ denoting the redshift of the host galaxy and where the Milky Way disk and halo contributions  $\textrm{DM}_{\textrm{MW,ISM}}$ and $\textrm{DM}_{\textrm{MW,halo}}$ are estimated using the NE2001~\citep{cordes2002ne2001} and~\citet{yamasaki2020galactic} models, respectively. The cosmic contribution is calculated using the \texttt{igm.average\_DM} function in the open source \href{https://github.com/FRBs/FRB/blob/main/frb/dm/igm.py}{FRB} repository with $f_\textrm{diffuse} = 0.844$.

Combined with dedicated observations of neutral gas in this galaxy merger, this system could provide the first observations of \msbold{both the neutral and ionized components of the gas surrounding a galaxy merger.}

If the HI column density $\Sigma$ were known, the ionization fraction could robustly be constrained along this sightline. While archival HI observations are not publicly available for this merging galaxy system,~\citet{tempel2014flux} suggests that it is ten times more luminous than the Milky Way ($2 \times 10^{11} L_\odot$), whose HI column density exceeds $\Sigma \gtrsim 10^{21}$ cm$^{-2}$ out to $15$ kpc ~\citep{kalberla2009hi}, which is a plausible upper limit for the displacement of the FRB from the center of its host galaxy. A more precise FRB localization would also make this estimate more quantitative.

This exercise could be extended to other FRB hosts in the literature which show some evidence of gas disturbance. However, the disturbance in those systems is not as striking as in the case of FRB 20231204A which exhibits a major galaxy merger:~\citet{michalowski2021asymmetric} uses HI line-asymmetry arguments to argue for evidence of recent gas inflow in a handful of FRB hosts, while~\citet{kaur2022fast} claim evidence for a minor merger. In the future, a sample of precisely localized FRBs as well as auxiliary observations characterizing the neutral gas of the star formation site ~\citep[e.g., ][]{leewadell2023host,yamanaka2024alma,glowacki2023wallaby} may be able to discriminate between the ISM and CGM contributions to star formation during galaxy mergers on the basis of their different ionization properties~\citep{sparre2023gas}.

\subsection{FRB 20231206A: an FRB in Abell 576}\label{sec:FRB20231206A}
The host galaxy of FRB 20231206A is a member of the galaxy cluster Abell 576~\citep{abell1989catalog,rines2000infall} and the burst has a large dispersion measure excess ($\dmh=295$ pc cm$^{-3}$). This X-ray cluster at $z \approx 0.03$ occupies $\sim 9$ square degrees on the sky and hosts over a thousand galaxies. A subset of the galaxies in the direction of the cluster appear at a significantly different redshift~\citep{rines2000infall} than the others; this has led to its identification as an infalling subcomponent with a relative velocity of $\sim 3000$ km~s$^{-1}$~\citep{dupke2007merger} relative to the X-ray core of the cluster. The host galaxy of FRB 20231206A -- one of the most luminous in our sample -- is part of this infalling sub-cluster. It has been proposed on the basis of the X-ray morphology of the cluster core that the sub-structure has already made its first passage through the core~\citep{kempner2004chandra}, but neither the morphological nor the redshift information can reveal whether the sub-structure is in front or behind the core as viewed from Earth.

The DM excess of FRB 20231206A suggests a way to break this degeneracy. At the redshift of the cluster center, the angular offset of the host galaxy of FRB 20231206A implies a transverse physical offset of $\approx 5$ Mpc from the cluster center, whose extent is quantified by its virial radius \msbold{$r_\textrm{200,crit} = 2.02$ Mpc} ~\citep[see \S3.1 of][]{rines2000infall}. Within the virial radius, the mass of the cluster is between $1.0-1.4 \times 10^{15} M_\odot$ ~\citep{girardi1998optical,rines2000infall}. We argue that the large (295 pc cm$^{-3}$) DM excess suggests the substructure FRB host lies behind the X-ray cluster rather than in front of it; it could then potentially constrain the ionized gas column of the ICM at large impact parameters (in this case, a gas density measurement at 2 $r_\textrm{200,crit}$) far beyond the typical reach of X-ray based gas fraction measurements which typically use cluster gas fraction measurements at smaller impact parameters to constrain $\Omega_b/\Omega_m$ ~\citep[e.g.,][]{ettori1999rosat, mantz2022cosmological}.
\subsection{FRB 20230203A: an edge-on FRB potentially within a rich optical cluster}\label{sec:FRB20230203A}
The sightline of FRB 20230203A has a notable DM excess of $\dmh = 203$ pc cm$^{-3}$. This may be attributable to the high inclination angle~\citep[$i\approx 80^\circ$;][]{kourkchi2020cosmicflows}, and therefore a larger ISM contribution, from the host galaxy. It may also be due to the intracluster medium of the rich, optically selected cluster NSC 100649+353929, which was discovered in the Northern Sky Optical Cluster Survey~\citep{2000AJ....119...12G,gal2003northern} and whose center is 3' away from the host galaxy of FRB 20230203A. We discuss this possibility acknowledging that the lack of spectroscopic redshifts for the cluster prevents us from drawing a definitive conclusion. 

In the radial direction, the cluster's photometric redshift ($z_{\textrm{phot}} = 0.1411$) is within 1500 km/s of the spectroscopic redshift of the host galaxy ($z_\mathrm{spec} = 0.1464$), suggesting that the host galaxy may be a cluster member. In the transverse direction, the FRB sightline passes through the virial radius of the cluster, which may be inferred using the richness-$r_{200}$ relation of~\citet{gal2009northern}. For rich optical clusters ($N_\textrm{gal} = 63$) like NSC 100649+353929, the $N_\textrm{gal}$-$R_{200}$ relation suggests a virial radius of 1.8 Mpc and an angular extent of $\sim12'$ at the redshift of the FRB. Though the host offset is unconstrained by our localization, the cluster's angular diameter encompasses the FRB sightline for reasonable FRB host offsets.

Another potential foreground candidate is SDSS CGB 73719, a photometrically identified galaxy group consisting of five members. The spectroscopic redshift for SDSS CGB 73719 is unknown and publicly available photometric redshift information from DESI DR9~\citep{zhou2023target} are inconclusive, but we note that the median redshift of groups in its parent catalog is $z = 0.09$, making it another potential contributor to the foreground DM budget of FRB 20230203A~\citep{mcconnachie2009compact}. Mass and redshift estimates from spectroscopic follow-up of this field will conclusively determine the host disk and foreground contributions to the DM budget of this sightline.

\subsection{FRB 20230703A: an FRB behind a gas-poor cluster}\label{sec:FRB20230703A}
We noticed that the line of sight to the host galaxy of FRB 20230703A~\citep{2023ApJS..268...14C} intersects seven galaxies in the NASA Extragalactic Database-Local Volume Sample (NED-LVS) catalog within their virial radii. Subsequent searches identified these as members of NSCS J121847+484410, a small cluster from the Northern Sky Optical Cluster Survey (NoSOCS)~\citep{2000AJ....119...12G}. This cluster is reported to have a virial mass $M_{200} = 3.6\times 10^{13} M_\odot$ (assuming an NFW profile),  and a heliocentric redshift $z = 0.0448$~\citep{2009MNRAS.392..135L}, compared with $z_h = 0.1184$.

It is notable that, despite seemingly passing through a cluster ICM, this FRB has \msbold{$\dmh=163~$pc~cm$^{-3}$. Assuming a reasonable electron density profile for the intracluster medium (a modified Navarro-Frenk-White profile with 75\% of the total cosmic baryons are in the hot phase~\citep{prochaska2019probing}) centered on the cluster's published position
, the expected DM contribution of this cluster is about $228$ pc~cm$^{-3}$. We defer to a future publication an analysis of this and other clusters (Lanman et al., in prep).}

\subsection{FRB 20230926A: radio emission from an FRB host}\label{sec:FRB20230926A}
We detected a radio source in VLASS with a significance of $4\sigma$ at the position of the FRB host of FRB 20230926A. The radio source has a luminosity of $L_\mathrm{3.0~\mathrm{GHz}} \sim 4\times10^{28}~\mathrm{erg~s}^{-1}~\mathrm{Hz}^{-1}$ (Table \ref{tab:gold_sample}) and is coincident with the extended radio source ILTJ175629.78+414836.4 (2.3 mJy; $>25\sigma$) detected in LoTSS at $144~\mathrm{MHz}$. While intriguing, the high chance coincidence probability between the radio burst and the radio source ($P_\mathrm{cc} \sim 0.1$) prevents us from claiming a statistically significant association \msbold{between the burst and the radio source. This source potentially adds to the growing sample of FRB hosts with radio-detected star formation~\citep[e.g.,][]{dong2024mapping}.}

\subsection{FRB 20231128A: a FRB with a luminous PRS candidate}\label{sec:FRB20231128A}
FRB 20231128A is a repeat burst of FRB 20191106C, which was first reported in~\citet{collaboration2023chime} with a marginal host galaxy association~\citep{Ibik2024host} confirmed now by this work. We found a point-like radio point in FIRST within the 1$\sigma$ localization ellipse at $13^h18^m18^s.77 +42^\circ59'33.75''$. The probability of chance coincidence ($P_\mathrm{cc}$) of the radio source of this flux with the localization ellipse of FRB\,20231128A was 0.008 (note that this was below the flux threshold used in a previous search described in~\citet{Ibik2024prs}). \msbold{The angular resolution of FIRST (5'') constrains the transverse size of the emission to $<10$ kpc at the distance of the host, and the FIRST flux (700$\mu$Jy, $4.9\sigma$) implies a specific luminosity of L$_{1.4 ~\mathrm{GHz}}$ = $1.95\times10^{29}~\mathrm{erg~s}^{-1}~\mathrm{Hz}^{-1}$.} At the position of the FIRST source, we identified a bright counterpart in LoTSS (ILT J131819.22+425958.9) whose peak flux of $2.2 \pm 0.1$ mJy ($35\sigma$) implies a specific luminosity of $L_{\mathrm{144~\mathrm{MHz}}} = 8\times10^{29}~\mathrm{erg~s}^{-1}~\mathrm{Hz}^{-1}$, and a spectral index of $\alpha = -0.60 \pm 0.05$. 

Radio emission from an AGN and star formation activity in the host galaxy are two possible explanations for the radio emission. However, the AGN hypothesis is disfavored based on the possible offset between the radio emission and the optical center of the host ($1.2 \pm 1''$, $2.3 \pm 1.9$ kpc), as well as the optical spectrum of the host galaxy, which classifies it as a star-forming galaxy~\citep{hardcastle2023lofar}. A background AGN also cannot be ruled out based on the data presented here; nevertheless, efforts are currently underway to characterize the source on milliarcsecond scales using the European VLBI Network, which will conclusively constrain its compactness and host galaxy offset, which will help discriminate between these possibilities.
\subsection{FRB 20231229A: a nearby FRB with HI in its host galaxy}\label{sec:FRB20231229A}
FRB 20231229A is the closest FRB in our sample at a distance of $\sim 90$ Mpc. We cross-matched its host galaxy to UGC 1234; it joins the small sample of five local-universe FRBs with known 21-cm emission, having a line profile that has been measured by ALFALFA~\citep{haynes2018arecibo}. \msbold{Its implied HI gas mass is $10^{10} M_\odot$; morphological classification of its HI emission line profile suggests no statistically significant signs of asymmetry~\citep{yu2022statistical}, in contrast to other FRB hosts with HI data~\citep{michalowski2021asymmetric}.}

\section{Conclusion}
The challenge of localizing FRBs and associating them with their host galaxies has been a bottleneck in the ongoing effort to understand their origins and utility as cosmic probes. In this paper, we have demonstrated the scalability and astrometric robustness of wide-field VLBI on a large sample of 81 FRBs, using data from the first operational CHIME/FRB Outrigger KKO~\citep{lanman2024chime}. We use a VLBI observation strategy and a calibrator grid developed in prior papers~\citep{leung2021synoptic,andrew2024vlbi} to present several-arcsecond accuracy localizations. Using PATH, we present secure host galaxy associations for 21 sources and compile redshifts for 19, of which 15 are newly obtained from our optical follow-up programs at the Lick, Keck, and Gemini Observatories. Many of our sources are at sufficiently low redshift $(z < 0.2)$ for exquisite multiwavelength characterization, but we reserve detailed spectral characterization of the host galaxies for future work.

The wide field of view of CHIME/FRB Outriggers will yield many low-redshift systems, which are rich targets for understanding the origin of FRBs. This is highly complementary to other detection experiments which have deeper redshift distributions. One source presented separately from this paper (FRB 20240209A) displays a large offset from its massive elliptical host, which is the most massive to date; see~\citet{shah2024repeating,eftekhari2024massive}. One sightline reveals a candidate PRS with a luminosity similar to that of FRB 20121102A; towards other sightlines, upper limits on persistent radio emission challenge the existence of PRS in a large low-redshift sample. More broadly, the observational techniques for FRB detection and localization, host galaxy association, and follow-up pave the way for upcoming large-scale FRB surveys such as CHORD and the DSA-2000~\citep{vanderlinde2019canadian,hallinan2019dsa}, which will revolutionize our understanding of FRBs and the universe they illuminate.

\begin{table*} 
\centering
\addtolength{\leftskip}{-6cm}
\begin{longtable}{lllllllccll}
\caption{A table of localizations for our gold sample of 21 FRBs which have secure host galaxy associations ($P(O|x) > 0.9$). We also include FRB 20230311A which has a secure redshift, but an ambiguous host (see \S\ref{sec:association}). All coordinates are provided in the ICRS frame, and localization contours are provided as ellipses with minor and major axis uncertainties provided as $b_{err}$ and $a_{err}$, measured in arcminutes, and angles measured in degrees east of north. Burst dispersion measures are provided in units of pc cm$^{-3}$ and have negligible uncertainties. The flux is defined as the peak flux of the burst in Janskys after applying the structure-maximizing DM found by DM-phase~\citep{seymour2019dm}; fluxes and fluences (in Jy ms) are quoted to 10\% accuracy, as done in~\citet{chime2024updating}. Upper limits on persistent radio emission are from VLASS at 3\,GHz in erg s$^{-1}$ Hz$^{-1}$; a dagger indicates that the VLASS cutout was not available. FRB 20230926A and 20231128A have associated persistent radio source counterparts which are detected at the $4-5\sigma$ level VLASS and FIRST respectively; both counterparts are unambiguously detected in LoTSS (see text). In the table entry for those sources, we report specific luminosities derived from VLASS at 3\,GHz and FIRST at 1.4\,GHz, respectively.} \label{tab:gold_sample} \\
\toprule
$\text{Name}$ & $\text{RA}_{\text{FRB}}$ & $\text{DEC}_{\text{FRB}}$ & $b_{\text{err}}$ & $a_{\text{err}}$ & Angle & $\text{DM}$ & $\text{Flux}$ & $\text{Fluence}$ & $\text{P(O} \vert \text{x)}$ & $L_\text{3GHz}$ \\
\midrule
\endfirsthead
\caption[]{A table of localizations for our gold sample of 21 FRBs which have secure host galaxy associations ($P(O|x) > 0.9$). We include FRB 20230311A which has a secure redshift, but an ambiguous host (see \S\ref{sec:association}). All coordinates are provided in the ICRS frame, and localization contours are provided as ellipses with minor and major axis uncertainties provided as $b_{err}$ and $a_{err}$, measured in arcminutes, and angles measured in degrees east of north. Burst dispersion measures are provided in units of pc cm$^{-3}$ and have negligible uncertainties. The flux is defined as the peak flux of the burst in Janskys after applying the structure-maximizing DM found by DM-phase~\citep{seymour2019dm}; fluxes and fluences (in Jy ms) are quoted to 10\% accuracy, as done in~\citet{chime2024updating}. Upper limits on persistent radio emission are from VLASS at 3\,GHz in erg s$^{-1}$ Hz$^{-1}$; a dagger indicates that the VLASS cutout was not available. FRB 20230926A and 20231128A have associated persistent radio source counterparts which are detected at the $4-5\sigma$ level VLASS and FIRST respectively; both counterparts are unambiguously detected in LoTSS (see text). In the table entry for those sources, we report specific luminosities derived from VLASS at 3\,GHz and FIRST at 1.4\,GHz, respectively.} \\
\toprule
$\text{Name}$ & $\text{RA}_{\text{FRB}}$ & $\text{DEC}_{\text{FRB}}$ & $b_{\text{err}}$ & $a_{\text{err}}$ & Angle & $\text{DM}$ & $\text{Flux}$ & $\text{Fluence}$ & $\text{P(O} \vert \text{x)}$ & $L_\text{3GHz}$ \\
\midrule
\endhead
\midrule
\multicolumn{11}{r}{Continued on next page} \\
\midrule
\endfoot
\bottomrule
\endlastfoot
FRB 20230203A & 151.66159 & 35.69410 & 0.033 & 0.300 & 10.15 & $420.1$ & $6.2$ & $173.0$ & 0.938 & $<4\times10^{29}$ \\
FRB 20230222A & 106.96036 & 11.22452 & 0.033 & 0.356 & 7.58 & $706.1$ & $4.5$ & $129.2$ & 0.985 & $<3\times10^{29}$ \\
FRB 20230222B & 238.73912 & 30.89870 & 0.033 & 0.192 & 9.89 & $187.8$ & $91.3$ & $37.5$ & 0.981 & $<2\times10^{29}$ \\
FRB 20230311A & 91.10966 & 55.94595 & 0.033 & 0.262 & 9.68 & $364.3$ & $9.5$ & $12.2$ & 0.774 & $<5\times 10^{29}$ \\
FRB 20230703A & 184.62445 & 48.72993 & 0.033 & 0.336 & 9.73 & $291.3$ & $9.4$ & $8.4$ & 0.973 & $<2\times10^{29}$ \\
FRB 20230730A & 54.66456 & 33.15930 & 0.033 & 0.204 & 9.91 & $312.5$ & $43.9$ & $37.1$ & 0.952 & $<8\times10^{29}$ \\
FRB 20230926A & 269.12488 & 41.81430 & 0.033 & 0.299 & 8.88 & $222.8$ & $26.8$ & $28.4$ & 0.996 & $ 4\times10^{28}$ \\
FRB 20231005A & 246.02800 & 35.44871 & 0.033 & 0.190 & 10.48 & $189.4$ & $221.2$ & $60.0$ & 0.985 & $<7\times10^{28}$ \\
FRB 20231007B & 307.53169 & 72.89235 & 0.033 & 0.195 & 7.29 & $202.4$ & $52.9$ & $138.9$ & 0.946 & - \\
FRB 20231011A & 18.24110 & 41.74910 & 0.033 & 0.192 & 13.73 & $186.3$ & $129.5$ & $58.3$ & 0.991 & $<1\times10^{29}$ \\
FRB 20231017A & 346.75429 & 36.65268 & 0.033 & 0.219 & 9.41 & $344.2$ & $33.6$ & $8.2$ & 0.918 & $<1\times10^{30}$ \\
FRB 20231025C & 71.97723 & 33.23799 & 0.033 & 0.258 & 9.23 & $510.4$ & $16.6$ & $3.9$ & 0.948 & - \\
FRB 20231025B & 270.78807 & 63.98908 & 0.033 & 0.427 & 9.95 & $368.7$ & $6.0$ & $5.2$ & 0.925 & $<2\times10^{30}$ \\
FRB 20231123A & 82.62325 & 4.47554 & 0.033 & 0.271 & 6.84 & $302.1$ & $52.7$ & $56.8$ & 0.956 & $<9\times10^{28}$ \\
FRB 20231128A & 199.57820 & 42.99271 & 0.033 & 0.498 & 9.90 & $331.6$ & $2.8$ & $67.7$ & 0.985 & $ 2\times10^{29}$ \\
FRB 20231201A & 54.58929 & 26.81767 & 0.033 & 0.338 & 0.52 & $169.4$ & $35.1$ & $10.8$ & 0.911 & - \\
FRB 20231202A & 65.03039 & 38.63328 & 0.033 & 0.220 & 9.05 & $437.7$ & $31.1$ & $33.8$ & 0.988 & - \\
FRB 20231204A & 207.99903 & 48.11600 & 0.033 & 0.436 & 9.49 & $221.0$ & $4.7$ & $25.5$ & 0.976 & $-^\dagger$ \\
FRB 20231206A & 112.44284 & 56.25627 & 0.033 & 0.462 & 11.33 & $457.7$ & $9.9$ & $179.4$ & 0.989 & $<5\times10^{28}$ \\
FRB 20231223C & 259.54465 & 29.49794 & 0.033 & 0.189 & 9.72 & $165.8$ & $68.1$ & $287.2$ & 0.953 & $<2\times10^{29}$ \\
FRB 20231229A & 26.46783 & 35.11292 & 0.033 & 0.190 & 10.05 & $198.5$ & $211.2$ & $95.5$ & 0.998 & $<8\times10^{26}$ \\
FRB 20231230A & 72.79761 & 2.39398 & 0.033 & 1.719 & 6.57 & $131.4$ & $6.6$ & $37.1$ & 0.918 & $<9\times10^{27}$ \\
\end{longtable}

\addtolength{\rightskip}{-4cm}
\label{tab:gold_sample}
\end{table*}

\begin{table*}
\caption{\msbold{A summary of the host galaxy candidates and optical follow-up conducted the systems listed in Table~\ref{tab:gold_sample}. In ``Observation'' we designate how the source was followed up, either with a reference to an archival redshift, within our dedicated optical follow-up programs at Lick, Gemini, and Keck, or not at all. A $\dagger$ in this column indicates that the sightline had sufficiently high extinction ($E(B-V) > 0.3$) to not warrant immediate follow up at this time.} Where relevant, we list the host redshift, the exact date of observation in UTC, the cataloged $r$-magnitude of the galaxy, and the exposure time in seconds.}
\begin{tabular}{lllccrr}
\hline
 TNS Name     & Host Candidate       & Observation  &   Redshift & Date &      $m_r$ &   Exposure \\
\hline 
 FRB 20230203A & J100639.32+354149.74 & Keck-2023B-3 &     0.1464 & 2023-12-14      &      17.92 &           900 \\
 FRB 20230222A & J070750.52+111322.36 & Lick-2023B-1 &     0.1223 & 2023-09-17      &      17.53 &          1200 \\
 FRB 20230222B & J155457.11+305359.55 & Lick-2023B-1 &     0.1100 & 2023-09-17      &      17.42 &          1200 \\
 FRB 20230311A & J060426.51+555641.63 & Keck-2023B-3 &     0.1918 & 2023-12-14      &      19.38 &          2700 \\
 FRB 20230703A & J121829.47+484330.12 & Keck-2023B-3 &     0.1184 & 2023-12-14      &      18.08 &           900 \\
 FRB 20230730A & J033839.58+330938.39 & Keck-2023B-3$\dagger$ &     0.2115 & 2023-12-14      &      21.41 &          2700 \\
 FRB 20230926A & J175629.80+414836.00 & Lick-2024A-1 &     0.0553 & 2024-05-09      &      15.46 &           900 \\
 FRB 20231005A & J162406.62+352649.16 & Lick-2024A-1 &     0.0713 & 2024-05-09      &      17.82 &           900 \\
 FRB 20231007B & J203007.06+725313.90 & Not observed$^\dagger$        &         -- &         --      &      18.76 &            -- \\
 FRB 20231011A & J011258.01+414456.38 & Keck-2023B-3 &     0.0783 & 2023-12-14      &      17.92 &           600 \\
 FRB 20231017A & J230700.80+363854.41 & Gemini-N-2024A-1 &     0.2450 & 2024-06-02       &      20.36 &          3600 \\
 FRB 20231025C & J044753.83+331415.10 & Not observed$^\dagger$        &         -- &              -- &      20.41 &            -- \\
 FRB 20231025B & J180308.65+635908.04 & Gemini-N-2024A-1    &     0.3238 &              2024-05-02  &      20.93 &            3600 \\
 FRB 20231123A & J053029.43+042827.78 & Lick-2024B-1 &     0.0729 & 2024-09-07      &      18.29 &          3600 \\
 FRB 20231128A & J131819.23+425958.97 & Lick-2024A-1 &     0.1079 & 2024-05-09      &      17.31 &           900 \\
 FRB 20231201A & J033821.02+264846.25 & Gemini-N-2024A-3 & 0.1119 & 2024-08-01      &      19.43 &          3600 \\
 FRB 20231202A & J042007.35+383758.32 & Not observed$^\dagger$        &         -- &              -- &      22.74 &            -- \\
 FRB 20231204A & J135159.86+480714.04 & Lick-2024A-1 &     0.0644 & 2024-05-09      &      15.40  &          900 \\
 FRB 20231206A & J072946.35+561511.68 & \citet{rines2000infall} & 0.0659 & -- & 16.12 & -- \\ 
 FRB 20231223C & J171810.66+292945.05 & Lick-2024A-1 & 0.1059 & 2024-05-09     &      16.49 & 900 \\
 FRB 20231229A & J014552.26+350628.19 & \citet{yu2022statistical} & 0.0190 & -- & 14.27 & -- \\
 FRB 20231230A & J045109.39+022205.02 & \citet{rines2003cairns}   & 0.0298 & -- & 14.65 & -- \\
\hline
\end{tabular}
\label{tab:z_table}
\end{table*}

\section{Acknowledgments}
We acknowledge that CHIME and the \kkoname\ Outrigger (KKO) are built on the traditional, ancestral, and unceded territory of the Syilx Okanagan people. \kkoname\ is situated on land leased from the Imperial Metals Corporation. We are grateful to the staff of the Dominion Radio Astrophysical Observatory, which is operated by the National Research Council of Canada. CHIME operations are funded by a grant from the NSERC Alliance Program and by support from McGill University, the University of British Columbia, and the University of Toronto. CHIME was funded by a grant from the Canada Foundation for Innovation (CFI) 2012 Leading Edge Fund (Project 31170) and by contributions from the provinces of British Columbia, Qu\'ebec, and Ontario. The CHIME/FRB Project was funded by a grant from the CFI 2015 Innovation Fund (Project 33213) and by contributions from the provinces of British Columbia and Qu\'ebec, and by the Dunlap Institute for Astronomy and Astrophysics at the University of Toronto. Additional support was provided by the Canadian Institute for Advanced Research (CIFAR), the Trottier Space Institute at McGill University, and the University of British Columbia. The CHIME/FRB baseband recording system is funded in part by a CFI John R. Evans Leaders Fund award to IHS.

The CHIME/FRB Outriggers program is funded by 
the Gordon and Betty Moore Foundation and by a National Science Foundation (NSF) grant (2008031).
The \kkoname~Outrigger is situated on land leased from the Imperial Metals Corporation.
FRB research at MIT is supported by an NSF grant (2008031).
FRB research at WVU is supported by an NSF grant (2006548, 2018490). We are grateful to Robert Kirshner for early support and encouragement in the Outriggers project.

The Pan-STARRS1 Surveys (PS1) and the PS1 public science archive have been made possible through contributions by the Institute for Astronomy, the University of Hawaii, the Pan-STARRS Project Office, the Max-Planck Society and its participating institutes, the Max Planck Institute for Astronomy, Heidelberg and the Max Planck Institute for Extraterrestrial Physics, Garching, The Johns Hopkins University, Durham University, the University of Edinburgh, the Queen's University Belfast, the Harvard-Smithsonian Center for Astrophysics, the Las Cumbres Observatory Global Telescope Network Incorporated, the National Central University of Taiwan, the Space Telescope Science Institute, the National Aeronautics and Space Administration under Grant No. NNX08AR22G issued through the Planetary Science Division of the NASA Science Mission Directorate, the National Science Foundation Grant No. AST-1238877, the University of Maryland, Eotvos Lorand University (ELTE), the Los Alamos National Laboratory, and the Gordon and Betty Moore Foundation.

The Fast and Fortunate for FRB Follow-up team acknowledges support from NSF grants AST-1911140, AST-1910471, and AST-2206490.

The Legacy Surveys consist of three individual and complementary projects: the Dark Energy Camera Legacy Survey (DECaLS; Proposal ID \#2014B-0404; PIs: David Schlegel and Arjun Dey), the Beijing-Arizona Sky Survey (BASS; NOAO Prop. ID \#2015A-0801; PIs: Zhou Xu and Xiaohui Fan), and the Mayall z-band Legacy Survey (MzLS; Prop. ID \#2016A-0453; PI: Arjun Dey). DECaLS, BASS, and MzLS together include data obtained, respectively, at the Blanco telescope, Cerro Tololo Inter-American Observatory, NSF’s NOIRLab; the Bok telescope, Steward Observatory, University of Arizona; and the Mayall telescope, Kitt Peak National Observatory, NOIRLab. 

Pipeline processing and analyses of the data were supported by NOIRLab and the Lawrence Berkeley National Laboratory (LBNL). The Legacy Surveys project is honored to be permitted to conduct astronomical research on Iolkam Du'ag (Kitt Peak), a mountain with particular significance to the Tohono O'odham Nation.

NOIRLab is operated by the Association of Universities for Research in Astronomy (AURA) under a cooperative agreement with the National Science Foundation. LBNL is managed by the Regents of the University of California under contract to the U.S. Department of Energy.

This project used data obtained with the Dark Energy Camera (DECam), which was constructed by the Dark Energy Survey (DES) collaboration. Funding for the DES Projects has been provided by the U.S. Department of Energy, the U.S. National Science Foundation, the Ministry of Science and Education of Spain, the Science and Technology Facilities Council of the United Kingdom, the Higher Education Funding Council for England, the National Center for Supercomputing Applications at the University of Illinois at Urbana-Champaign, the Kavli Institute of Cosmological Physics at the University of Chicago, Center for Cosmology and Astro-Particle Physics at the Ohio State University, the Mitchell Institute for Fundamental Physics and Astronomy at Texas A\&M University, Financiadora de Estudos e Projetos, Fundacao Carlos Chagas Filho de Amparo, Financiadora de Estudos e Projetos, Fundacao Carlos Chagas Filho de Amparo a Pesquisa do Estado do Rio de Janeiro, Conselho Nacional de Desenvolvimento Cientifico e Tecnologico and the Ministerio da Ciencia, Tecnologia e Inovacao, the Deutsche Forschungsgemeinschaft and the Collaborating Institutions in the Dark Energy Survey. The Collaborating Institutions are Argonne National Laboratory, the University of California at Santa Cruz, the University of Cambridge, Centro de Investigaciones Energeticas, Medioambientales y Tecnologicas-Madrid, the University of Chicago, University College London, the DES-Brazil Consortium, the University of Edinburgh, the Eidgenossische Technische Hochschule (ETH) Zurich, Fermi National Accelerator Laboratory, the University of Illinois at Urbana-Champaign, the Institut de Ciencies de l’Espai (IEEC/CSIC), the Institut de Fisica d’Altes Energies, Lawrence Berkeley National Laboratory, the Ludwig Maximilians Universitat Munchen and the associated Excellence Cluster Universe, the University of Michigan, NSF’s NOIRLab, the University of Nottingham, the Ohio State University, the University of Pennsylvania, the University of Portsmouth, SLAC National Accelerator Laboratory, Stanford University, the University of Sussex, and Texas A\&M University.

BASS is a key project of the Telescope Access Program (TAP), which has been funded by the National Astronomical Observatories of China, the Chinese Academy of Sciences (the Strategic Priority Research Program “The Emergence of Cosmological Structures” Grant \# XDB09000000), and the Special Fund for Astronomy from the Ministry of Finance. The BASS is also supported by the External Cooperation Program of the Chinese Academy of Sciences (Grant \# 114A11KYSB20160057), and the Chinese National Natural Science Foundation (Grant \# 12120101003, \# 11433005).

The Legacy Survey team makes use of data products from the Near-Earth Object Wide-field Infrared Survey Explorer (NEOWISE), which is a project of the Jet Propulsion Laboratory/California Institute of Technology. NEOWISE is funded by the National Aeronautics and Space Administration.

The Legacy Surveys imaging of the DESI footprint is supported by the Director, Office of Science, Office of High Energy Physics of the U.S. Department of Energy under Contract No. DE-AC02-05CH1123, by the National Energy Research Scientific Computing Center, a DOE Office of Science User Facility under the same contract; and by the U.S. National Science Foundation, Division of Astronomical Sciences under Contract No. AST-0950945 to NOAO.

The Photometric Redshifts for the Legacy Surveys (PRLS) catalog used in this paper was produced thanks to funding from the U.S. Department of Energy Office of Science, Office of High Energy Physics via grant DE-SC0007914.

This work has made use of data from the European Space Agency (ESA) mission
{\it Gaia} (\url{https://www.cosmos.esa.int/gaia}), processed by the {\it Gaia}
Data Processing and Analysis Consortium (DPAC,
\url{https://www.cosmos.esa.int/web/gaia/dpac/consortium}). Funding for the DPAC
has been provided by national institutions, in particular, the institutions
participating in the {\it Gaia} Multilateral Agreement.

This research has made use of the CIRADA cutout service at \texttt{cutouts.cirada.ca}, operated by the Canadian Initiative for Radio Astronomy Data Analysis (CIRADA). CIRADA is funded by a grant from the Canada Foundation for Innovation 2017 Innovation Fund (Project 35999), as well as by the Provinces of Ontario, British Columbia, Alberta, Manitoba, and Qu\'ebec, in collaboration with the National Research Council of Canada, the US National Radio Astronomy Observatory and Australia’s Commonwealth Scientific and Industrial Research Organisation. 
\vspace{5mm}
\facilities{CHIME, Shane (KAST), Gemini (GMOS), Keck I (DEIMOS),
}

\software{
\texttt{Astropy} \citep{collaboration2013astropy,collaboration2018astropy},
\texttt{difxcalc}~\citet{gordon2016difxcalc},
\texttt{pyfx}~\citep{leung2024vlbi},
\texttt{PypeIt}~\citep{prochaska2020pypeit},
\texttt{FFFF-PZ}~\citep{coulter2022yse,Coulter2023}
}
\appendix
\section{Astrometric Calibration}\label{sec:accuracy}
\subsection{CHIME Baseband Localization: Accuracy of Quoted Declination Uncertainties}
Since the CHIME-only baseband localization plays an important role in establishing the north-south position of the source, we validate the baseband pipeline using a sample of $100$ single pulses from $18$ distinct pulsars that emitted single pulses captured by the baseband system at CHIME whose positions have been independently determined using e.g. our Very Long Baseline Array program (Curtin et al., in prep). The pulses cover a wide range of detected S/N values ($5 \leq \text{S/N} \leq 100$) and declinations ($+6^\circ \leq \delta \leq +80^\circ$). This range completely covers the range of S/N ratios spanned by the FRB data and the range of declinations of all but two of our host galaxies (FRB 20231123A and FRB 20230222A).

Unfortunately, the lack of bright, single-pulse emitting pulsars in the range $-3^\circ\leq \delta \leq +6^\circ$ limits our ability to robustly test baseband localisations within this region of CHIME's primary beam. The Crab pulsar -- PSR B0531+21 -- as well as B1642$-$03 -- are omitted from the sample of pulsars used to calibrate the baseband localization. Baseband localisations of the Crab are systematically offset by $\sim 2 ~\mathrm{arcmin}$ West of of CHIME's primary beam. We suspect this offset arises from confusion with the surrounding, radio-bright nebula whose angular extent of roughly $5~\mathrm{arcmin}$ in declination at $5~\mathrm{GHz}$~\citep{Bietenholz_2004} makes it semi-resolved by the short internal baselines of CHIME. We do not observe this behavior in any other pulsar at similar declinations to the Crab and therefore conclude that the Crab is an outlier. We have also identified instances for which pulses originating from PSR B1642$-$03 are poorly baseband localized for reasons that are not yet entirely well understood. We suspect that imperfect characterization of CHIME's thermal expansion is likely the driving factor at these very low declinations~\citep{michilli2021analysis}. 

\begin{figure*}
    \centering
    \includegraphics[width=1.0\linewidth]{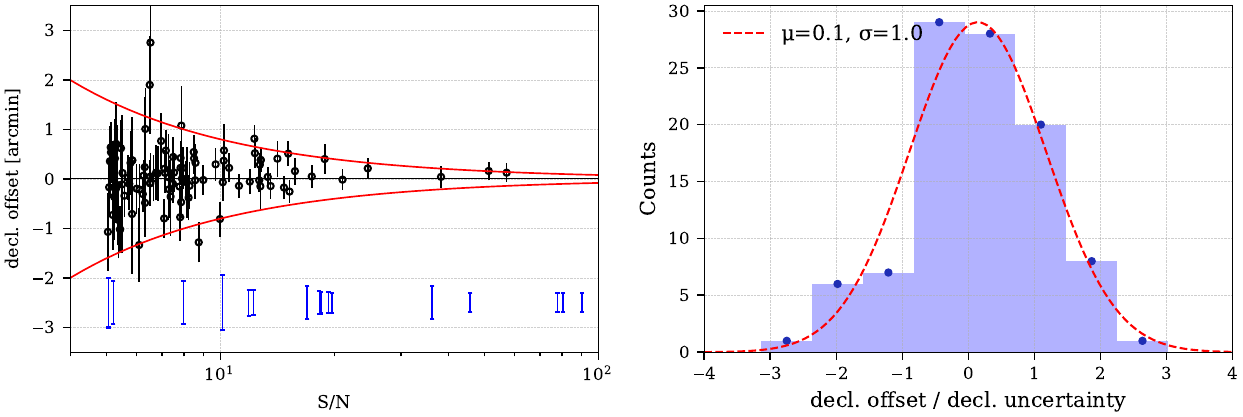}
    \caption{Validation of declination uncertainties outputted by the CHIME/FRB baseband localisation pipeline~\citep{michilli2021analysis}. \textit{Left panel:} Declination offsets as a function of detection S/N for a sample of 100 pulses from 18 unique pulsars covering a declination range $+6^\circ\leq\delta\leq+80^\circ$. Offsets and their associated uncertainties are represented by open-face, black markers and errorbars, respectively. For reference, the measured S/N  and declination uncertainties associated with FRBs presented in this work are indicated by the blue error bars at a fiducial decl. offset of $-2.5$ arcmin. Theoretical uncertainty for a burst detected at zenith by CHIME/FRB as a function of S/N is depicted by the two red curves~\citep{masui2019algorithms,michilli2021analysis}. \textit{Right:} Histogram of measured declination offsets normalized by their outputted uncertainties. As expected, the normalized values are well described by a Gaussian distribution with a mean of 0.1 and a standard deviation of 1.0 (overplotted dashed, red line).}
    \label{fig:bb_pulsars}
\end{figure*}

For the remaining set of 100 pulses, we plot a summary of the baseband localizations performed to validate the uncertainties along the declination axis in Fig.~\ref{fig:bb_pulsars}. In the left panel, we plot the measured declination offset along with its associated uncertainty as a function of detection S/N for each pulse in black. We overplot the theoretical uncertainty for an FRB detected at $\nu = 600~\mathrm{MHz}$ at zenith by CHIME/FRB in red. This uncertainty increases as the angle from the zenith increases~\citep[$\sigma_\theta\propto 1/\mathrm{cos}\ \theta$][]{michilli2021analysis}; hence some values tend away from their theoretical prediction. For the FRBs presented in this paper, we plot measured S/N and associated declination uncertainties in blue, offset downward by $-2.5^\circ$, for visual comparison with pulsars with a similar range of S/N values. In the right panel, we plot a histogram of the declination offsets normalised by their uncertainties. Assuming that the baseband localisation pipeline is well-calibrated, we expect the distribution of values to be Gaussian distributed with a mean of 0 and a standard deviation of 1; we measure a mean of 0.1 and standard deviation of 1.0, suggesting that the quoted uncertainties are well-calibrated to the astrometric error in the declination direction over the aforementioned range of detected S/N and declinations. 

\subsection{VLBI Localization: Declination-dependent astrometric accuracy and impact of calibrator selection}
\begin{figure*}
    \centering	\includegraphics[width=1.01\linewidth]{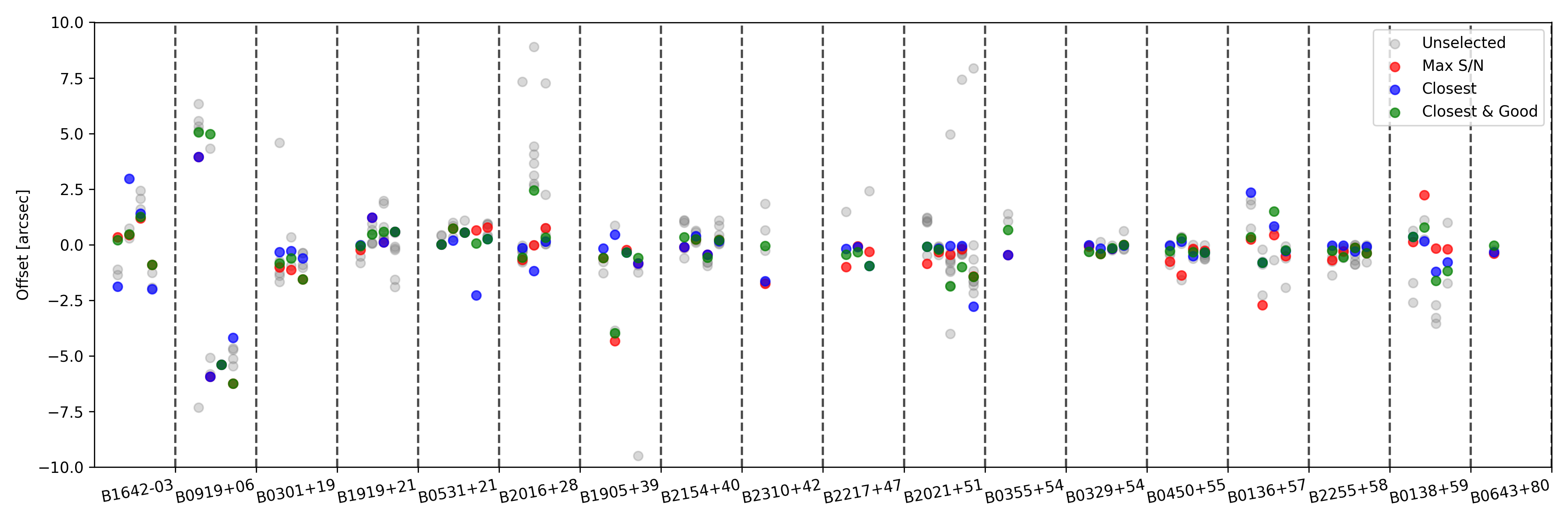}
	\caption{Offsets between measured and true pulsar positions, along the CHIME-KKO baseline direction, for a sample of pulsars. Each point is the offset of the measured position from the true position using a different in-beam calibrator. Each column shows offsets for a given baseband acquisition, grouped by target pulsar. The color of the point indicates which calibrator would be selected by each selection strategy -- red for maximum S/N, blue for the closest in declination, and green for our chosen strategy of selecting the closest calibrator that is within 2\,ns from the median. Gray points indicate calibrators that do not qualify under any of these strategies.}
	\label{fig:psr_cal_test}
\end{figure*}

We perform a similar validation check of our VLBI localization procedure, using 62 single pulse localizations of 18 distinct pulsars at a variety of declinations (see Fig.~\ref{fig:psr_cal_test}). Here, we use the abundance of astrometric calibrators from the Radio Fundamental Catalogue and Very Long Baseline Array (VLBA) calibrator grid~\citep{ma1998international,petrov2021wide} at 600 MHz~\citep{andrew2024vlbi} to explore the impact of the often ad-hoc choice of VLBI calibrator on the resulting astrometry. We find few-percent variations in the sample-averaged RMS localization accuracy. 

For instance, selecting the closest calibrator minimizes the impact of ionospheric and beam variations, but choosing the calibrator that maximizes the S/N of the target after applying delay and phase solutions can ensure a more robust detection of the FRB. However, under either rule, it is possible to choose a calibrator that is a confused source. Since we expect all sources to have a similar total instrumental delay, we can reduce the odds of that happening by comparing the delays of all calibrators in the field, which we use to identify outliers. We thus adopt the following strategy for calibrator selection: we select only those whose measured delays are within 2~ns of the median delay calculated over all calibrators. From the remaining set we choose the calibrator closest in declination to the baseband localization of the FRB.

\begin{figure}
    \centering
    \includegraphics[width=0.5\linewidth]{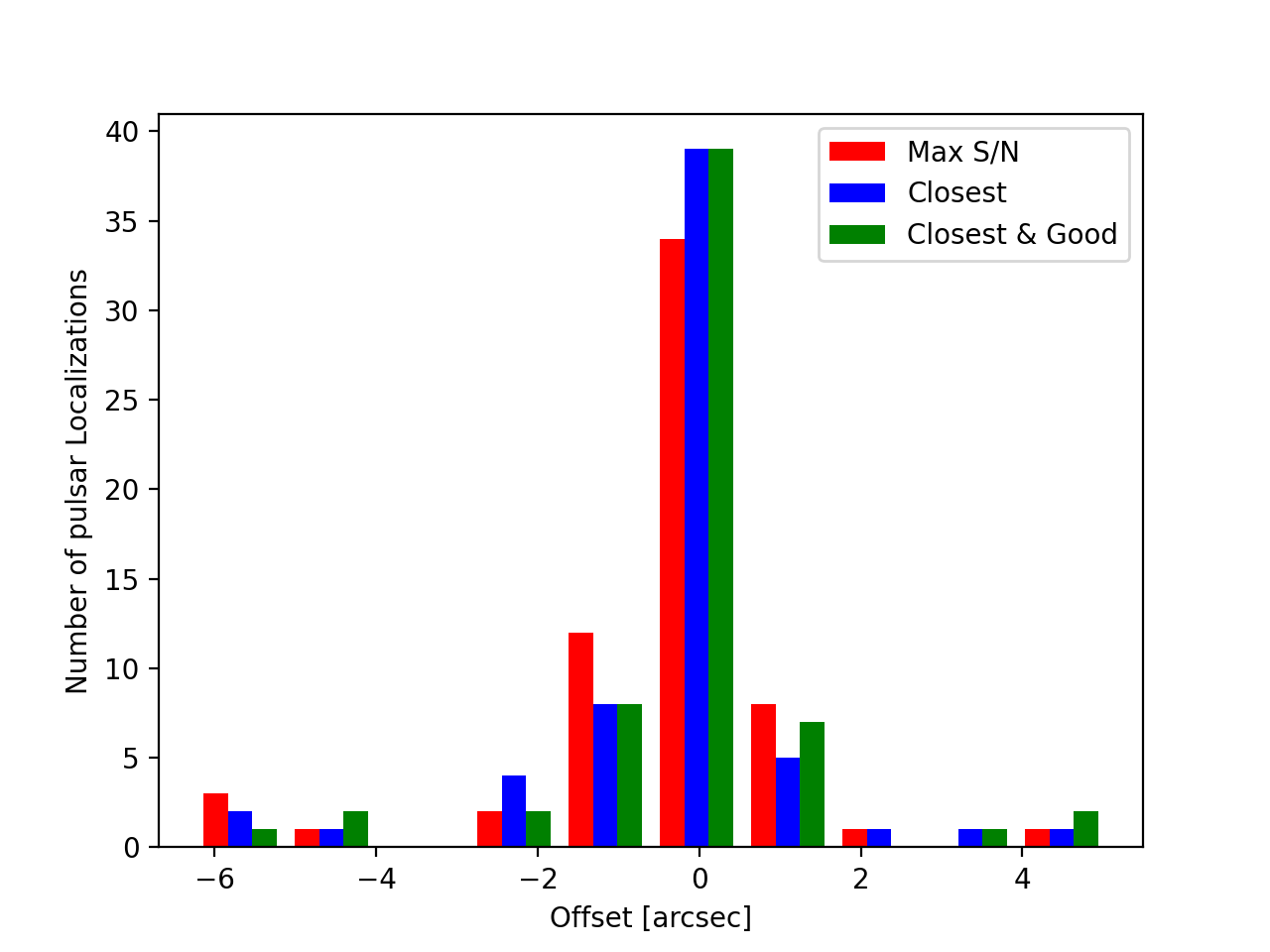}
    \caption{The offset distribution of the pulsar localizations in our sample of test pulsars, plotted here including low-declination sources. At the $1\sigma$ level the distribution has RMS values of 1.5'', 1.6'', and 1.54'' for the ``Closest'', ``max-S/N'', and ``Closest \& good'' strategies respectively. On the basis of these statistics and the overall false-positive fraction of the host associations in our sample, we choose a 1$\sigma$ astrometric error of 2''. At higher declinations, if the localizations of B0919+06 are omitted, they are 0.919'', 0.936'', and 0.943'' respectively, further establishing that this is a conservative choice.}
    \label{fig:psr_offsets}
\end{figure}

To compare this with other strategies, we perform localizations using all detected VLBI calibrators for a set of pulsars with known positions, and compare the residuals between the measured and true positions. Figure~\ref{fig:psr_cal_test} shows the offsets for a set of between 1 and 5 observations of 18 pulsars. For a fixed x-coordinate on the plot, each column of points represents a single observation. Each point in the column represents the measured astrometric offset for a particular in-beam calibrator. The gray points are the calibrators that were not chosen, while the colored points indicate the offsets measured with the calibrator chosen using one of three selection criteria. ``Closest \& Good'' refers to the strategy described in the previous paragraph, while ``S/N'' and ``Closest'' refer to using the calibrator that maximizes the S/N of the phase-referenced target visibilities, and the closest calibrator on the sky in declination, respectively.
Fig.~\ref{fig:psr_offsets} shows the offset distribution for pulsars localized as a function of which strategy is used. Their RMS values are quite similar, with a few outliers originating from pulsars at low declinations. This result is encouraging for next-generation FRB telescopes, which will rely on similar techniques to pinpoint FRBs: these include CHIME's successor CHORD~\citep{vanderlinde2019canadian}, as well as other compact arrays being built as HIRAX~\citep{newburgh2016hirax}, BURSTT~\citep{lin2022burstt}, and CASPA~\citep{luo2024fast}.  

We have conservatively inflated our $1\sigma$ astrometric error bars to $2''$ for the whole sample on the basis of the analysis summarized in Fig.~\ref{fig:psr_offsets}. Therefore, we regard the host associations in this paper as highly secure. In addition, for the two systems at declinations lower than $21^\circ$ (FRB 20231123A and FRB 20230222A), the host association is still secure ($P(O|x) > 0.9$) when the astrometric uncertainty is inflated further yet to $2.5''$, though we believe this to be unnecessarily conservative.

We note that any individual host in the sample may be misidentified with a false-positive rate of $\sim 10\%$ as described in \S\ref{sec:association}. The effect on population-level statistics of the full sample due to astrometric errors is likely to be minimal due to the small number of low-declination sources.

\begin{longtable}{lllllllccll}
\caption{The remaining FRB localizations, in the same format as Table~\ref{tab:gold_sample}.} \label{fig:full_sample} \\
\toprule
$\text{Name}$ & $\text{RA}_{\text{FRB}}$ & $\text{DEC}_{\text{FRB}}$ & $b_{\text{err}}$ & $a_{\text{err}}$ & Angle & $\text{DM}$ & $\text{Flux}$ & $\text{Fluence}$ & $\text{P(O} \vert \text{x)}$ \\
\midrule
\endfirsthead
\caption[]{The remaining FRB localizations, in the same format as Table~\ref{tab:gold_sample}.} \\
\toprule
$\text{Name}$ & $\text{RA}_{\text{FRB}}$ & $\text{DEC}_{\text{FRB}}$ & $b_{\text{err}}$ & $a_{\text{err}}$ & Angle & $\text{DM}$ & $\text{Flux}$ & $\text{Fluence}$ & $\text{P(O} \vert \text{x)}$ \\
\midrule
\endhead
\midrule
\multicolumn{10}{r}{Continued on next page} \\
\midrule
\endfoot
\bottomrule
\endlastfoot
FRB 20221209A & 297.36956 & 59.40618 & 0.033 & 0.215 & 7.03 & $792.2$ & $388.2$ & $489.6$ & 0.709 \\
FRB 20230314A & 310.28879 & 6.95568 & 0.033 & 0.188 & 7.33 & $429.0$ & $298.6$ & $139.4$ & 0.615 \\
FRB 20230317A & 178.89504 & 72.35246 & 0.033 & 0.279 & 9.98 & $696.7$ & $10.8$ & $97.5$ & 0.421 \\
FRB 20230409A & 148.54862 & 61.14785 & 0.033 & 0.550 & 7.31 & $408.8$ & $28.4$ & $72.7$ & 0.684 \\
FRB 20230410A & 214.77745 & 25.45794 & 0.033 & 0.361 & 9.19 & $643.4$ & $7.9$ & $12.8$ & 0.717 \\
FRB 20230411A & 280.05738 & 47.49506 & 0.033 & 0.198 & 9.46 & $551.2$ & $45.5$ & $16.7$ & 0.032 \\
FRB 20230414A & 304.47601 & 84.89969 & 0.033 & 0.238 & 6.01 & $501.0$ & $73.3$ & $11.8$ & 0.012 \\
FRB 20230414B & 11.10581 & 51.81368 & 0.033 & 0.219 & 10.42 & $433.9$ & $23.2$ & $7.6$ & 0.001 \\
FRB 20230615A & 78.12933 & 61.66923 & 0.033 & 0.573 & 8.77 & $695.9$ & $4.3$ & $19.2$ & 0.857 \\
FRB 20230616A & 184.61458 & 28.35665 & 0.033 & 0.283 & 8.60 & $918.3$ & $31.9$ & $83.8$ & 0.231 \\
FRB 20230702A & 132.12469 & 26.38950 & 0.033 & 0.254 & 8.42 & $598.9$ & $29.8$ & $52.6$ & 0.538 \\
FRB 20230720A & 4.45444 & 21.27125 & 0.033 & 0.228 & 8.26 & $533.0$ & $29.0$ & $433.2$ & 0.375 \\
FRB 20230802A & 339.28450 & 30.78872 & 0.033 & 0.229 & 9.05 & $305.3$ & $22.8$ & $69.3$ & 0.382 \\
FRB 20230805A & 36.01767 & 52.69549 & 0.033 & 0.440 & 10.36 & $638.5$ & $7.2$ & $15.6$ & 0.240 \\
FRB 20230828A & 22.28703 & 34.42965 & 0.033 & 0.208 & 10.18 & $377.3$ & $50.7$ & $338.9$ & 0.000 \\
FRB 20230911A & 49.64797 & 2.66713 & 0.033 & 0.189 & 6.67 & $351.4$ & $368.8$ & $773.5$ & 0.687 \\
FRB 20230913A & 131.90375 & 87.99511 & 0.033 & 0.203 & -6.17 & $251.1$ & $336.8$ & $92.0$ & 0.000 \\
FRB 20230918A & 161.05073 & 61.12086 & 0.033 & 0.194 & 8.44 & $366.0$ & $88.6$ & $58.2$ & 0.185 \\
FRB 20230923A & 206.61410 & 45.40801 & 0.033 & 0.230 & 9.46 & $918.3$ & $23.3$ & $34.0$ & 0.747 \\
FRB 20230924A & 193.36694 & 42.57986 & 0.033 & 0.373 & 10.02 & $991.1$ & $8.0$ & $9.7$ & 0.457 \\
FRB 20230928A & 104.40755 & 60.57290 & 0.033 & 0.186 & 7.43 & $533.8$ & $224.8$ & $9942.7$ & 0.438 \\
FRB 20231005B & 344.38055 & 15.22388 & 0.033 & 0.313 & 8.08 & $640.9$ & $16.3$ & $33.5$ & 0.267 \\
FRB 20231006B & 109.61997 & 84.77614 & 0.033 & 0.267 & -1.92 & $1252.3$ & $68.2$ & $20.6$ & 0.262 \\
FRB 20231007A & 185.38497 & 79.88514 & 0.033 & 0.409 & 4.24 & $819.7$ & $13.7$ & $36.7$ & 0.873 \\
FRB 20231009A & 211.75689 & 17.21141 & 0.033 & 0.851 & 8.04 & $391.0$ & $9.1$ & $46.1$ & 0.227 \\
FRB 20231014A & 343.76040 & 54.49522 & 0.033 & 0.329 & 10.66 & $860.9$ & $12.2$ & $40.5$ & 0.034 \\
FRB 20231018A & 113.95251 & 67.61565 & 0.033 & 0.420 & 12.94 & $546.7$ & $23.4$ & $805.0$ & 0.680 \\
FRB 20231019A & 19.10358 & 20.62750 & 0.033 & 0.520 & 8.21 & $246.7$ & $11.5$ & $53.6$ & 0.392 \\
FRB 20231019B & 216.53182 & 44.50831 & 0.033 & 0.253 & 9.50 & $909.4$ & $176.1$ & $162.9$ & 0.070 \\
FRB 20231020A & 76.12081 & 72.07889 & 0.033 & 0.663 & 11.04 & $267.3$ & $7.0$ & $18.3$ & 0.571 \\
FRB 20231028A & 249.46599 & 80.08290 & 0.033 & 2.534 & 9.73 & $585.9$ & $32.4$ & $66.2$ & 0.223 \\
FRB 20231028B & 151.36185 & 54.37889 & 0.033 & 0.187 & 9.10 & $333.6$ & $74.9$ & $64.1$ & 0.891 \\
FRB 20231101A & 135.39422 & 0.94179 & 0.033 & 0.280 & 6.41 & $412.1$ & $33.1$ & $215.8$ & 0.395 \\
FRB 20231102A & 228.83742 & 56.79919 & 0.033 & 0.250 & 9.85 & $906.2$ & $15.2$ & $12.0$ & 0.350 \\
FRB 20231104B & 135.10516 & 46.33993 & 0.033 & 0.600 & 9.24 & $382.9$ & $3.7$ & $18.8$ & 0.392 \\
FRB 20231110A & 8.89710 & 15.83719 & 0.033 & 0.384 & 8.34 & $448.1$ & $7.2$ & $81.1$ & 0.405 \\
FRB 20231110B & 171.61679 & 35.80670 & 0.033 & 0.529 & 9.07 & $726.7$ & $6.0$ & $152.8$ & 0.338 \\
FRB 20231118A & 312.39187 & 37.92352 & 0.033 & 0.343 & 10.07 & $473.1$ & $7.3$ & $109.1$ & 0.000 \\
FRB 20231118B & 213.67978 & 72.35021 & 0.033 & 0.280 & 4.14 & $317.4$ & $180.1$ & $99.0$ & 0.877 \\
FRB 20231123C & 4.08948 & 20.02933 & 0.033 & 0.227 & 8.32 & $304.5$ & $59.6$ & $8.9$ & 0.332 \\
FRB 20231126A & 21.81603 & 89.28894 & 0.033 & 0.985 & -6.85 & $673.3$ & $23.3$ & $15.5$ & 0.122 \\
FRB 20231127A & 36.83517 & 58.90870 & 0.033 & 0.238 & 11.15 & $412.5$ & $17.5$ & $185.7$ & 0.666 \\
FRB 20231203A & 113.72603 & 52.73416 & 0.033 & 1.026 & 9.34 & $335.2$ & $15.0$ & $10.6$ & 0.250 \\
FRB 20231209A & 180.90951 & 50.07879 & 0.033 & 0.365 & 9.61 & $1232.1$ & $11.8$ & $36.3$ & 0.505 \\
FRB 20231210A & 13.13264 & 16.30105 & 0.033 & 0.828 & 3.11 & $482.4$ & $24.3$ & $11.0$ & 0.509 \\
FRB 20231210B & 59.66792 & 27.02746 & 0.033 & 0.378 & 9.16 & $505.9$ & $8.9$ & $17.3$ & 0.653 \\
FRB 20231214A & 143.54611 & 6.36881 & 0.033 & 0.219 & 5.12 & $243.8$ & $83.8$ & $19.2$ & 0.774 \\
FRB 20231219A & 284.74183 & 89.20394 & 0.033 & 0.620 & -8.30 & $603.7$ & $13.3$ & $12.3$ & 0.011 \\
FRB 20231223B & 78.69951 & 48.75465 & 0.033 & 1.403 & 8.81 & $499.4$ & $10.7$ & $15.0$ & 0.260 \\
FRB 20231223D & 144.00728 & 82.10684 & 0.033 & 0.705 & -3.66 & $729.0$ & $68.0$ & $131.6$ & 0.208 \\
FRB 20231223A & 206.97352 & 16.32940 & 0.033 & 0.344 & 8.40 & $866.6$ & $16.9$ & $38.7$ & 0.608 \\
FRB 20231224A & 131.45986 & 72.91279 & 0.033 & 0.377 & 9.71 & $683.0$ & $20.1$ & $13.2$ & 0.082 \\
FRB 20231224B & 172.97016 & 82.87272 & 0.033 & 0.289 & 2.58 & $373.2$ & $271.9$ & $374.1$ & 0.607 \\
FRB 20231225C & 211.72144 & 74.44913 & 0.033 & 1.830 & 6.57 & $515.6$ & $22.5$ & $81.7$ & 0.217 \\
FRB 20231225B & 54.87781 & -3.06619 & 0.033 & 1.242 & 5.87 & $227.7$ & $10.4$ & $65.6$ & 0.813 \\
FRB 20231225A & 135.60297 & 33.91857 & 0.033 & 0.792 & 9.87 & $646.6$ & $14.7$ & $25.3$ & 0.478 \\
FRB 20231225D & 330.33719 & 33.95085 & 0.033 & 0.411 & 8.90 & $473.1$ & $8.8$ & $75.1$ & 0.777 \\
FRB 20231230B & 355.77282 & 78.08919 & 0.033 & 0.275 & 6.54 & $476.8$ & $78.3$ & $46.1$ & 0.827 \\
FRB 20231231A & 342.40187 & 16.42678 & 0.033 & 0.189 & 8.58 & $1247.5$ & $204.4$ & $281.0$ & 0.373 \\
FRB 20240210B & 222.52186 & 81.56315 & 0.033 & 0.290 & 4.84 & $630.4$ & $46.5$ & $241.1$ & 0.552 \\
FRB 20240210C & 325.60494 & 2.36546 & 0.033 & 1.197 & 6.68 & $380.9$ & $44.2$ & $369.6$ & 0.503 \\
\end{longtable}

\bibliographystyle{aasjournal}
\bibliography{references}



\end{CJK*}
\end{document}